\documentclass[a4paper, 11pt]{article}

\usepackage{interference}

\usepackage[pagewise]{lineno} 

\usepackage{setspace} 

\usepackage{multirow}

\usepackage{authblk}

\usepackage{booktabs}
\usepackage{graphicx}
\usepackage[table,xcdraw]{xcolor}

\hypersetup{
    colorlinks=true,
    citecolor=blue,
    linkcolor=red,
    urlcolor=blue,
    filecolor=blue
}

\usepackage{algorithm}
\usepackage{algpseudocode}

\usepackage{amsthm}

\theoremstyle{plain} 
\newtheorem{theorem}{Theorem}
\newtheorem{proposition}{Proposition}
\newtheorem{lemma}{Lemma}

\newtheorem{assumption}{Assumption}

\theoremstyle{definition} 
\newtheorem{definition}{Definition}

\theoremstyle{remark} 
\newtheorem{example}{Example}

\allowdisplaybreaks

\newcommand{\blind}{1}

\usepackage[margin = 1in]{geometry}

\title{\textbf{Nonparametric Causal Survival Analysis with Clustered Interference}}

    \author[1]{Chanhwa Lee  \thanks{Corresponding author: Chanhwa Lee. Email: chanhwaacademic@gmail.com.}}
    \author[2]{Donglin Zeng}
    \author[3]{Michael Emch}
    \author[4]{John D. Clemens}
    \author[5]{Michael G. Hudgens}

    \affil[1]{\small Google}
    \affil[2]{Department of Biostatistics, University of Michigan Ann Arbor}
    \affil[3]{Departments of Geography and Epidemiology, University of North Carolina at Chapel Hill}
    \affil[4]{Fielding School of Public Health, University of California Los Angeles}
    \affil[5]{Department of Biostatistics, University of North Carolina at Chapel Hill}

\begin{document}

\def\spacingset#1{\renewcommand{\baselinestretch}%
{#1}\small\normalsize} \spacingset{1}


\maketitle

\begin{abstract}
    Inferring treatment effects on a survival time outcome based on data from an observational study is challenging due to the presence of censoring and possible confounding. An additional challenge occurs when a unit's treatment affects the outcome of other units, i.e., there is interference. In some settings, it is reasonable to assume interference only occurs within clusters of units, i.e., there is clustered interference. In this paper, nonparametric methods are developed which can accommodate confounding, censored outcomes, and clustered interference. The approach avoids parametric assumptions and permits inference about counterfactual scenarios corresponding to stochastic policies which modify the propensity score distribution, and thus may have application across diverse settings. The proposed nonparametric cross-fitting estimators allow for flexible data-adaptive estimation of nuisance functions and are consistent and asymptotically normal with parametric convergence rates. Simulation studies demonstrate the finite sample performance of the proposed estimators, and the methods are applied to a cholera vaccine study in Bangladesh.
\end{abstract}

\noindent%

\spacingset{1.2} 

\section{Introduction}

Drawing inference about the effect of a treatment or exposure from observational data is challenging due to potential confounding. When the outcome of interest is a survival time (i.e., time to event), the presence of censoring introduces an additional complication. A third analytical challenge may occur when there is \textit{interference} \citep{cox58}, i.e., the treatment of a unit (individual) affects other units’ outcomes. Interference can occur where units interact with each other; e.g., an individual receiving a cholera vaccine may protect their family members from cholera \citep{ali2005herd}, an initiative that encourages students to attend classes may influence the attendance rates of their siblings \citep{barrera11}, and the academic performance of a student may be affected by their friends’ alcohol consumption \citep{qu22efficient}. This paper presents methods that accommodate confounding, censored outcomes, and interference.

In some settings it may be reasonable to assume there is \textit{clustered (or partial) interference} \citep{sobel06, hudgens08}, i.e., units can be partitioned into clusters such that a unit’s potential outcomes may depend on the others’ treatment in the same cluster but not on others in different clusters. Clusters may be defined based on the spatial or temporal separation of units, such as schools \citep{hong2006evaluating}, households \citep{park2022efficient}, or villages \citep{kilpatrick24gformula}. 

Under clustered interference, various causal estimands have been proposed to characterise treatment effects under different counterfactual scenarios. For instance, it may be of interest to compare the risk of cholera in a village when 70\% of individuals within the village are vaccinated compared to when 30\% are vaccinated, or to estimate the risk of cholera when an individual is unvaccinated and 50\% of their neighbors are vaccinated. Such estimands can be represented by the expected potential outcome when units receive treatment according to a specific treatment allocation policy. The Type B policy \citep{tchetgen12}, where units independently select treatment with the same probability, commonly serves as the basis for defining target causal estimands in the presence of interference due to its simplicity and ease of interpretation  \citep{perez14, liu19, park2022efficient}. However, the Type B policy may not be relevant in settings where the propensity for treatment differs across individuals. \citet{Papadogeorgou19, barkley20, kilpatrick24gformula}, and \citet{lee2025efficient} proposed inferential methods for more general policies based on a shift or modification of the distribution of propensity scores, which allow for heterogeneous propensity scores depending on individual- and cluster-level characteristics. However, these methods do not allow for censored outcomes. 

On the other hand, various methods have been developed which permit causal inference from observational data with censored outcomes \citep[e.g.,][]{robins2000correcting, rotnitzky2005inverse, ozenne2020estimation, cui2023estimating}. However, these approaches do not allow for interference between individuals. While \citet{chakladar2022inverse} allows for confounding, censored outcomes, and clustered interference, their approach utilizes parametric models and is limited to Type B estimands.

In this paper, flexible inference procedures are developed that account for confounding, clustered interference, and right censoring. The methods are general in the sense that they can be applied to any stochastic policy that modifies the propensity score distribution and, thus, are potentially relevant across diverse settings. The proposed nonparametric sample splitting estimators permit flexible data-adaptive estimation of nuisance functions, and are consistent and asymptotically normal, converging at the usual parametric rate. Furthermore, under mild conditions, the proposed methods yield estimators for the counterfactual survival function which weakly converges to a Gaussian process.

The proposed methods are evaluated in simulation studies and then applied to a cholera vaccine study in Bangladesh \citep{clemens1988field}. Prior analyses of the vaccine’s effect \citep[e.g.,][]{perez14} suggest possible interference within baris (i.e., clusters of patrilineal-related households). However, these analyses have either relied on parametric assumptions, failed to account for right censoring, or targeted unrealistic counterfactual policies. The analysis presented here overcomes these limitations.

\section{Causal estimands with clustered interference}

\subsection{Data structure}

Consider a setting where the outcome of interest is a survival time 
(e.g., time until cholera infection) 
that is subject to right censoring. 
The primary goal is to draw inference about the effect of a binary treatment on the survival time distribution in the observational setting 
(i.e., treatment is not randomly assigned), 
and there may be interference within clusters. 
Assume data are observed from $m$ clusters, 
and let $N_i$ denote the size of cluster $i \in \{1, \dots, m\}$.
Let $(\Tij, \Aij, \Xij) \in \bR^{+} \times \{0,1\} \times \bR^{p}$ 
represent the time to event, binary treatment status, and pre-treatment covariates 
for unit $j \in \{1, \dots, N_i\}$ in cluster $i$.
The covariates $\Xij$ may include shared variables within cluster $i$, 
such as urbanity of the cluster, or population density.
Define
$[ \xi_{ij} ]_{j=1}^{N_i}
:=
(\xi_{i1}, \dots, \xi_{iN_i})^\top$
as a vector of $\xi_{ij}$'s.
Let $\Ti = [ \Tij]_{j=1}^{N_i}$, 
$\Ai = [ \Aij]_{j=1}^{N_i}$, 
$\BX_i = [\Xij^\top]_{j=1}^{N_i}$,
and denote the full data for cluster $i$ by
$\Zi = (\Ti, \Ai, \BX_i, N_i)$.
Also, define $\cAni = \{0,1\}^{n_i}$ and $\cXni$ as the support of $\Ai$ and $\BX_i$ given $N_i = n_i$.
In the presence of right censoring, either the event time or censoring time is observed, depending on which occurs first.
Let 
$\Cij \in \bR^{+}$,
$\Yij = \min\{\Tij, \Cij\}$,
$\Deltaij = \indicator(\Tij \le \Cij)$
denote the censoring time, observed time, and event indicator,
where $\mathbbm{1}(\cdot)$ is the indicator function.
Further, define $\Yi = [\Yij]_{j=1}^{N_i}$ and
$\Deltai = [\Deltaij]_{j=1}^{N_i}$.
The observed data (under censoring) for cluster $i$ is then denoted by
$\Oi = (\Yi, \Deltai, \Ai, \BX_i, N_i)$.
Assume $(\BO_1, \dots, \BO_m)$ is an independent and identically distributed random sample from a super population of clusters.

Under clustered interference, a unit's potential event time may depend not only on its treatment but also on the treatments of others in the same cluster.
The potential event time for unit $j$ in cluster $i$ when cluster $i$ receives $\ai = [\aij]_{j=1}^{N_i} \in \cANi$ is denoted by $\Tij(\ai)$, and let
$\Ti(\ai) = [\Tij(\ai)]_{j=1}^{N_i}$.
Sometimes, $\ai$ is expressed as
$(\aij, \aimj)$, 
where $\aij$ is the treatment status of unit $j$, 
and 
$\aimj = (a_{i1}, \dots, a_{i(j-1)}, a_{i(j+1)}, \dots,
\allowbreak
a_{iN_i})^\top \in \cANim$
is the treatment status of all other units in cluster $i$.
Accordingly, $\Tij(\ai)$ is sometimes written as 
$\Tij(\aij, \aimj)$.
If there is no interference, 
$\Tij(\aij, \aimj) 
= 
\Tij(\aij, \Ba'_{i(-j)})$, 
i.e., potential outcomes for unit $j$ do not depend on the treatments of others, $\aimj$.

\subsection{Target causal estimands}
\label{sec:estimands}

Define the target causal estimand as the weighted average of a transformation $\cR$ of the potential event time $\Ti(\ai)$, expressed as:
\begin{align}
	\Psi(\cR; \Bw)
	=
	\textstyle
	E \left\{ 
		\suma
		\Bw(\ai, \BX_i, N_i)^\top
		\cR(\Ti(\ai))
	\right\},
	\label{eq:estimand}
\end{align}
where $\Bw(\ai, \BX_i, N_i)$ is a user-specified weight vector of length $N_i$, which may depend on $\ai, \BX_i, N_i$.
Here, if the transformation $\cR$ is applied to a vector $\Ti$, it is assumed to be applied element-wise, 
i.e., $\cR(\Ti(\ai)) = [\cR(\Tij(\ai))]_{j=1}^{N_i}$ and $\cR(\Ti) = [\cR(\Tij)]_{j=1}^{N_i}$.
The choice of transformation $\cR$ and weight function $\Bw$ determines the causal estimand of interest, tailored to the scientific question being addressed. 
For instance, selecting $\cR(T) = T$ and 
$\Bw(\ai, \BX_i, N_i) 
=
N_i^{-1}
\indicator(\ai = \boldsymbol{1}_{N_i})
\boldsymbol{1}_{N_i}$,
where $\boldsymbol{1}_{N_i}$ is the length $N_i$ column vector of ones,
leads to
$\Psi(\cR; \Bw)
=
\textstyle
E \big\{ 
	\sumjline
	\Tij(\boldsymbol{1}_{N_i})
\big\}$,
which is the mean event time when all units receive treatment.
Similarly, selecting 
$\Bw(\ai, \BX_i, N_i) 
=
N_i^{-1}
\{
	\indicator(\ai = \boldsymbol{1}_{N_i}) - \indicator(\ai = \boldsymbol{0}_{N_i})
\}
\boldsymbol{1}_{N_i}$ yields
$
\Psi(\cR; \Bw)
=
\textstyle
E \big[
	\sumjline
    \allowbreak
	\{\Tij(\boldsymbol{1}_{N_i})-\Tij(\boldsymbol{0}_{N_i})\}
\big]$,
which reduces to the average treatment effect (ATE) on the event time under no interference setting if $N_i \equiv 1$.
Other choices of $\cR$ could be $\cR(T) = \min\{T, h\}$ for the restricted mean survival time up to time $h$, or $\cR(T) = \indicator(T \le \tau)$ for the risk of an event by time $\tau$. 
Additional choices for the weight function $\Bw$ are discussed below.


This flexibility in choosing $\cR$ and $\Bw$ allows for (\ref{eq:estimand}) to include a wide range of target causal estimands, encompassing many previously proposed estimands in the presence of clustered interference \citep{tchetgen12, park2022efficient, lee2025efficient}. These estimands are formulated as weighted averages of potential outcomes, 
where the weight function $\Bw$ is a function of a treatment allocation policy \citep{munoz12}.
Mathematically,
a treatment allocation policy $Q$ is a probability distribution $Q(\cdot|\mathbf{X}_i, N_i): \cANi \mapsto [0,1]$,
representing the counterfactual scenario where a cluster of size $N_i$ with covariates $\mathbf{X}_i$ receives treatment $\mathbf{a}_i \in \mathcal{A}(N_i)$ with probability $Q(\mathbf{a}_i|\mathbf{X}_i, N_i)$. 
For example, in a cluster with $N_i = 2$ units, there are $2^2 = 4$ possible treatment allocations:
$\ai = (a_{i1}, a_{i2}) \in \mathcal{A}(2) = \{(0,0), (0,1), (1,0), (1,1)\}$. One example policy could be
$
\big(
	Q(0,0|\mathbf{X}_i, N_i), 
	Q(0,1|\mathbf{X}_i, N_i), 
    \allowbreak
	Q(1,0|\mathbf{X}_i, N_i), 
	Q(1,1|\mathbf{X}_i, N_i)
\big)
=
\big(
	0.1,
	0.3,
	0.2,
	0.4
\big),
$
where both units are untreated with probability 0.1, unit 1 is untreated and unit 2 is treated with probability 0.3, and so forth. A deterministic policy where $Q(1,1|\mathbf{X}_i, N_i) = 1$ and $Q(\ai|\mathbf{X}_i, N_i) = 0$ otherwise ($\ai \ne (1,1)$) corresponds to the scenario where every unit receives treatment. However, deterministic policies may not always be practically relevant since treatments are often not uniformly applied in real-world settings.
%
%
In contrast, stochastic policies allow units to be treated with probabilities between 0 and 1, potentially depending on covariates.

The following definitions describe typical causal estimands under clustered interference based on treatment allocation policies,
which can be expressed in the form of $\Psi(\cR; \Bw)$, 
given by \eqref{eq:estimand} with appropriate choices of $\cR$ and $\Bw$, as summarized in Table~\ref{tab:estimand}.

\begin{definition}[Causal estimands under clustered interference]
	The \textit{expected average potential outcome (EAPO) under policy $Q$} is defined by
	\begin{align*}
		\mu(\cR; Q) 
		= 
		\textstyle
		E \left\{ 
			N_i^{-1}
			\sum_{j=1}^{N_i}
			\suma 
			\cR(\Tij(\ai)) 
			Q(\ai | \BX_i, N_i) 
		\right\}.
	\end{align*}
	Here, $\cR(\Tij(\ai))$ is averaged over the counterfactual distribution of $\Ai$ under policy $Q$, 
	then averaged over units in a cluster, 
	and finally averaged over clusters in the (super-) population.
	Next, the \textit{EAPO when treated under policy $Q$} is defined as
	\begin{align*}
		\mu_{1}(\cR; Q) 
		= 
		\textstyle
		E \Big\{ 
			N_i^{-1} 
			\sum_{j=1}^{N_i}
			\sum_{\aimj \in \mathcal{A}(N_i-1)} 
			\cR(\Tij(1, \aimj))
			Q(\aimj | \BX_i, N_i) 
		\Big\},
	\end{align*}
	where 
	$Q(\aimj | \BX_i, N_i) 
	=
	\sum_{a \in \{0,1\}}
	Q(a, \aimj | \BX_i, N_i)$
	is the probability of all units in cluster $i$ other than $j$ receiving treatment $\mathbf{a}_{i(-j)}$ under policy $Q$. 
	Similarly, the \textit{EAPO when untreated under policy $Q$}, $\mu_{0}(\cR; Q)$, is defined by substituting $\Tij(1, \aimj)$ with $\Tij(0, \aimj)$ in $\mu_{1}(\cR; Q)$.
	Under no interference,
	$\Tij(a, \aimj)$ does not depend on $\aimj$ and may be written simply as $\Tij(a)$ for $a \in \{0,1\}$, 
	implying 
	$
	\mu_{a}(\cR; Q) 
	\equiv
	E \big\{ 
		N_i^{-1} 
		\sum_{j=1}^{N_i}
		\cR(\Tij(a))
	\big\}
	$
	regardless of $Q$. 
	Thus, changes in $\mu_a(\cR; Q)$ with respect to $Q$ imply the presence of interference.
	
\end{definition}

\begin{definition}[Causal effects under clustered interference]
	To quantify the effect of the treatment under clustered interference, 
	causal effects are defined as contrasts of the EAPOs as follows.
	The \textit{direct effect} is defined by $DE(\cR; Q) = \mu_1(\cR; Q) - \mu_0(\cR; Q)$, which quantifies the effect of a unit receiving treatment under policy $Q$. 
	For two policies $Q$ and $Q'$, the \textit{overall effect} is defined as $OE(\cR; Q, Q') = \mu(\cR; Q) - \mu(\cR; Q')$, contrasting the EAPOs under two policies. 
	The \textit{spillover effect when treated} is defined by $SE_1(\cR; Q, Q') = \mu_1(\cR; Q) - \mu_1(\cR; Q')$, which compares EAPOs when treated under policy $Q$ versus $Q'$. 
	Similarly, the \textit{spillover effect when untreated} is defined as $SE_0(\cR; Q, Q') = \mu_0(\cR; Q) - \mu_0(\cR; Q')$.
\end{definition}

\begin{table}[H]
	\centering
	\caption{Causal estimands $\Psi(\cR; \Bw)$ and corresponding weight functions $\Bw$ ($a \in \{0,1\}$)}
	\label{tab:estimand}
	\renewcommand{\arraystretch}{1.1} 
	\resizebox{0.9\textwidth}{!}{%
		\begin{tabular}{cc}
			\toprule
			Estimand $\Psi(\cR; \Bw)$                         & Weight function $\Bw(\aXNi)$                                         \\ \midrule
			$\mu(\cR; Q)$                                     & $N_i^{-1} \vecj{Q(\ai | \BX_i, N_i)}$                                \\
			$\mu_a(\cR; Q)$                                   & $N_i^{-1} [\indicator(a_{ij} = a) Q(\aimj| \BX_i, N_i)]_{j=1}^{N_i}$ \\
			$DE(\cR; Q) = \mu_1(\cR; Q)-\mu_0(\cR; Q)$        & $N_i^{-1} \vecj{(2a_{ij} - 1) Q(\aimj| \BX_i, N_i)}$                 \\
			$SE_a(\cR; Q, Q') = \mu_a(\cR; Q)-\mu_a(\cR; Q')$ & $N_i^{-1} \vecj{\indicator(a_{ij} = a) (Q-Q')(\aimj| \BX_i, N_i)}$   \\
			$OE(\cR; Q, Q') = \mu(\cR; Q) - \mu(\cR; Q')$     & $N_i^{-1} \vecj{(Q-Q')(\ai | \BX_i, N_i)}$                           \\ \bottomrule
		\end{tabular}
	}
\end{table}

\vspace{-0.2cm}

\subsection{Example policies}
\label{sec:example_policies}

This section outlines example policies $Q$ commonly used in the clustered interference setting. The methods developed in this paper are general and can be applied to stochastic policies beyond these specific examples under mild conditions.

\vspace{0.1cm}

\begin{example}[Type B Policy]

	\citet{tchetgen12} proposed the Type B policy,
	under which units select treatment independently with the same probability $\alpha$ such that 
	$Q_{\scriptscriptstyle \textup{B}}(\mathbf{a}_i|\mathbf{X}_i, N_i; \alpha) = \prod_{j=1}^{N_i} \alpha^{a_{ij}} (1-\alpha)^{1-a_{ij}}$.
	For observational data with right censored outcomes, 
	\cite{chakladar2022inverse}
	suggested parametric inverse probability of censoring weighted estimators for Type B policy estimands.

\end{example}

\vspace{0.1cm}

\begin{example}[Parametric Propensity Score Shift policy]

	Aside from randomized studies, individuals typically do not receive treatment with the same probability, such that the Type B policy may not be particularly relevant in many settings.
	Alternatively, \citet{Papadogeorgou19} and \citet{barkley20} proposed
	Parametric Propensity Score Shift (PPSS) policies that allow for the different probability of receiving treatment among units. 
	Under PPSS policies, a parametric model is assumed for $\Ai | \XNi$ such as:
	$
	\text{pr}(\Ai | \XNi)
	=
	\int \prod_{j=1}^{N_i}
		\{g(\beta_0 + \Xij^\top \boldsymbol{\beta} + u)\}^{\Aij}
		\allowbreak
		\{1-g(\beta_0 + \Xij^\top \boldsymbol{\beta} + u)\}^{1-\Aij}
	dF(u),
	$
	where $g$ is a link function and $u$ is a random effect with distribution $F$. The policy distribution is then obtained by shifting the intercept from $\beta_0$ to $\gamma$:
	$
	Q_{\scriptscriptstyle \textup{PPSS}}(\ai | \XNi; \gamma)
	=
	\int \prod_{j=1}^{N_i}
		\{g(\gamma + \Xij^\top \boldsymbol{\beta} + u)\}^{\aij}
		\{1-g(\gamma + \Xij^\top \boldsymbol{\beta} + u)\}^{1-\aij}
	dF(u).
	$
	While these estimands may be more relevant than the Type B policy in many settings, their interpretation is unclear when the parametric propensity score models are misspecified.

\end{example}

\vspace{0.1cm}

\begin{example}[Cluster Incremental Propensity Score policy]
	
	To address the limitation of the PPSS policy, \citet{lee2025efficient} proposed the Cluster Incremental Propensity Score (CIPS) policy, which does not rely on parametric models. 
	The CIPS policy distribution is defined as
	$
	Q_{\scriptscriptstyle \textup{CIPS}}(\mathbf{a}_i | \mathbf{X}_i, N_i; \delta) 
	=
	\prod_{j=1}^{N_i} 
		(\pi_{ij,\delta})^{a_{ij}}
		(1 - \pi_{ij,\delta})^{1-a_{ij}},
	$
	where $\pi_{ij} = \text{pr}(A_{ij} = 1 | \mathbf{X}_i, N_i)$ is the factual propensity score, and $\pi_{ij, \delta} = \delta \pi_{ij} / (\delta \pi_{ij} + 1 - \pi_{ij})$ is the counterfactual propensity score. The parameter $\delta > 0$ controls the shift in treatment odds, with $\pi_{ij, \delta} / (1-\pi_{ij, \delta}) \allowbreak = \delta \{\pi_{ij} / (1-\pi_{ij})\}$, making treatment receipt more likely when $\delta > 1$ and less likely when $\delta < 1$.
	Thus, this policy maintains the ranking of units within clusters based on their propensity scores; i.e., if $\pi_{ij} < \pi_{ik}$, then $\pi_{ij,\delta} < \pi_{ik,\delta}$.
	The CIPS policy does not require parametric modeling of $\pi_{ij}$, allowing utilization of flexible nonparametric methods to estimate $\pi_{ij}$.

\end{example}

\vspace{0.1cm}

\begin{example}[Treated Proportion Bound policy]

	The Treated Proportion Bound (TPB) policy \citep{lee2025efficient} considers the counterfactual scenario where a minimum proportion  $\rho \in [0,1]$ of individuals in each cluster is treated.
	Specifically, the TPB policy distribution is defined as 
	$Q_{\scriptscriptstyle \textup{TPB}}(\mathbf{a}_i | \mathbf{X}_i, N_i; \rho) = \mathbbm{1}(\overline{\mathbf{a}}_i \ge \rho) \text{pr}(\mathbf{a}_i | \mathbf{X}_i, N_i) / \{\sum_{\overline{\mathbf{a}}'_i \ge \rho} \text{pr}(\mathbf{a}'_i | \mathbf{X}_i, N_i)\}$, 
	where $\overline{\mathbf{a}}_i = N_i^{-1} \sum_{j=1}^{N_i} a_{ij}$ is the proportion of treated units in cluster $i$. Under this policy, treatment assignments with $\overline{\mathbf{a}}_i < \rho$ are excluded, guaranteeing a minimum treatment coverage of $\rho$ in each cluster. 
	Unlike Type B and CIPS policies, the TPB policy is constructed based on cluster-level treatment probabilities rather than individual-level propensity scores.
	
\end{example}

\vspace{0.1cm}

For a policy $Q_{\scriptscriptstyle <\textup{NAME}>}(\cdot|\mathbf{X}_i, N_i;\theta)$ indexed by a parameter $\theta$ (e.g., $\alpha \in (0,1)$ in the Type B policy), 
the target estimand $\Psi(\cR; \Bw)$ will sometimes be denoted as $\Psi_{\scriptscriptstyle <\textup{NAME}>}(\cR; \theta)$ to emphasize the dependence of $\Bw$ on $\theta$ via $Q$, or simply $\Psi(\cR; \theta)$ when the target estimand is clear based on the context. 
Likewise, $\mu(\cR; Q)$ will be denoted by $\mu_{\scriptscriptstyle <\textup{NAME}>}(\cR; \theta)$, with similar shorthand for other estimands, e.g., $\mu_{\scriptscriptstyle <\textup{NAME}>, \scriptstyle 1}(\cR; \theta)$ and $OE_{\scriptscriptstyle <\textup{NAME}>}(\cR; \theta, \theta')$. 
When $\cR(T; \tau) = \indicator(T \le \tau)$, the estimand is written as $\Psi(\tau; \theta)$, highlighting its dependence on $\tau$ and $\theta$, e.g., 
$\mu_{\scriptscriptstyle <\textup{NAME}>}(\tau; \theta)$, 
$DE_{\scriptscriptstyle <\textup{NAME}>}(\tau; \theta)$, 
$SE_{\scriptscriptstyle <\textup{NAME}>, a}(\tau; \theta, \theta')$.

\subsection{Assumptions and identifiability}

The following enumerates the sufficient conditions for identifying the causal estimands discussed above, 
which are standard in the clustered interference literature 
\citep{tchetgen12, liu19, park2022efficient, qu22efficient, kilpatrick24gformula}:
\begin{assumption}[Identifying conditions]
    \label{assump:ident}
    The following conditions hold.

    \textup{(I1)} {Consistency}: $\Ti = \suma \Ti(\ai)\indicator(\Ai = \ai)$.

    \textup{(I2)} {Conditional Exchangeability}: $\Ti(\ai) \indep \Ai | \mathbf{X}_i, N_i \text{ for all } \ai \in \cANi$.

    \textup{(I3)} {Weight-Conditional Positivity}:  
	$\text{pr}(\Ai = \ai \mid \mathbf{X}_i, N_i) > 0$ if $\Bw(\ai, \BX_i, N_i) \neq \textbf{0}_{N_i}$.

    \textup{(I4)} {Conditional Independent Censoring}: $C_{ij} \indep T_{ij} | \Ai, \BX_i, N_i$.

    \textup{(I5)} {Noncensoring Positivity}: $\text{pr}(\Deltaij = 1| \Ai, \BX_i, N_i) > 0 $.

\end{assumption}

Conditions (I1) -- (I3) extend individual-level causal assumptions to cluster-level data, enabling inference under clustered interference. Selecting covariates $\BX$ to satisfy conditional exchangeability often requires domain expertise. 
Note condition (I3) does not require positivity for treatment allocations $\ai$ where the weight $\Bw(\aXNi)$ equals $\textbf{0}_{N_i}$, 
because such allocations $\ai$ do not contribute to the estimand $\Psi(\cR; \Bw)$ and thus observing data with $(\Ai = \aXNi)$ is unnecessary.
Thus, (I3) is less restrictive than the traditional positivity assumptions which typically require all treatment combinations to have a non-zero probability of being observed 
(see \cite{kennedy19} for a similar condition in the absence of interference).
Condition (I4) assumes censoring is uninformative within cluster-level strata, though not necessarily marginally, and (I5) ensures a non-zero probability of observing event times. 

Lemma~\ref{lemma:ident} establishes identifiability of causal estimands using both full and observed data under these assumptions. The first part of Lemma \ref{lemma:ident} regarding identifiability from the full data is shown in \cite{lee2025efficient},
and 
the second part is shown in the supplementary material Section A.1.

\begin{lemma}
\label{lemma:ident}
	Under \textup{(I1) -- (I3)}, 
	the target causal estimand $\Psi(\cR;\Bw)$ can be expressed as
	\begin{align*}
		\Psi(\cR; \Bw)
		=
		\textstyle
		E \left[ 
			\suma
			\Bw(\ai, \BX_i, N_i)^\top
			E\{\cR(\Ti) | \Ai = \ai, \BX_i, N_i\}
		\right]
		.
	\end{align*}
	Thus, $\Psi(\cR;\Bw)$ is identifiable from the full data $\Zi = (\Ti, \Ai, \BX_i, N_i), i = 1,\dots,m$.

	In addition, if \textup{(I1) -- (I5)} hold, 
	then 
	\begin{align*}
		\Psi(\cR;\Bw)
		=
		E \bigg[ 
		  \textstyle
		  \suma
		  \Bw(\ai, \BX_i, N_i)^\top
		  \bigg[
			E \bigg\{
			  \displaystyle
			  \frac
				{\Deltaij \cR(\Yij) }
				{\SCij{\Yij}}
			  \bigg| \Ai = \ai, \BX_i, N_i 
			\bigg\}
		  \bigg]_{\raisebox{0.5ex}{$\scriptstyle j=1$}}^{\raisebox{-0.5ex}{$\scriptstyle N_i$}}
		\bigg]
		,
	\end{align*}
	where 
	$\SCij{r} 
	=
	\textup{pr}(\Cij > r | \AXNi)$.
	Thus, $\Psi(\cR;\Bw)$ is identifiable from the observed data $\Oi = (\Yi, \Deltai, \Ai, \BX_i, N_i), i = 1,\dots,m$.
	
\end{lemma}

\section{Methods}

\subsection{Semiparametric efficient estimation using full data under no censoring}

In the absence of censoring and interference, consistent, asymptotically normal, and semiparametric efficient estimators for various causal estimands (e.g., ATE) can be constructed using the efficient influence functions (EIFs) \citep{tsiatis06,kennedy19,hines22}. These estimators typically achieve the parametric $m^{-1/2}$ convergence rate (where $m$ is the number of observations), 
provided the nuisance functions are estimated at a rate faster than $m^{-1/4}$. 
This flexibility allows the use of nonparametric and machine learning methods for nuisance function estimation, reducing the risk of model misspecification. 

\cite{lee2025efficient} utilized EIFs to develop such estimators in the presence of clustered interference whern there is no censoring.
The basic approach is as follows.
Let $P_{\AXN}$ denote the true distribution of $(\Ai, \BX_i, N_i)$,
and assume it belongs to the statistical model 
$\cM_{\AXN}$,
which is a collection of distributions sharing the same support as $P_{\AXN}$.
Often, the weight function $\Bw$ is defined as a functional of $P_{\AXN}$ (e.g., CIPS and TPB policies), denoted as the mapping $P_{\AXN} \mapsto \Bw(\axn; P_{\AXN})$ for each fixed $\axn$. 
In other words, the value of $\Bw(\axn; P_{\AXN})$ could vary even for the same $\axn$ if the underlying data distribution $P_{\AXN}$ changes. 
For brevity, the dependence of $\Bw$ on $P_{\AXN}$ is often omitted when the context is clear.
The following condition is assumed.

\vspace{-0.2cm}

\begin{assumption}[Conditional Influence Function of $\Bw$]
	\label{assump:fulldata}
	\

	\textup{(C1)}
	For any fixed $\Ba, \Bx, n$,
	there exists a function 
	$\Ai, \BX_i, N_i \mapsto \Bphi(\Ai, \BX_i, N_i; \Ba)$ 
	such that 
	(i)
	$E\left\{
		\Bphi(\mathbf{A}_i, \mathbf{X}_i, N_i; \mathbf{a})
		\mid
		\mathbf{X}_i, N_i
	\right\} = \textbf{0}_{N_i}$
	and
	(ii)
	for any regular parametric submodel $P_{\AXN, \varepsilon} \in \cM_{\AXN}$ with parameter $\varepsilon \in (-t,t)$ for some $t > 0$, satisfying $P_{\AXN, \varepsilon=0} = P_{\AXN}$ and likelihood score $l_\varepsilon'(\cdot)$,
	it holds that
	\begin{align*}
		\frac
			{\partial \Bw(\Ba,\Bx,n; P_{\AXN, \varepsilon})}
			{\partial \varepsilon}
		\Big|_{\varepsilon=0}
		=
		E
		\left\{
			\Bphi(\mathbf{A}_i, \mathbf{X}_i, N_i; \mathbf{a})
			l_\varepsilon'(\mathbf{A}_i, \mathbf{X}_i, N_i)
			\mid
			\mathbf{X}_i = \mathbf{x}, N_i = n
		\right\}
		.
	\end{align*}
	
\end{assumption}

\vspace{0.4cm}

Condition (C1) aligns with the definition of Conditional Influence Function (CIF) in \cite{chernozhukov2024conditional}, treating $\Bw(\Ba,\Bx,n; P_{\AXN})$ as pathwise differentiable given covariates, if not marginally. 
This approach facilitates constructing the EIF of the target estimand $\Psi(\cR; \Bw)$ as a functional of the observed data distribution. 
If $\Bw$ is independent of $P_{\AXN}$ (e.g., Type B policy), then $\Bphi = \textbf{0}_{N_i}$. For policies like CIPS or TPB, where $\Bw$ depends on $P_{\AXN}$, $\Bphi$ also becomes a functional of $P_{\AXN}$.

Under assumptions (I1) -- (I3), and (C1), Theorem 1 of \cite{lee2025efficient} provides the EIF of $\Psi(\cR; \Bw)$ given by
$\phiF(\cR; \Zi)  - \Psi(\cR; \Bw)$,
where
$\phiF(\cR; \Zi)  = \sumjline \varphi_{ij}^{F}(\cR; \mathbf{Z}_i)$,
with
$\varphi_{ij}^{F}(\cR; \mathbf{Z}_i) := \suma \OR_{ij}(\cR; \Zi, \ai) + \text{BC}_{ij}(\cR; \Zi)$,
and

\vspace{-0.2cm}
\begin{align*}
	\OR_{ij}(\cR; \Zi, \ai) &= \big\{ w_j(\aXNi) + \phi_j(\AXNi; \ai) \big\} E\{\cR(\Tij) | \Ai = \aXNi\}, \\
	\text{BC}_{ij}(\cR; \Zi) &= \frac{w_j(\AXNi)}{\text{pr}(\AbarXNi)} \big[\cR(\Tij) - E\{\cR(\Tij) | \AXNi\}\big].
\end{align*}
Here, $\Bw(\mathbf{a}_i, \mathbf{X}_i, N_i) = [N_i^{-1}w_j(\mathbf{a}_i, \mathbf{X}_i, N_i)]_{j=1}^{N_i}$ and $\Bphi(\mathbf{A}_i, \mathbf{X}_i, N_i; \mathbf{a}_i) = [N_i^{-1}\phi_j(\mathbf{A}_i, \mathbf{X}_i, N_i; \mathbf{a}_i)]_{j=1}^{N_i}$. 
The first term, $\suma \OR_{ij}(\cR; \Zi, \ai)$ (outcome regression), represents a weighted average of the conditional expectation of the outcome $\cR(\Tij)$, while the second term, $\text{BC}_{ij}(\cR; \Zi)$ (bias correction), represents a weighted residual.
If the function $\varphi^{F}(\cR; \mathbf{Z}_i)$ is known, the full data estimator can be constructed based on the EIF such that
$\widetilde{\Psi}^F(\cR; \Bw) = \sumiline \phiF(\cR; \Zi)$,
while in practice $\varphi^{F}(\cR; \mathbf{Z}_i)$ may be unknown and thus need to be estimated.
Refer to \cite{lee2025efficient} for more details.

\subsection{Observed data estimating equation with right censored outcomes}
\label{sec:esteq}

In the presence of censoring, however, the full data $\Zi$ is generally not always observed, and using complete data only (i.e., $\Deltaij = 1$) would result in biased estimation.
In the no interference setting, 
inverse probability of censoring weighting (IPCW) 
\citep[e.g.,][]{robins2000correcting, rotnitzky2005inverse}
or augmented IPCW estimators 
\citep[e.g.,][]{ozenne2020estimation, cui2023estimating}
can be used to adjust for censoring;
however, such estimators typically do not allow for interference.
In the clustered interference setting,
\cite{chakladar2022inverse} proposed parametric IPCW estimators for Type B policy estimands,
but these estimators are not consistent if the parametric models are mis-specified.
To overcome these limitations, 
below IPCW-type estimators are proposed which are 
(i) nonparametric, (ii) account for censoring, confounding, and clustered interference, and (iii) permit inference about generic counterfactual policies $Q$.

Motivated by \cite{tsiatis06} Chapter 10.4, 
the following observed data estimating equation (assuming known nuisance functions) is proposed:
\begin{align}
	0 =
	\sum_{i=1}^{m}
	\sumj
		\Bigg[
			&
			\frac{\Delta_{ij} \{\phiF_{ij}(\cR; \Zi) - \Psi(\cR;\Bw)\} }{\SCij{\Yij}}
			\nonumber
			\\
			&+
			\int_{0}^{\infty}
				\frac
					{E \big\{ 
						\phiF_{ij}(\cR; \Zi) - \Psi(\cR;\Bw)
						\mid \Tij \ge r, \AXNi
					\big\}}
					{\SCij{r}}
				d\MijC
		\Bigg],	
		\label{eq:esteq}
\end{align}
where 
$\MijC
= 
\indicator(\Yij \le r, \Deltaij = 0)
+
\int_{0}^{\min\{\Yij, r\}}
d \{\log{\SCij{u}}\}$
is a mean zero martingale for the censoring process. 
The first term uses IPCW, $\Delta_{ij}/\SCij{\Yij}$, to adjust for censoring by upweighting uncensored data, while the second term eliminates first-order bias, ensuring the estimator is multiply robust (i.e., consistent if some, but not all, nuisance functions are correctly estimated).
Solving the observed data estimating equation results in the estimator
$
\widetilde{\Psi}(\cR; \Bw)
=
\sumiline
	\phiP(\cR; \mathbf{O}_i)
$,
where
$
\phiP(\cR; \mathbf{O}_i)
=
\sumjline
	\phiP_{ij}(\cR; \mathbf{O}_i)
$
and
$\phiP_{ij}(\cR; \mathbf{O}_i) 
= 
\suma \OR_{ij}(\cR; \Oi, \ai)
\allowbreak
+
\allowbreak
\IPCWBC_{ij}(\cR; \Oi)
\allowbreak
+
\allowbreak
\AUG_{ij}(\cR; \Oi)$,
with
\begin{align*}
\resizebox{.95\textwidth}{!}{
	$\begin{aligned}
		\OR_{ij}(\cR; \Oi, \ai) 
		=&
		\big\{ w_j(\aXNi) + \phi_j(\AXNi; \ai) \big\}
		E\{\cR(\Tij) | \Ai = \aXNi\}
		, 
		\\
		\IPCWBC_{ij}(\cR; \Oi) 
		=&
		\frac
			{w_j(\AXNi)}
			{\text{pr}(\AbarXNi)}
		\left[
			\frac{\Deltaij}
			{\SCij{\Yij}}
			\cR(\Yij)
			- 
			E\{\cR(\Tij) | \AXNi\}
		\right]
		, 
		\\
		\AUG_{ij}(\cR; \Oi) 
		=&
		\frac
			{w_j(\AXNi)}
			{\text{pr}(\AbarXNi)}
		\int_{0}^{\infty}
			\frac
				{E\{\cR(\Tij)\indicator(\Tij \ge r) | \AXNi\}}
				{\SCij{r} \STij{r}}
		d\MijC
		.
	\end{aligned}$
}
\end{align*}
\noindent
Here,
$\STij{r} 
=
\text{pr}(\Tij > r | \AXNi)$ 
is the event time survival function.
Both $\phiP_{ij}(\cR; \Oi)$ 
and 
$\phiF_{ij}(\cR; \Zi)$
include an outcome regression term ($\OR_{ij}$)
and a bias correction term ($\IPCWBC_{ij}$ and $\text{BC}_{ij}$),
with $\phiP_{ij}(\cR; \Oi)$ using IPCW to avoid using $\Tij$ directly.
Additionally, 
$\phiP_{ij}(\cR; \Oi)$ includes an augmentation term ($\AUG_{ij}$), enabling the estimator to achieve an $m^{-1/2}$ convergence rate under mild conditions, even with nonparametric nuisance function estimators and censoring (see Section \ref{sec:theory}).
Proposition \ref{prop:unbias} establishes the unbiasedness of the estimating function $\phiP(\cR; \Oi)$ when all nuisance functions are known. In practice, these nuisance functions must be estimated, as discussed in the next section.

\begin{proposition}
	\label{prop:unbias}
	Assume $\phiP(\cR; \Oi)$ is a known function of $\Oi$. 
	If \textup{(I1)} -- \textup{(I5)} and \textup{(C1)} hold, 
	then
	$E \big\{
		\phiP(\cR; \Oi)
	\big\}
	=\Psi(\cR; \Bw)$.
\end{proposition}

\subsection{Nuisance functions}
\label{sec:nuis}

To derive estimators based on the proposed estimating function $\phiP(\cR; \Oi)$, the following nuisance functions must be estimated:
(i) the event time distribution function, 
$\BF^T(r|\AXNi) 
= 
\big[
	F_{ij}^T(r|\AXNi) 
\big]_{j=1}^{n_i}$;
(ii) the censoring time survival function,
$\BS^C(r|\AXNi) 
= 
\big[
	S_{ij}^C(r|\AXNi) 
\big]_{j=1}^{n_i}$;
(iii) the cluster treatment probability,
$H(\AXNi) 
= 
\text{pr} \big(
	\mathbf{A}_{i} | \mathbf{X}_i, N_i
\big)$;
(iv) the weight function,
$\Bw(\mathbf{a}_i, \mathbf{X}_i, N_i)$;
and
(v) the CIF of $\Bw$,
$\Bphi(\mathbf{A}_i, \mathbf{X}_i, N_i; \mathbf{a}_i)$.
Here,
$\FTij{r} 
=
\text{pr}(\Tij \le r | \AXNi)$ 
is the event time distribution function.
Let $\boldsymbol{\eta} = (\BF^T, \BS^C, H, \Bw, \Bphi)$ denote the vector of nuisance functions.
Typically, $\Bw$ and $\Bphi$ are specific to the chosen policy $Q$ and target estimand and are often functions of $H$, so they do not need to be estimated separately (see examples in supplementary material Section B). 
Note the quantities appearing in $\phiP(\cR; \Oi)$, such as
$E\{\cR(\Tij) | \AXNi\}
= 
\int_0^\infty
	\cR(t) 
dF_{ij}^T(t|\AXNi)$
and
$E\{\cR(\Tij)\indicator(\Tij \ge r) | \AXNi\}
=
\int_r^\infty
	\cR(t) 
dF_{ij}^T(t|\AXNi)$
can be estimated once $\BF^T$ is estimated. 
Additionally, $\AUG$ can be approximated using the Stieltjes integral with respect to the forward difference of
$\MijChat
= 
\indicator(\Yij \le r, \allowbreak \Deltaij = 0)
+
\int_{0}^{\min\{\Yij, r\}}
d \{\log{\widehat{S}^C_{ij}(u|\AXNi)}\}$.

Various methods have been proposed for estimating $H$ in settings where treatment selection within clusters might be correlated, and cluster-level confounding can occur \citep{chang2022flexible, salditt2023parametric}. Generalized linear mixed model (GLMM) may be used but risk model misspecification.
Alternatively, nonparametric data-adaptive approaches that allow for clustered data may be used, permitting nonsmooth and nonlinear data-generating processes.
In this work, random effect Bayesian additive regression trees (BART) \citep{dorie2022dbarts}
are primarily utilized due to ease of implementation with available statistical software, 
although other nonparametric methods for clustered data such as generalized mixed effects regression trees \citep{hajjem2017generalized} or random effect generalized boosted modeling \citep{salditt2023parametric} could also be used.

For $\BF^T$ and $\BS^C$, methods such as mixed effects proportional hazard models \citep{vaida2000proportional}, shared frailty models \citep{munda2012parfm}, and survival trees \citep{fan2006trees} can be used for clustered survival data. In this work, random survival forests \citep{ishwaran2023random} are employed with the following specifications: 
(i) $F_{ij}^T(r|\AXNi) = \bP(\Tij \le r | \Aij, \Aimj, \Xij, \Ximj) =: \cF_{*}^T(r | \Aij, \Aimj, \Xij, \Ximj)$; 
(ii) $S_{ij}^C(r|\AXNi) = \bP(\Cij > r | \Aij, \Aimj, \Xij, \Ximj) =: \cS_{*}^C(r | \Aij, \Aimj, \Xij, \Ximj)$, 
where $\Aimj = (\sum_{k \ne j} A_{ik}) / (N_i-1)$
and  $\Ximj = (\sum_{k \ne j} \BX_{ik}) / (N_i-1)$.
This specification is closely related to the conditional stratified interference assumption in \cite{park2024minimum} and
weak stratified interference assumption in \cite{kilpatrick24gformula},
and is sometimes referred to as exposure mapping \citep{aronow2017estimating, park2022efficient},
where the individual-level outcome (event time) is assumed to be a function of a lower-dimensional summary $\big( \Aij, \Aimj \big)$ of cluster-level exposure $\Ai$.
Note under this specification $\Tij$ and $T_{ik}$ $(j \ne k)$ both potentially depend on $A_{il}$ and $\BX_{il}$ for $l \in \{1,\dots,N_i\}$,
thus $\Tij$ and $T_{ik}$ may be marginally dependent,
allowing the model to capture within cluster correlation.
Here, the functions $\cF_*^T, \cS_*^C: [0,\infty) \times \{0,1\} \times [0,1] \times \mathbb{R}^p \times \mathbb{R}^p \mapsto [0,1]$ have fixed domains not varying by cluster sizes, unlike $\BF^T, \BS^C: [0,\infty) \times \cAni \times \cXni \times \mathbb{N} \mapsto [0,1]^{n_i}$, 
and thus standard regression methods can be utilized.
Nevertheless, 
the theoretical results derived in this paper do not necessitate this particular specification for the nuisance functions;
for instance,
$\cS_*^C$ may not depend on
$\overline{\mathbf{A}}_{i(-j)}$,
and thus $\overline{\mathbf{A}}_{i(-j)}$ can be excluded from the model.
Even though we mainly utilize random survival forest \citep{ishwaran2023random} for estimation of $\cF_*^T$ and $\cS_{*}^C$, other flexible data-adaptive methods could be used that are robust to model misspecification as long as they satisfy the mild conditions given in Section \ref{sec:theory}.

\subsection{Estimators}

Let
$\phiP(\cR; \Oi) = \phiP(\cR; \Oi, \Beta)$ denote the proposed estimating function for estimating $\Psi(\cR;\Bw)$ in Section \ref{sec:esteq},
where now the dependence on the nuisance functions $\Beta$ is made notationally explicit.
Based on the estimating function $\varphi$,
one might consider the estimator
$
\check{\Psi} (\cR; \Bw)
=
\sumiline
	\phiP(\cR; \mathbf{O}_i, \Betahat),
$
where $\Betahat$ is trained on the whole data $\{\BO_1, \dots, \BO_m\}$.
However, 
due to the repeated use of the sample in training nuisance functions $\Beta$ and evaluating the estimating function $\varphi$,
certain conditions on the complexity of nuisance function estimators (e.g., Donsker class) are necessary
for the estimator $\check{\Psi} (\cR; \Bw)$ to exhibit favorable large sample properties.
These conditions fundamentally restrict the complexity of the nuisance function estimators available, 
which rules out a number of flexible nonparametric estimators, such as tree-based methods, neural networks, or complex ensemble estimators \citep{kennedy19}.

To overcome such shortcomings, cross-fitting estimators \citep{chernozhukov18} can be constructed, 
which are consistent, asymptotically normal, and multiply robust,
with less restrictive assumptions about the complexity of the nuisance function estimators.
The proposed nonparametric cross-fitting (NCF) estimator $\widehat{\Psi}(\cR; \Bw)$ and its variance estimator $\widehat{\sigma}(\cR; \Bw)^2$ are constructed as described in Algorithm \ref{alg:estimator}, where $\bPmk \big\{ f(\BO) \big\} = \sumikline f(\Oi)$ for any function $f$ of $\BO$.

\begin{algorithm}
    \caption{Construction of the proposed NCF estimator.} \label{alg:estimator}
        \begin{spacing}{1.1}
        \begin{algorithmic}
            \Require 
            Cluster-level data $(\mathbf{O}_1, \dots, \mathbf{O}_m)$;
            number of folds $K$
            \Ensure 
            Estimator $\widehat{\Psi}(\mathcal{R};  \mathbf{w})$ and variance estimator $\widehat{\sigma} (\mathcal{R}; \mathbf{w})^2$ 
            \State 
            Randomly partition the data into $K$ disjoint folds;  $G_i \in \{1, \dots, K\}$: fold assignment
            \For{$k = 1, \dots, K$}
                \State 
                Set $m_k = \sum_{i=1}^{m} \indicator(G_i = k)$ 
                \State
                Train $\Beta$ on $\{\Oi: G_i \neq k\}$, 
				and denote it as
				$\Betahatk
				=
				\big(
					\BFThatk,
					\BSChatk,
					\Hhatk, 
					\Bwhatk, 
					\Bphihatk
				\big)$
                \State
                Evaluate $\{\phiP(\cR; \Oi, \Betahatk): G_i = k\}$ and average to yield
                $\bPmk \big\{ 
                    \phiP(\cR; \BO, \Betahatk) 
                \big\}$
            \EndFor
            \State
            Compute estimator:
            $\widehat{\Psi}(\cR; \Bw)
            =
            \sumkline
            \bPmk \big\{ 
                \phiP(\cR; \BO, \Betahatk) 
            \big\}$
            \State
            Compute variance estimator:
            $\widehat{\sigma}(\cR; \Bw)^2 
            =
            \sumkline
            \mathbb{P}_m^k 
            \Big[ 
                \big\{ 
                    \phiP(\cR; \BO, \Betahatk) - \widehat{\Psi}(\cR; \Bw)
                \big\}^2
            \Big]$
        \end{algorithmic}
        \end{spacing}
\end{algorithm}

The essence of the cross-fitting procedure lies in the division of the sample into a nuisance function training fold and an evaluation fold, preventing the double use of the sample.
Specifically,
the proposed NCF estimator is given as follows:
\begin{align*}
        \widehat{\Psi}(\cR; \Bw) 
        =&
        \sumk
        \sumik
        \Bigg[
            \suma
                \ORhatk(\cR; \Oi, \ai) 
                +
                \IPCWBChatk(\cR; \Oi)
                +
                \AUGhatk(\cR; \Oi)
        \Bigg]
\end{align*}
where
$\ORhatk(\cR; \Oi, \ai) = \sumjline \widehat{\textup{OR}}_{ij,(k)}(\cR; \Oi, \ai)$
is obtained by substituting the nuisance functions with $\Betahatk$,
and $\IPCWBChatk$ and $\AUGhatk$ are obtained in the similar manner.

Three practical modifications to the proposed estimators are as follows:

1. \textit{Subsampling}: The term
$\suma 
\allowbreak
\ORhatk(\tau; \Oi, \ai)$ may be approximated by 
$
r^{-1}
\sum_{q=1}^{r}
\ORhatk
\allowbreak
\big(\tau; \Oi, \ai^{(q)}\big)
/ \allowbreak
\Hhatk\big(\ai^{(q)}, \XNi \big)
$,
where $\mathbf{a}_i^{(q)}$ $(q=1,\dots,r)$ is a random sample from $\Hhatk\big(\cdot, \XNi \big)$,
which reduces the number of summands from $2^{N_i}$ to $r \in \mathbb{N}$ (subsampling degree). 
The finite sample performance of the proposed estimators demonstrates that the bias and confidence interval coverage of the estimators are insensitive to $r$, but the variance of the estimators decreases in $r$ (See supplementary material Section C.4),
which is supported by the large sample theory in supplementary material Section A.9.

2. \textit{Bounding}: To address extreme inverse probability weights $\Bwhatk (\AXNi) / \Hhatk (\AXNi)$ in $\IPCWBC$ and $\AUG$, which can lead to unbounded estimators, a H\'{a}jek-style estimator can be used by replacing $m_k$ in $\bPmk$ with 
$
\sum_{i: G_i = k} 
	\Bwhatk (\AXNi)^\top \textbf{1}_{N_i} 
	/ 
	\Hhatk (\AXNi).
$
This ensures bounded estimates for naturally bounded estimands (e.g., $\mu(\tau; Q) \in [0,1]$ for $\cR(T) = \indicator(T \le \tau)$) and improves finite sample performance without affecting asymptotic properties (see supplementary material Sections A.10 and C.3).

3. \textit{Split-Robust}: To reduce dependence on a single sample split, repeat cross-fitting $S$ times and use the median of the $S$ estimates as the final split-robust estimator \citep{chernozhukov18}. Greater values of $S$ are advised to decrease variability. In the simulation studies presented in Section \ref{sec:simul}, a value of $S=1$ performed adequately, while in the data analysis in Section \ref{sec:application}, stable results were produced with $S=15$.

Estimators that utilize all three of these modifications are referred to as split-robust, bounded, subsampling nonparametric cross-fitting (SBS-NCF) estimators.
Algorithm S1 in supplementary material Section A.8 describes step-by-step construction of the SBS-NCF estimators, which are evaluated in the simulation studies and real data analysis presented below.

\section{Theoretical results}
\label{sec:theory}

\subsection{Large sample properties}

Here, the large sample properties of the NCF estimator are presented.
Let $\norm{f}_{L_2(P)} = \big\{ \int f(\mathbf{o})^2 dP(\mathbf{o}) \big\}^{1/2}$ denote $L_2(P)$ norm, 
treating $f$ as fixed even when it is estimated from the sample and thus random,
where $P$ is the observed data distribution such that $\BO = (\BY, \boldsymbol{\Delta}, \AXN) \sim P$.
Let $\norm{\cdot}_2$ be the Euclidean vector norm.
Also, let
$\lambdaTij{u}
=
- d \log{\STij{u}} / du$ 
denote the conditional event hazard function,
and let
$\Blambda^T(u|\axni) 
\allowbreak
= 
\big[
    \lambda_{ij}^T(u|\AaXxNni) 
\big]_{j=1}^{n_i}$.
The estimator
$\BlambdaThatk(u|\axni)
=
\big[
    \widehat{\lambda}_{ij,(k)}^T(u|\AaXxNni) 
\big]_{j=1}^{n_i}$
is obtained from the forward difference of 
$-\textup{log}\big(\widehat{S}_{ij,(k)}^T\big)$.
The following is assumed in the derivation of the large sample properties of the NCF estimators.

\begin{assumption}[Conditions for large sample properties]
	\label{assump:nuis}
	For all 
	$i = 1,\dots,m$,
	$j = 1,\dots, \allowbreak N_i$,
	$\ai, \mathbf{a}'_i \in \mathcal{A}(N_i)$,
	and $k = 1,\dots,K$,
	some $d, D > 0$ exist such that 

	\vspace{0.1cm}

	\textup{(L1)} Boundedness: 
	If $\Bwhatk(\aXNi) \ne \textbf{0}_{N_i}$, then $\Hhatk(\aXNi) > d$;
	
	\noindent
	$S^T_{ij}(\Yij|\AXNi),
	\widehat{S}^T_{ij}(\Yij|\AXNi),
	S^C_{ij}(\Yij|\AXNi),
	\widehat{S}^C_{ij}(\Yij|\AXNi)
	> d$;
	
	\noindent
    $\normt{\Bphi(\Ba', \XNi; \Ba)},
    \normt{\Bphihatk(\Ba', \XNi; \Ba)},
    \normt{\Bw(\aXNi)},
    \normt{\Bwhatk(\aXNi)}
    \le D$.

	\vspace{0.2cm}
    \textup{(L2)} Convergence rates: For some $r_{H}, r_{\Blambda^T}, r_{\BS^C}, r_{\Bphi}, r_{\Bw} = O(1)$, 
	
	\noindent
	\scalebox{0.95}{
    $
    \normp{
            \big(\Hhatk-H\big)(\AXN)
    }
    =
    O_{P}(r_H)
    $; 
    $
    \normp{
        \int_{0}^{\infty}
            \normt{
                \big(
                    \BlambdaThatk
                    -
                    \Blambda^T
                \big)(u|\AXN)
            }
        du
    }
    =
    O_{P}(r_{\Blambda^T})
    $};

	\noindent
    \scalebox{0.95}{
    $
    \normp{
        \int_{0}^{\infty}
            \normt{
                \big(
                    \BSChatk
                    -
                    \BS^C
                \big)(u|\AXN)
            }
        du
    }
    =
    O_{P}(r_{\BS^C})
    $;
    $
    \normp{
		\sumaN
            \normt{
                \big(
                    \Bphihatk - \Bphi
                \big)
                (\mathbf{A}, \mathbf{X}, N; \mathbf{a})
            }
    }
    =
    O_{P}(r_{\Bphi})
    $; 
    } 

	\noindent
    \scalebox{0.95}{
    $
    \normp{
		\sumaN
            \normt{
                \big(
                \Bwhatk
                -
                \Bw
                \big)
                (\mathbf{a}, \mathbf{X}, N)
                +
                \sum_{\mathbf{a}' \in \mathcal{A}(N)}
                    \Bphihatk(\mathbf{a}', \mathbf{X}, N; \mathbf{a})
                    H(\mathbf{a}', \mathbf{X}, N)
            }
    }
    =
    O_{P}(r_{\Bw}^2)
    $.
    }

	\vspace{0.2cm}
	
	\textup{(L3)} Finite cluster size: $\text{pr}(N_i \le n_{\max}) = 1$ for some $n_{\max} \in \mathbb{N}$.

\end{assumption}

Condition (L1) bounds the nuisance functions and their estimators, 
while (L2) specifies the convergence rates of nuisance functions estimators.
The last condition in (L2) describes the convergence rate of the second order remainder term in the von Mises expansion \citep{hines22} of $\Bw$ conditional on $\mathbf{X}, N$, where CIF $\Bphi$ acts as a pathwise derivative conditionally.
Condition (L3) ensures that the cluster sizes are bounded.
The following theorem provides the consistency and asymptotic normality of $\widehat{\Psi}(\cR;\Bw)$, 
as well as the consistency of the variance estimator $\widehat{\sigma}(\cR;\Bw)^2$.

\begin{theorem}
\label{thm:largesample}
Under \textup{(I1) -- (I5)}, \textup{(C1)}, and \textup{(L1) -- (L3)}, the following hold as $m \to \infty$.
\vspace{-0.3cm}
\begin{enumerate}
        
    \item[\textup{(i)}] 
    \textup{\underline{(Consistency)}}
    If $\rw = o(1)$ 
    and 
    $\rT(\rH + \rphi + \rC) = o(1)$, 
    then $\widehat{\Psi}(\cR;\Bw) \overset{p}{\to} \Psi(\cR;\Bw)$.
    
    \item[\textup{(ii)}] 
    \textup{\underline{(Asymptotic Normality)}}
    If 
    $\rphi + \rH + \rT + \rC = o(1)$, 
    $\rw = o(m^{-1/4})$, 
    and $\rT(\rH + \rphi + \rC) = o(m^{-1/2})$, 
    then
    $
    m^{1/2}\{\widehat{\Psi}(\cR;\Bw) - \Psi(\cR;\Bw)\}
    \overset{d}{\to}
    N\big(0,\sigma(\cR; \Bw)^2\big)
    $,
    where 
    $
    \sigma(\cR; \Bw)^2
    \allowbreak
    =
    \textup{var}
        \big\{ 
            \varphi(\cR; \mathbf{O}, \boldsymbol{\eta}) 
        \big\}
    $.
    
    \item[\textup{(iii)}] 
    \textup{\underline{(Consistent Variance Estimator)}}
    If $\rw + \rphi + \rH + \rT + \rC = o(1)$,
    then $\widehat{\sigma}(\cR;\Bw) \overset{p}{\to} \sigma(\cR; \Bw)$.
    Thus, if all conditions hold,
    $
    m^{1/2}\{\widehat{\Psi}(\cR;\Bw) - \Psi(\cR;\Bw)\}/\widehat{\sigma}(\cR;\Bw)
    \overset{d}{\to}
    N(0,1)
    $.
        
    \end{enumerate}

\end{theorem}
Note that if $\Bw$ is consistently estimated, $\Blambda^T$ need not be consistent as long as $H$, $\Bphi$, and $\BS^C$ are consistent, and vice versa, for $\widehat{\Psi}(\cR;\Bw)$ to be consistent. For asymptotic normality, a convergence rate of $o(m^{-1/4})$ for all nuisance estimators suffices, which is slower than the parametric $m^{-1/2}$ rate. Many modern data-adaptive methods meet these conditions. For instance, random (survival) forests achieve rates as fast as $m^{-1/(p+2)}$ in $L_2(P)$, where $p$ is the number of continuous components in $\BX$, under regularity conditions \citep{biau2012analysis, cui2022consistency}. Kernel regression for dependent data can yield $m^{-2/(4+p)}$ rates with optimal bandwidths \citep{sun2019counting, park2022efficient}.
The efficiency loss due to censoring is captured by the variance difference between $\phiP(\cR; \Oi, \Beta)$ and $\phiF(\cR; \Zi, \Beta)$:
$\text{var} \Big( 
	\bP(\AbarXNi)^{-1} \Bw(\AXNi)^\top 
	\big[ 
		\cR(\Yij) \big\{ \Deltaij/\SCij{\Yij} - 1 \big\} 
	\big]_{j=1}^{N_i} 
	+ \AUG(\cR; \Oi) 
\Big)$,
which reduces to 0 in the absence of censoring. Finally, the consistency of the variance estimator ensures valid inference for $\Psi(\cR;\Bw)$.

The results in Theorem \ref{thm:largesample} apply to any policy $Q$ satisfying (C1). Table \ref{tab:examples} outlines sufficient conditions for the consistency and asymptotic normality of $\widehat{\Psi}(\cR;\Bw)$ and the consistency of $\widehat{\sigma}(\cR;\Bw)$ for example policies in Section \ref{sec:example_policies}. 
For the CIPS policy, conditional independence of $A_{ij}$'s is assumed,
and the individual-level propensity score $\pi(j, \xn) = \bP(A_j = 1 | \BX = \Bx, N = n)$ is estimated instead of directly estimating $H$. 
The estimator of $H$ can then be constructed as $\widehat{H}(\aXNi) = \prod_j \widehat{\pi}(j,\XNi)^{\aij} \{1-\widehat{\pi}(j,\XNi)\}^{1-\aij}$.
Assume the convergence rate of the $\pi$ estimator is
$ 
\big|\big|
	\sum_{j=1}^{N}
		\lvert
			(\widehat{\pi} - \pi) (j, \mathbf{X}, N)
		\rvert
\big|\big|_{L_2(P)}
\allowbreak
=
O_{P}(r_{\pi})
$
for some $r_{\pi} = O(1)$.

\begin{table}[H]
    \captionsetup{width=\textwidth}
    \caption{Sufficient conditions for the large sample properties of proposed estimators}
    \renewcommand{\arraystretch}{1.3} 
    \label{tab:examples}
    \centering
        \resizebox{\textwidth}{!}{%
        \begin{tabular}{cccc}
        \toprule
        Policy & Consistency & Asymptotic Normality & Consistent Variance Estimator \\
        \toprule
        Type B & $\rT(\rH+\rC) = o(1)$ & $\rT(\rH+\rC) = o(m^{-1/2})$ & $\rH + \rT + \rC = o(1)$  \\
        CIPS & $\rpi = o(1)$, $\rT \rC = o(1)$ & $\rpi = o(m^{-1/4})$, $\rT(\rpi + \rC) = o(m^{-1/2})$ & $\rpi + \rT + \rC = o(1)$  \\
        TPB & $\rH = o(1)$, $\rT \rC = o(1)$ & $\rH = o(m^{-1/4})$, $\rT(\rH + \rC) = o(m^{-1/2})$ & $\rH + \rT + \rC = o(1)$ \\
        \bottomrule
        \end{tabular}%
        }
        \captionsetup{width=0.9\textwidth}
    \end{table}

Since the Type B policy does not depend on the observed data distribution,
i.e., $\Bw$ is a constant functional of $P$,
the CIF $\Bphi$ equals $\textbf{0}_{N_i}$. 
This makes the proposed Type B policy estimators consistent if either 
(i) $\Blambda^T$ or
(ii) $H$ and $\BS^C$,
but not necessarily both,
are consistently estimated.
In contrast, for the CIPS (or TPB) policy, $\pi$ (or $H$) needs to be consistently estimated since the policy distribution is constructed based on the factual propensity. 
However, as long as $\pi$ (or $H$) is consistently estimated, 
only either $\Blambda^T$ or $\BS^C$ needs to be consistently estimated.

\subsection{Weak convergence over policies}

Considering a collection of policies 
$Q(\cdot|\mathbf{X}_i, N_i;\theta)$ indexed by $\theta \in \allowbreak \Theta \allowbreak \subset \allowbreak \bR$,
it is possible to obtain more general theoretical results, as presented in the following theorem.

\begin{theorem}
	\label{thm:weakconv_theta}

	Let $\Bw(\axn; \theta)$ and $\Bphi(\mathbf{a}', \xn; \mathbf{a}; \theta)$ denote the nuisance functions for the policy with index $\theta$.
	Assume the uniform convergence of nuisance functions:
	$\sup_{\theta \in \Theta}
	\big|\big|
		\sumaN
			\big|\big|
				\big(
					\Bphihatk - \Bphi
				\big)
				(\mathbf{A}, \mathbf{X}, N; \mathbf{a}; \theta)
			\big|\big|_2
	\big|\big|_{L_2(P)}
	=
	O_P(r_{\Bphi})$
	and
	$\sup_{\theta \in \Theta}
	\allowbreak
	\big|\big|
	\allowbreak
		\sumaN
		\allowbreak
			\big|\big|
			\allowbreak
				\big(
				\Bwhatk
				-
				\Bw
				\big)
				(\mathbf{a}, \mathbf{X}, N; \theta)
				\allowbreak
				+
				\allowbreak
				\sum_{\mathbf{a}' \in \mathcal{A}(N)}
				\allowbreak
					\Bphihatk(\mathbf{a}', \mathbf{X}, N; 
					\allowbreak
					\mathbf{a}; \theta)
					\allowbreak
					H(\mathbf{a}', \mathbf{X}, N)
					\allowbreak
			\big|\big|_{2}
	\big|\big|_{L_2(P)}
	\allowbreak
	=
	\allowbreak
	O_P(r_{\Bw}^2)$.
	Further, assume the function classes
	$\cF_{\Bw} = \{\Bw(\axn;\theta): \theta \in \Theta\}$
	and
	$\cF_{\Bphi} = \{\Bphi(\Ba',\xn;\Ba, \theta): \theta \in \Theta\}$
	are Donsker classes,
	and all conditions in Theorem \ref{thm:largesample} hold.
	Let 
	$\Psi(\cR; \theta)$
	and
	$\varphi(\cR, \theta; \mathbf{O}, \boldsymbol{\eta})$
	denote the target estimand
	and the proposed estimating function under the policy with index $\theta$, respectively,
	and
	$
    \sigma(\cR; \theta)^2
    \allowbreak
    =
    \textup{var}
        \big\{ 
            \varphi(\cR, \theta; \mathbf{O}, \boldsymbol{\eta}) 
        \big\}
    $.
	Then,
	$m^{1/2}\big\{
		\widehat{\Psi}(\cR; \cdot)
		- 
		\Psi(\cR; \cdot)
	\big\} 
	/ 
	\widehat{\sigma}(\cR; \cdot)
	\rightsquigarrow
	\mathbb{G}(\cdot)$
	in $\ell^{\infty} (\Theta)$
	as $m \to \infty$,
	where $\ell^{\infty} (\Theta)$ is a function space with the finite supremum norm over $\Theta$,
	and
	$\mathbb{G}(\cdot)$ is a mean zero Gaussian process with covariance function
	$
	E \{ 
		\mathbb{G}(\theta_1) 
		\mathbb{G}(\theta_2) 
	\} \allowbreak
	= 
	\textup{Cov} [
		\{\varphi(\cR, \theta_1; \mathbf{O}, \boldsymbol{\eta}) 
		- 
		\Psi(\cR; \theta_1)\}
		/ 
		\sigma(\cR; \theta_1)
		,
        \allowbreak
		\{\varphi(\cR, \theta_2; \mathbf{O}, \boldsymbol{\eta}) 
		- 
		\Psi(\cR; \theta_2)\}
		/ 
		\sigma(\cR; \theta_2)
	]$.

\end{theorem}

For Type B and CIPS policies, 
$\Bw$ and $\Bphi$ are Lipschitz continuous with respect to the policy index $\theta$. 
For TPB policy, 
$\Bw$ and $\Bphi$ are step functions with a finite number of jumps under the finite cluster size assumption (L3). 
Thus, $\cF_{\Bw}$ and $\cF_{\Bphi}$ are Donsker classes, and Theorem \ref{thm:weakconv_theta} applies to these example policies.

Based on Theorem \ref{thm:weakconv_theta}, a uniform confidence band (UCB) for $\Psi(\cR; \theta)$ over $\Theta$ can be constructed. The $(1-\alpha)$ UCB is obtained by finding the critical value $c_{\alpha}$ satisfying:
\begin{align*}
	\text{pr}\left\{
		\sup_{\theta \in \Theta}
		\left|
			\frac
				{\widehat{\Psi}(\cR; \theta)
				- 
				\Psi(\cR; \theta)}
				{\widehat{\sigma}(\cR; \theta) / m^{1/2}}
		\right| 
		\le 
		c_{\alpha} 
	\right\}
	=
	1-\alpha+o(1),
\end{align*}
where the supremum above can be approximated using the multiplier bootstrap process \citep{chernozhukov2014gaussian, belloni2015uniform, kennedy19}, as detailed in Theorem \ref{thm:uniform_inference}.

\begin{theorem}
	\label{thm:uniform_inference}
	Let $\widehat{c}_{\alpha}$ be the $(1-\alpha)$ sample quantile of the supremum of the multiplier bootstrap process:
	\begin{align*}
		\sup_{\theta \in \Theta}
		\left|
			\sumk
			\sumik
			\left[
				\xi_i
				\left\{
				\frac
					{\varphi(\cR, \theta; \Oi, \Betahatk) - \widehat{\Psi}(\cR; \theta)}
					{\widehat{\sigma}(\cR; \theta) / m^{1/2}}
				\right\}
			\right]
		\right| 
		,
	\end{align*}
	where $(\xi_1, \dots, \xi_m)$ are iid Rademacher random variables (i.e., $\xi_i \in \{-1, 1\}$ with equal probability). Then, under the conditions of Theorem \ref{thm:weakconv_theta},
	\begin{align*}
		\textup{pr}\left\{
			\widehat{\Psi}(\cR; \theta)
			-
			\widehat{c}_{\alpha}
			\frac{\widehat{\sigma}(\cR; \theta)}{m^{1/2}}
			\le
			\Psi(\cR; \theta)
			\le
			\widehat{\Psi}(\cR; \theta)
			+
			\widehat{c}_{\alpha}
			\frac{\widehat{\sigma}(\cR; \theta)}{m^{1/2}}
			\textup{, for all } \theta \in \Theta
		\right\}
		=
		1-\alpha+o(1).
	\end{align*}
\end{theorem}

\vspace{0.3cm}
The multiplier bootstrap method is computationally efficient as it avoids recomputing nuisance function estimators, requiring only the generation of multipliers $\xi_i$'s. The UCB is particularly useful for assessing the presence of interference in a given application.
For instance, changes in EAPO when untreated, $\mu_0(\cR;\theta)$, with respect to $\theta$ indicate the presence of interference over the given policy index range $\Theta$; thus one can test for interference by checking if the confidence band contains a horizontal line.
Furthermore, the UCB can control false discovery rates when multiple hypotheses are tested over $\Theta$, i.e., checking multiple locations to see if the point-wise confidence intervals exclude the null value simultaneously.

When the counterfactual risk is of interest, with $\cR(T; \tau) = \indicator(T \le \tau)$, 
Theorems \ref{thm:largesample} and \ref{thm:weakconv_theta} can be extended to cases where the estimator depends on the time of interest $\tau$.
These results show that the proposed estimators for the counterfactual risks weakly converge to a Gaussian process, as in standard survival analysis.

\begin{theorem}
\label{thm:weakconv_tau_theta}
Assume all conditions in Theorem \ref{thm:weakconv_theta} hold.
Let $\Psi(\tau; \theta)$ and $\varphi(\tau, \theta; \mathbf{O}, \boldsymbol{\eta})$ denote the target estimand and the proposed estimating function when $\cR(T; \tau) = \indicator(T \le \tau)$ under the policy with index $\theta$, respectively,
and
$
\sigma(\tau; \theta)^2
\allowbreak
=
\textup{var}
    \big\{ 
        \varphi(\tau, \theta; \mathbf{O}, \boldsymbol{\eta}) 
    \big\}
$.
Then,
$m^{1/2}\big\{
	\widehat{\Psi}(\cdot; \cdot)
	- 
	\Psi(\cdot; \cdot)
\big\}
\big/ 
	\widehat{\sigma}(\cdot; \cdot)
\rightsquigarrow
\mathbb{G}(\cdot, \cdot)$
in $\ell^{\infty} ([0,\infty) \times \Theta)$
as $m \to \infty$,
where $\mathbb{G}(\cdot, \cdot)$ is a mean zero Gaussian random field with covariance function
$E \{ 
	\mathbb{G}(\tau_1, \theta_1) 
	\mathbb{G}(\tau_2, \theta_2) 
\} 
= 
\textup{Cov} [
		\{\varphi(\tau_1, \theta_1; \mathbf{O}, \boldsymbol{\eta}) 
		- 
		\Psi(\tau_1; \theta_1)\}
		/ 
		\sigma(\tau_1; \theta_1)
		,
        \allowbreak
		\{\varphi(\tau_2, \theta_2; \mathbf{O}, \boldsymbol{\eta}) 
		- 
		\Psi(\tau_2; \theta_2)\}
		/ 
		\sigma(\tau_2; \theta_2)
	]
$.

\end{theorem}

In conclusion, under mild conditions, the proposed NCF estimators are consistent and asymptotically normal with $m^{-1/2}$ rate of convergence, even when nonparametric data-adaptive methods are used for nuisance function estimation. 
The large sample properties of the modified SBS-NCF estimators are similar to those of the (unmodified) NCF estimators and are detailed in supplementary material Sections A.9 and A.10.

\vspace{-0.2cm}

\section{Simulation study}
\label{sec:simul}

The finite sample performance of the proposed estimators was evaluated using $D = 1000$ simulated datasets. Each dataset consisted of $m = 200$ clusters, with cluster sizes $N_i$ sampled from $\{5, \dots, 20\}$. For each cluster, five cluster-level covariates $X_{i,c1}, \dots, X_{i,c5} \overset{iid}{\sim} \text{N}(0,1)$ were generated. For unit $j$, individual-level covariates $X_{ij1}, \dots, X_{ij5} \overset{iid}{\sim} \text{N}(0,1)$ and $X_{ij6}, \dots, X_{ij10} \overset{iid}{\sim} \text{Bernoulli}(0.5)$ were generated.
The treatment status was sampled as $A_{ij} \sim \text{Bernoulli}(\pi_{ij})$, where 
$\pi_{ij} = \Phi\big(-0.1 - 0.2X_{ij1} + 0.2X_{ij2}^2 + 0.1\mathbbm{1}(X_{ij1} > 0)X_{ij6} - 0.3\max\{X_{i,c1}, 0.5\} + b_i\big)$,
$b_i \overset{iid}{\sim} \text{N}(0,0.5)$ is a cluster-level random effect, and $\Phi$ is the standard normal cumulative distribution function.
The event time $T_{ij}$ and censoring time $C_{ij}$ were generated from Gamma distributions $\text{Gamma}(a,s)$ with a scale parameter $s = 2$ and shape parameters $a = 0.1 + 0.3A_{ij} + 0.3\text{sin} (1.57\overline{\mathbf{A}}_{i(-j)}) X_{ij1}^2 + 0.1A_{ij}\overline{\mathbf{A}}_{i(-j)} + 0.1X_{ij2}^2\max\{X_{i,c1}, 0.1\} + 0.1\mathbbm{1}(X_{i,c1}X_{i,c2} < 0.5)$ for $T_{ij}$
and $a = 0.2 + 0.5A_{ij} + 0.5\overline{\mathbf{A}}_{i(-j)}X_{ij2}^2 + 0.1\max\{X_{ij1}, 0.1\} + 0.1\mathbbm{1}(X_{i,c2} < 0.5)$ for $C_{ij}$,
respectively.
Finally, the observed time $\Yij = \min\{\Tij, \Cij\}$ and event indicator $\Deltaij = \indicator(\Tij \le \Cij)$ were computed for inclusion in the simulated datasets.

The finite sample performance of the SBS-NCF estimators was evaluated with $K=2$, $r=100$, and $S=1$. 
The nuisance function $H$ was estimated using random effect BART \citep[\texttt{dbarts} R package,][]{dorie2022dbarts}, while $\cF_{*}^T$ and $\cS_{*}^C$ were estimated using random survival forests \citep[\texttt{randomForestSRC} R package,][]{ishwaran2023random}. 
The target parameters were Type B policy estimands for $\alpha \in \{0.3, 0.45, 0.6\}$ with $\cR(T; \tau) = \indicator(T \le \tau)$ at $\tau \in \{0.2, 0.4\}$. Causal effects were compared to the baseline $\alpha' = 0.45$. Results are shown in Table \ref{tab:simulTypeB}, with additional results for TPB policy estimands in supplementary material Section C.1.
The SBS-NCF estimators performed well, with minimal empirical bias (Bias), close agreement between average standard error (ASE) and empirical standard error (ESE), 
and 95\% point-wise confidence interval (Cov) and uniform confidence band (UCov) coverages near nominal levels. 
These results demonstrate that the proposed nonparametric estimator can yield valid statistical inference in finite samples.

\vspace{-0.2cm}
\begin{table}[ht]
	\centering
	\captionsetup{width=\textwidth}
	\caption{Simulation results for 
	the SBS-NCF estimators
	for Type B policy with $\alpha \in \{0.3, 0.45, 0.6\}$, $\alpha' = 0.45$, and $\tau \in \{0.2, 0.4\}$}
	\vspace{-0.1cm}
	\renewcommand{\arraystretch}{1}
	\label{tab:simulTypeB}
	\resizebox{\textwidth}{!}{%
	\begin{tabular}{ccccccccccccccccc}
	\hline
	\multirow{2}{*}{$\alpha$} &  & \multirow{2}{*}{Estimand}                                                   &  & \multicolumn{6}{c}{$\tau = 0.2$}      &  & \multicolumn{6}{c}{$\tau = 0.4$}      \\ \cline{5-10} \cline{12-17} 
							  &  &                                                                             &  & Truth & Bias & ASE & ESE & Cov & UCov &  & Truth & Bias & ASE & ESE & Cov & UCov \\ \hline
	\multirow{7}{*}{0.3}      &  & $\mu_{\scriptscriptstyle \textup{B}}(\tau; \alpha)$                         &  & 45.1  & 0.1  & 2.0 & 2.1 & 94  & 95   &  & 55.5  & -0.1 & 2.1 & 2.1 & 95  & 95   \\
							  &  & $\mu_{\scriptscriptstyle \textup{B}, \scriptstyle 1}(\tau; \alpha)$         &  & 25.2  & -0.9 & 2.8 & 2.8 & 93  & 92   &  & 36.7  & -1.1 & 3.1 & 3.2 & 93  & 91   \\
							  &  & $\mu_{\scriptscriptstyle \textup{B}, \scriptstyle 0}(\tau; \alpha)$         &  & 53.7  & 0.3  & 2.5 & 2.6 & 94  & 93   &  & 63.6  & 0.2  & 2.5 & 2.6 & 95  & 94   \\
							  &  & $DE_{\scriptscriptstyle \textup{B}}(\tau; \alpha)$                          &  & -28.5 & -1.2 & 3.6 & 3.6 & 94  & 90   &  & -26.9 & -1.3 & 3.9 & 4.0 & 95  & 90   \\
							  &  & $SE_{\scriptscriptstyle \textup{B}, \scriptstyle 1}(\tau; \alpha, \alpha')$ &  & 7.5   & 0.1  & 1.5 & 1.6 & 95  & 94   &  & 7.5   & 0.1  & 1.5 & 1.6 & 96  & 95   \\
							  &  & $SE_{\scriptscriptstyle \textup{B}, \scriptstyle 0}(\tau; \alpha, \alpha')$ &  & 2.7   & -0.1 & 2.0 & 2.0 & 96  & 96   &  & 3.2   & 0.0  & 2.3 & 2.3 & 96  & 95   \\
							  &  & $OE_{\scriptscriptstyle \textup{B}}(\tau; \alpha, \alpha')$                 &  & 3.7   & -0.3 & 1.8 & 1.9 & 94  & 93   &  & 3.6   & -0.3 & 1.9 & 1.9 & 95  & 93   \\ \hline
	\multirow{4}{*}{0.45}     &  & $\mu_{\scriptscriptstyle \textup{B}}(\tau; \alpha)$                         &  & 37.6  & 0.0  & 1.4 & 1.5 & 95  & 95   &  & 48.1  & -0.2 & 1.6 & 1.6 & 96  & 95   \\
							  &  & $\mu_{\scriptscriptstyle \textup{B}, \scriptstyle 1}(\tau; \alpha)$         &  & 22.5  & -0.8 & 1.8 & 1.9 & 91  & 92   &  & 33.5  & -1.1 & 2.1 & 2.1 & 92  & 91   \\
							  &  & $\mu_{\scriptscriptstyle \textup{B}, \scriptstyle 0}(\tau; \alpha)$         &  & 50.0  & 0.7  & 2.1 & 2.2 & 93  & 93   &  & 60.0  & 0.5  & 2.2 & 2.2 & 93  & 94   \\
							  &  & $DE_{\scriptscriptstyle \textup{B}}(\tau; \alpha)$                          &  & -27.6 & -1.5 & 2.7 & 2.7 & 91  & 90   &  & -26.5 & -1.6 & 2.9 & 3.0 & 91  & 90   \\ \hline
	\multirow{7}{*}{0.6}      &  & $\mu_{\scriptscriptstyle \textup{B}}(\tau; \alpha)$                         &  & 31.3  & -0.2 & 1.6 & 1.6 & 95  & 95   &  & 41.5  & -0.4 & 1.7 & 1.8 & 95  & 95   \\
							  &  & $\mu_{\scriptscriptstyle \textup{B}, \scriptstyle 1}(\tau; \alpha)$         &  & 20.5  & -0.7 & 1.8 & 1.8 & 94  & 92   &  & 31.0  & -1.1 & 2.1 & 2.2 & 93  & 91   \\
							  &  & $\mu_{\scriptscriptstyle \textup{B}, \scriptstyle 0}(\tau; \alpha)$         &  & 47.4  & 0.7  & 2.7 & 2.9 & 93  & 93   &  & 57.3  & 0.6  & 2.9 & 3.1 & 95  & 94   \\
							  &  & $DE_{\scriptscriptstyle \textup{B}}(\tau; \alpha)$                          &  & -27.0 & -1.4 & 3.1 & 3.5 & 91  & 90   &  & -26.3 & -1.7 & 3.5 & 3.8 & 90  & 90   \\
							  &  & $SE_{\scriptscriptstyle \textup{B}, \scriptstyle 1}(\tau; \alpha, \alpha')$ &  & -6.4  & -0.2 & 1.2 & 1.3 & 95  & 94   &  & -6.5  & -0.3 & 1.3 & 1.4 & 96  & 95   \\
							  &  & $SE_{\scriptscriptstyle \textup{B}, \scriptstyle 0}(\tau; \alpha, \alpha')$ &  & -2.0  & 0.0  & 1.4 & 1.4 & 96  & 96   &  & -2.5  & 0.0  & 1.7 & 1.7 & 95  & 95   \\
							  &  & $OE_{\scriptscriptstyle \textup{B}}(\tau; \alpha, \alpha')$                 &  & -2.6  & 0.0  & 2.0 & 2.2 & 95  & 93   &  & -2.7  & 0.1  & 2.1 & 2.4 & 94  & 93   \\ \hline
	\end{tabular}
	}
    \caption*{\small
    Truth: true value of the estimand ($\times$100),
    Bias: average bias of estimates ($\times$100), 
    ASE: average standard error estimates ($\times$100), 
    ESE: empirical standard error ($\times$100), 
    Cov: 95\% point-wise confidence interval coverage,
	UCov: 95\% uniform confidence band coverage.}
	\vspace{-0.5cm} 
\end{table}

\begin{figure}[h]
	\centering
	\includegraphics[width=\textwidth]{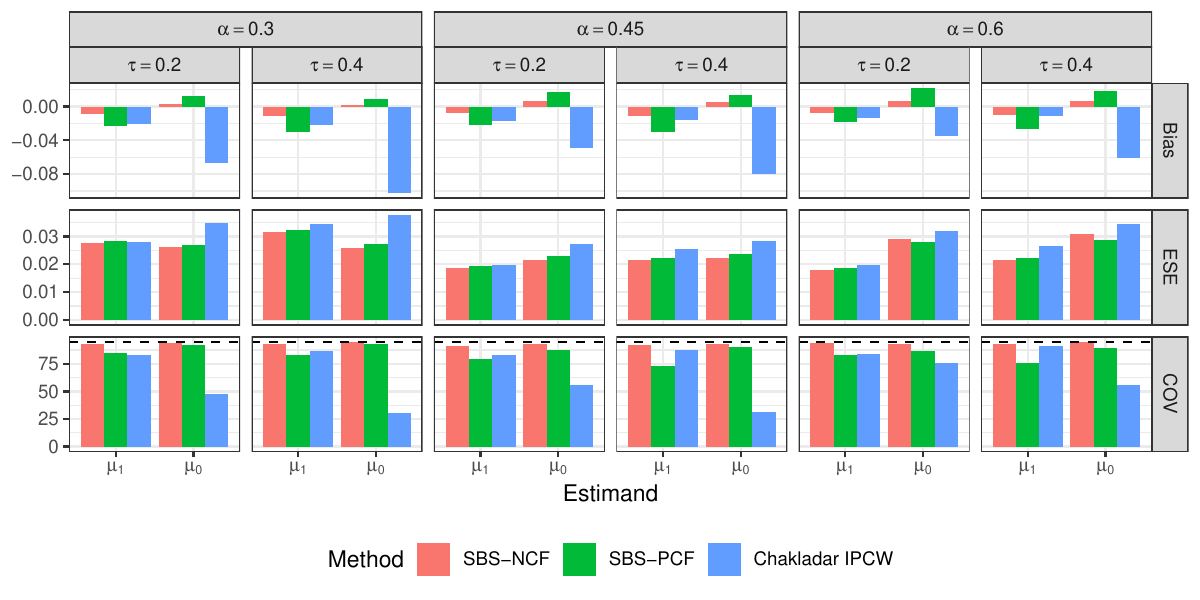}
	\caption{Finite sample performance of the SBS-NCF, SBS-PCF, and \cite{chakladar2022inverse} IPCW estimators for Type B policy estimands; 
	Bias: average bias of estimates, ESE: empirical standard error, Cov: 95\% point-wise confidence interval coverage.}
	\label{fig:NCF_PCF_IPCW_comparison}
\end{figure}

Second, the SBS-NCF estimators were compared with the parametric IPCW estimators proposed by \citet{chakladar2022inverse}. For additional comparison, parametric cross-fitting (SBS-PCF) estimators were constructed using the same procedure as the SBS-NCF estimators, except $H$ was estimated via a GLMM, and $\cF_{*}^T$ and $\cS_{*}^C$ were estimated using Cox regression. The target parameters were Type B policy estimands $\mu_{\scriptstyle \text{B},1}(\tau; \alpha)$ and $\mu_{\scriptstyle \text{B},0}(\tau; \alpha)$ for $\alpha \in \{0.3, 0.45, 0.6\}$ at $\tau \in \{0.2, 0.4\}$. Results, shown in Fig.~\ref{fig:NCF_PCF_IPCW_comparison}, indicate that the SBS-NCF estimators performed well, while the SBS-PCF and Chakladar IPCW estimators exhibited poor performance due to misspecified nuisance functions.

%
%

Finally, additional simulation studies were conducted, and the results are described in supplementary material Section C.
First, the SBS-NCF estimators were assessed across varying the number of clusters $m$, showing reduced bias, lower empirical standard error (ESE), and improved 95\% CI coverage as $m$ increased,
as supported by Theorem \ref{thm:largesample}.
Second, the unbounded and bounded NCF estimators were compared. 
Both estimators exhibited good finite sample performance, 
while the bounded estimators had smaller bias and ESE in scenarios where inverse probability weights $\Bwhatk (\AXNi) / \Hhatk (\AXNi)$ were possibly extreme (e.g., Type B policy with $\alpha = 0.6$ where $\{\bP(\Aij = 1 | \XNi)\} \approx 0.4$ in the simulation setting).
Third, the effect of subsampling degree $r$ on the finite sample performance of the SBS-NCF estimators was investigated; 
the bias was insensitive to $r$, 
but the empirical SE tended to decrease as $r$ increased and stabilized around $r=100$,
while the 95\% CI coverage achieved the nominal level regardless of $r$.
Lastly, the finite sample performance of the SBS-NCF estimators 
was evaluated for different levels of correlation between treatment within clusters (by varying $\sigma_b = \text{sd}(b_i)$) and different distributions of $N_i$;
the SBS-NCF estimators performed well regardless of these variations,
since the methods can accommodate correlated treatment selection
and are agnostic to the distribution of $N_i$.

\section{Application to cholera vaccine data}
\label{sec:application}

In this section, the proposed methods are applied to a cholera vaccine study conducted in Bangladesh \citep{clemens1988field}. The inferential goal is to quantify the effect of the vaccine on the risk of cholera over time, accounting for confounding, censoring, and interference. Eligible individuals decided whether to participate in the study, and those who did participate were randomised to receive one of two types of cholera vaccine or a placebo; as in previous analyses \citep[e.g.,][]{perez14}, no distinction is made between the two cholera vaccines in this analysis.
Baseline covariates and outcome data were available for all trial-eligible individuals, including those who did not participate in the trial, necessitating adjustment for potential confounding for the analysis of all individuals in the data set.
The time to cholera incidence (event time) is subject to censoring due to the end of study follow-up, emigration from the study location, or death, making it essential to address censoring in the analysis.
Additionally, prior research on the effects of the cholera vaccine suggests possible interference within baris, i.e., clusters of patrilineally-related households \citep{ali2005herd, root2011role, perez14, chakladar2022inverse}. However, these earlier analyses relied on parametric models and focused solely on Type B policy estimands.


Given spatial and patrilineal separation between baris, no interference is assumed between individuals in different baris.
Included in the analysis are 5,625 baris of size ranging from 2 to 239,
with a total of 112,154 individuals, of whom 48,763 were vaccinated.  
There were 458 incident cases of cholera,
with a mean event time of 256 days (IQR: [183, 364] days),
and 
111,696 individuals' event times were censored 
(IQR: [397, 431] days).
Refer to supplementary material Section D.1 for details about the observed data distribution.
The SBS-NCF estimators were employed to evaluate 
Type B policies for $\alpha \in [0.3, 0.6]$ 
and TPB policies for $\rho \in [0, 0.5]$
over time (days) period $\tau \in [0,450]$ with $K=5$, $r=100$, $S=15$. 
As in prior analyses \citep{perez14, liu19, barkley20, chakladar2022inverse},
baseline covariates age and distance to the closest river were selected to adjust for confounding. 
The nuisance function $H$ was estimated using the random effect BART,
while $\cF_{*}^T$ and $\cS_{*}^C$ were estimated by random survival forests.



\begin{figure}[h]
    \centerline{\includegraphics[width = \textwidth]{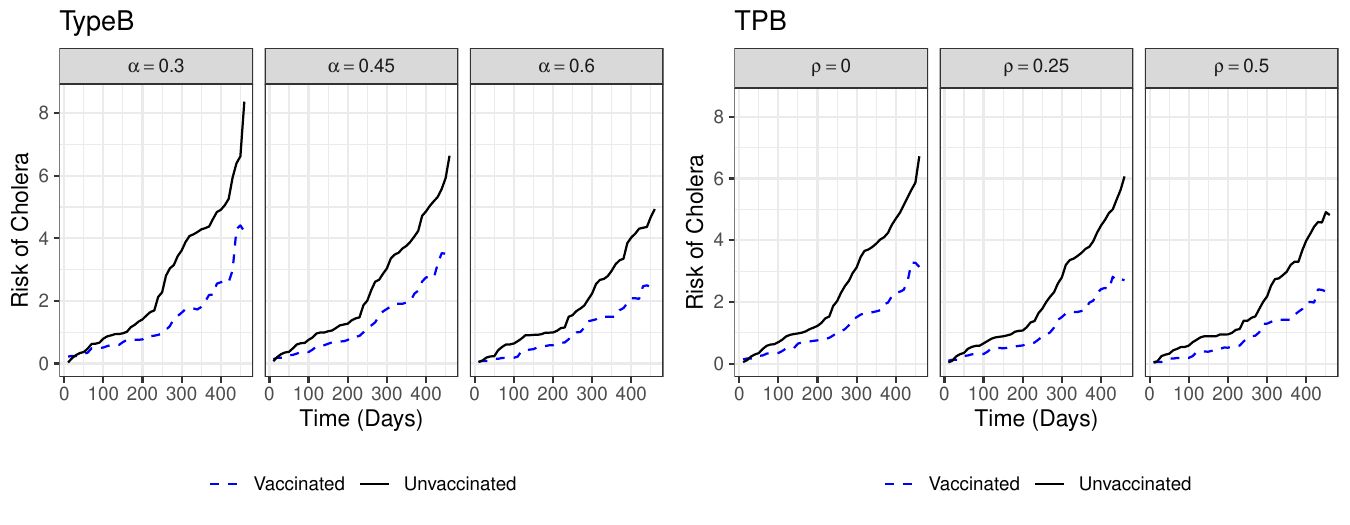}}
    \caption{
    Estimated risk ($\times1000$) of cholera over time when unvaccinated (black solid line, $\mu_0$) or vaccinated (blue dashed line, $\mu_1$) 
    under Type B policies (left) and TPB policies (right).}
\label{fig:risk_combined}
\end{figure}

Figure \ref{fig:risk_combined} presents the estimated risk of cholera when unvaccinated ($\mu_0$) or vaccinated ($\mu_1$) under
Type B policy with $\alpha \in \{0.3, 0.45, 0.6\}$ 
and TPB policy with $\rho \in \{0, 0.25, 0.5\}$.
The estimated risk when unvaccinated is higher than when vaccinated, suggesting a beneficial direct effect of vaccination. 
The estimated risk of cholera when unvaccinated decreases as $\alpha$ or $\rho$ increases, suggesting unvaccinated individuals may benefit from other individuals in their baris being vaccinated. 
The estimated risk of cholera when vaccinated also decreases as $\alpha$ or $\rho$ increases,
although the decrease is modest compared to when unvaccinated.


\begin{figure}[!h]
    \centerline{\includegraphics[width = \textwidth]{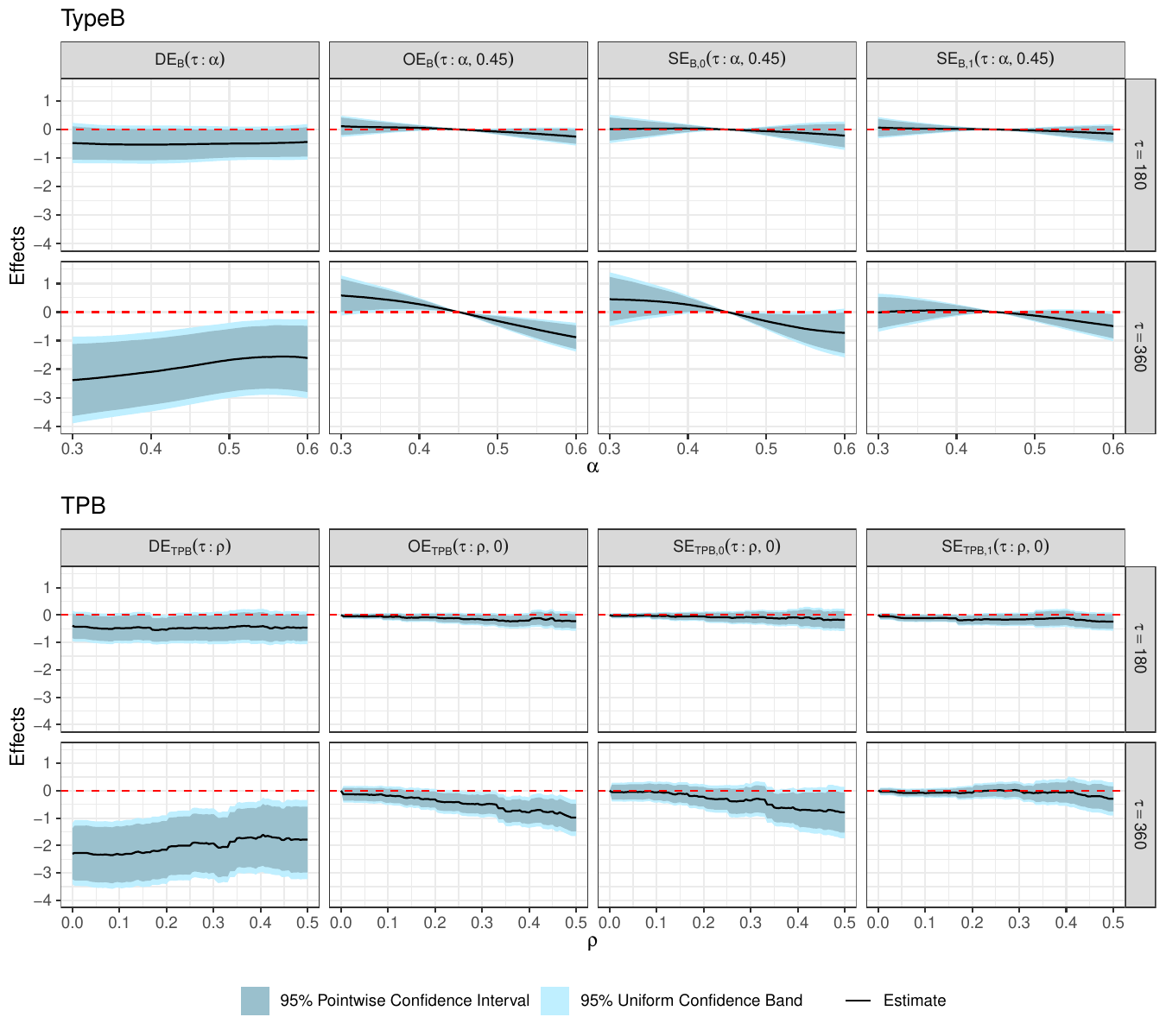}}
    \caption{
    Estimated causal effects ($\times1000$)
    $DE$,  
    $OE$,  
    $SE_0$,  
    $SE_1$
    under 
    Type B policies (top)
    and
    TPB policies (bottom)
    at $\tau \in \{180, 360\}$ days.
	Black line indicate the point estimates,
	light blue shaded area indicates the 95\% point-wise confidence intervals,
	and dark blue shaded area indicates the 95\% uniform confidence bands.
    Red dashed horizontal line indicates the null value of 0.}
\label{fig:effects_combined}
\end{figure}

Estimates of the direct, overall, and spillover effects when untreated and treated at $\tau=180$ days and $\tau=360$ days are shown in Fig. \ref{fig:effects_combined} (see supplementary material Section D.2 for figures with more time points).
Type B policies with
$\alpha \in [0.3, 0.6]$ were compared to $\alpha' = 0.45$ since the observed vaccine coverage was nearly 45\%,
and TPB policies with
$\rho \in [0,0.5]$ were compared to the factual (observed) scenario, which corresponds to $\rho' = 0$.
The direct effect estimates are always negative for both policies, 
with 95\% CIs across $\alpha \in [0.3, 0.6]$ for $\tau=360$ excluding the null value of zero,
suggesting beneficial direct vaccine effects on the one-year risk of cholera. 
Furthermore, 95\% UCBs also exclude the null value of zero, 
providing simultaneous evidence that the direct effect is negative across the entire coverage range.
For example, 
$\widehat{DE}_{\scriptstyle \text{B}}(\tau = 360; \alpha = 0.3) = -2.4$ (95\% CI: [-3.7, -1.2]), indicating that we would expect approximately 2.4 fewer cases of cholera by one year per 1000 vaccinated individuals compared to 1000 unvaccinated individuals if individuals received the vaccine with probability of 0.3.
The direct effect estimates are farther from 0 at $\tau = 360$ than at $\tau = 180$, implying the individual beneficial effect of the vaccine, as measured by the risk difference, is greater one year after vaccination.
The direct effect estimates decrease (in magnitude) when $\alpha$ or $\rho$ increases,
suggesting the individual benefit of being vaccinated decreases as vaccine coverage increases.


Spillover effect estimates for untreated are positive when $\alpha < \alpha' = 0.45$,
and are negative when $\alpha > \alpha' = 0.45$ and $\rho > \rho' = 0$, 
indicating that the unvaccinated individuals' risk of cholera is inversely related to the vaccination coverage,
and the magnitude is larger at the later time point.
For instance,
$\widehat{SE}_{\scriptstyle \text{TPB},0}(\tau = 360; \rho = 0.5, \rho' = 0) = -0.8$ (95\% CI: [-1.6, -0.1])
indicates that if at least 50\% of individuals were vaccinated then 0.8 fewer cases of cholera per 1000 unvaccinated by one year would be expected compared to the factual scenario (no restriction on the proportion vaccinated).
This result suggests the presence of interference; in particular, an unvaccinated individual may benefit from other individuals within their baris receive the vaccine.
On the other hand, the estimates of spillover effect when treated tend to change less with $\alpha$ or $\rho$ with corresponding CIs including the null value of zero, 
implying that once vaccinated
there is minimal additional protective effect from other vaccinated individuals.
Estimates of the overall effect display a similar trend as the spillover effects when untreated whereby the risk of cholera is lower the higher the vaccine coverage, as quantified by negative estimates at higher vaccine coverage, e.g.,
$\widehat{OE}_{\scriptstyle \text{B}}(\tau = 360; \alpha = 0.6, \alpha' = 0.45) = -0.9$ (95\% CI: [-1.3, -0.5]).

Note that even though the results generally align for Type B and TPB policies, 
the interpretation of the estimates differs since the treatment allocation mechanism of each policy is different.
For example, even with $\alpha=0.9$ for the Type B policy, there is still a nonzero probability of every unit being unvaccinated $(1-0.9)^{N_i} > 0$, 
whereas the TPB policy rules out such a possibility unless $\rho = 0$.
The choice of the policy evaluated should be pertinent to the research question, and the interpretation of the estimation result depends on the target policy.

In summary, the cholera vaccine protects individuals from cholera,
and unvaccinated individuals may benefit from spillover effects from vaccinated individuals with the magnitude of such benefit increasing with vaccine coverage.
These results generally align with the prior research \citep{perez14, barkley20, chakladar2022inverse}.
However, 
unlike previous analyses, the results here utilize nonparametric data-adaptive estimation which does not require correctly-specified parametric models
and consider the TPB policy in addition to the classical Type B policy.

\section{Discussion}

There are various potential directions for future research associated with this paper.
Firstly, the asymptotic regimen of the proposed methods depends on the number of clusters $m$ growing large. 
Subsequent research could focus on developing methodologies that are more adept at handling scenarios with a small number of clusters, e.g., where the asymptotics may rely on the size of clusters growing large.
Secondly, extensions of the method to encompass multivalued (more than two options) or continuous treatments could be explored. 
Lastly, 
extensions could be developed that do not require cluster interference and instead allow for more general interference structures.


\begin{center}
{\large\bf SUPPLEMENTARY MATERIAL}
\end{center}

The online supplementary material contains proofs of theorems, additional simulation results, and details of the real data analysis.

\begin{center}
{\large\bf DATA AVAILABILITY STATEMENT}
\end{center}

The cholera vaccine data are not publicly available. Synthetic data resembling the real data are provided in the GitHub repository (\url{https://github.com/chanhwa-lee/NPSACI}) for replicating the analysis in the paper. The repository also contains the source code for reproducing all numerical results.

\if1\blind
{
    \begin{center}
    {\large\bf ACKNOWLEDGEMENTS \& FUNDING INFORMATION}
    \end{center}
    
    The authors thank 
    the Causal Inference Research Lab at the University of North Carolina at Chapel Hill for helpful discussions.    
    Chanhwa Lee and Michael G. Hudgens were supported by NIH R01 AI085073. 
    Donglin Zeng was supported by R01 GM124104. 
    The content is solely the responsibility of the authors and does not necessarily represent the official views of the NIH.
    The authors report there are no competing interests to declare.
    
} \fi

    
    
        
    
    

\spacingset{0.8} 
\bibliographystyle{agsm}
\bibliography{Bibliography.bib}

@Book{cox58,
   author =   {Cox, D. R.},
   title =    {Planning of Experiments},
   publisher =    {Wiley, New York},
   year =     {1958},
 }

@article{hudgens08,
    author = {Michael G Hudgens and M. Elizabeth Halloran},
    title = {Toward Causal Inference With Interference},
    journal = {J. Am. Statist. Assoc.},
    volume = {103},
    number = {482},
    pages = {832-842},
    year  = {2008},
    publisher = {Taylor & Francis}
}

@article{sobel06,
    author = {Michael E Sobel},
    title = {What Do Randomized Studies of Housing Mobility Demonstrate? {C}ausal inference in the face of interference},
    journal = {J. Am. Statist. Assoc.},
    volume = {101},
    number = {476},
    pages = {1398-1407},
    year  = {2006},
    publisher = {Taylor & Francis}
}

@article{barkley20,
  title={Causal inference from observational studies with clustered interference, with application to a cholera vaccine study},
  author={Barkley, Brian G and Hudgens, Michael G and Clemens, John D and Ali, Mohammad and Emch, Michael E},
  journal={Ann. Appl. Statist.},
  volume={14},
  number={3},
  pages={1432--1448},
  year={2020},
  publisher={Institute of Mathematical Statistics}
}

@article{tchetgen12,
    author = {Tchetgen Tchetgen, Eric J and Tyler J VanderWeele},
    title ={On causal inference in the presence of interference},
    journal = {Stat. Methods Med. Res.},
    volume = {21},
    number = {1},
    pages = {55-75},
    year = {2012}
}

@article{Papadogeorgou19,
    author = {Papadogeorgou, Georgia and Mealli, Fabrizia and Zigler, Corwin M.},
    title = {Causal inference with interfering units for cluster and population level treatment allocation programs},
    journal = {Biometrics.},
    volume = {75},
    number = {3},
    pages = {778-787},
    year = {2019}
}

@article{kennedy19,
    author = {Edward H. Kennedy},
    title = {Nonparametric Causal Effects Based on Incremental Propensity Score Interventions},
    journal = {J. Am. Statist. Assoc.},
    volume = {114},
    number = {526},
    pages = {645-656},
    year  = {2019},
    publisher = {Taylor & Francis}
}

@article{perez14,
    author = {Perez-Heydrich, Carolina and Hudgens, Michael G. and Halloran, M. Elizabeth and Clemens, John D. and Ali, Mohammad and Emch, Michael E.},
    title = {Assessing effects of cholera vaccination in the presence of interference},
    journal = {Biometrics.},
    volume = {70},
    number = {3},
    pages = {731-741},
    year = {2014}
}

@article{park2022efficient,
  title={Efficient semiparametric estimation of network treatment effects under partial interference},
  author={Park, Chan and Kang, Hyunseung},
  journal={Biometrika.},
  volume={109},
  number={4},
  pages={1015--1031},
  year={2022},
  publisher={Oxford University Press}
}

@article{liu19,
  title={Doubly robust estimation in observational studies with partial interference},
  author={Liu, Lan and Hudgens, Michael G and Saul, Bradley and Clemens, John D and Ali, Mohammad and Emch, Michael E},
  journal={Stat.},
  volume={8},
  number={1},
  pages={e214},
  year={2019},
  publisher={Wiley Online Library}
}

@article{chernozhukov18,
    author = {Chernozhukov, Victor and Chetverikov, Denis and Demirer, Mert and Duflo, Esther and Hansen, Christian and Newey, Whitney and Robins, James},
    title = "{Double/debiased machine learning for treatment and structural parameters}",
    journal = {Econometrics J.},
    volume = {21},
    number = {1},
    pages = {C1-C68},
    year = {2018},
    month = {01},
    issn = {1368-4221}
}

@book{tsiatis06,
  title={Semiparametric Theory and Missing Data},
  author={Tsiatis, Anastasios A},
  year={2006},
  publisher={New York, NY: Springer}
}

@incollection{kennedy16,  
    title     = {Semiparametric Theory and Empirical Processes in Causal Inference},
    author    = {Kennedy, Edward H.},                       
    pages     = {141--167},                                 
    crossref  = {kennedybook16}                      
}

@book{kennedybook16, 
    year      = {2016},                                     
    editor    = {He, Hua and Wu, Pan and {Chen, DG}},
    title     = {Statistical Causal Inferences and Their Applications in Public Health Research}, 
    booktitle = {Statistical Causal Inferences and Their Applications in Public Health Research}, 
    author    = {Kennedy, Edward H.},            
    publisher = {Springer International Publishing},              
    address   = {Cham}                   
}

@article{hines22,
    author = {Oliver Hines and Oliver Dukes and Karla Diaz-Ordaz and Stijn Vansteelandt},
    title = {Demystifying Statistical Learning Based on Efficient Influence Functions},
    journal = {Am. Stat.},
    volume = {76},
    number = {3},
    pages = {292-304},
    year  = {2022},
    publisher = {Taylor & Francis}
}

@article{park2024minimum,
  title={Minimum resource threshold policy under partial interference},
  author={Park, Chan and Chen, Guanhua and Yu, Menggang and Kang, Hyunseung},
  journal={J. Am. Statist. Assoc.},
  volume={119},
  number={548},
  pages={2881--2894},
  year={2024},
  publisher={Taylor \& Francis}
}

@article{munoz12,
  title={Population intervention causal effects based on stochastic interventions},
  author={Mu{\~n}oz, Iv{\'a}n D{\'\i}az and Van Der Laan, Mark},
  journal={Biometrics.},
  volume={68},
  number={2},
  pages={541--549},
  year={2012},
  publisher={Wiley Online Library}
}

@article{barrera11,
  title={Improving the design of conditional transfer programs: Evidence from a randomized education experiment in {C}olombia},
  author={Barrera-Osorio, Felipe and Bertrand, Marianne and Linden, Leigh L and Perez-Calle, Francisco},
  journal={Am. Econ. J. Appl. Econ.},
  volume={3},
  number={2},
  pages={167--95},
  year={2011}
}

@article{kennedy22,
  title={Semiparametric doubly robust targeted double machine learning: a review},
  author={Kennedy, Edward H},
  journal={arXiv preprint arXiv:2203.06469},
  year={2022}
}

@article{majerek2005conditional,
  title={Conditional strong law of large number},
  author={Majerek, Dariusz and Nowak, Wioletta and Zieba, Wieslaw},
  journal={Int. J. Pure Appl. Math},
  volume={20},
  number={2},
  pages={143--156},
  year={2005}
}

@article{qu22efficient,
  title = {Efficient Treatment Effect Estimation in Observational Studies under Heterogeneous Partial Interference},
  author = {Qu, Zhaonan and Xiong, Ruoxuan and Liu, Jizhou and Imbens, Guido},
  journal = {arXiv:2107.12420v3},
  year = {2022}
}

@article{kilpatrick24gformula,
author = {Kilpatrick, Kayla W. and Lee, Chanhwa and Hudgens, Michael G.},
title = {G-formula for observational studies under stratified interference, with application to bed net use on malaria},
journal = {Stat. Med.},
volume = {43},
number = {15},
year = {2024},
pages = {2852-2868},
eprint = {https://onlinelibrary.wiley.com/doi/pdf/10.1002/sim.10102}
}

@article{lee2025efficient,
  author = {Chanhwa Lee and Donglin Zeng and Michael G. Hudgens and},
  title = {Efficient Nonparametric Estimation of Stochastic Policy Effects with Clustered Interference},
  journal = {J. Am. Statist. Assoc.},
  volume = {120},
  number = {549},
  pages = {382--394},
  year = {2025},
  publisher = {ASA Website},
  doi = {10.1080/01621459.2024.2340789},
  URL = {https://doi.org/10.1080/01621459.2024.2340789},
  eprint = {https://doi.org/10.1080/01621459.2024.2340789}
}

@article{chakladar2022inverse,
  title={Inverse probability weighted estimators of vaccine effects accommodating partial interference and censoring},
  author={Chakladar, Sujatro and Rosin, Samuel and Hudgens, Michael G and Halloran, M Elizabeth and Clemens, John D and Ali, Mohammad and Emch, Michael E},
  journal={Biometrics.},
  volume={78},
  number={2},
  pages={777--788},
  year={2022},
  publisher={Wiley Online Library}
}

@article{cui2023estimating,
  title={Estimating heterogeneous treatment effects with right-censored data via causal survival forests},
  author={Cui, Yifan and Kosorok, Michael R and Sverdrup, Erik and Wager, Stefan and Zhu, Ruoqing},
  journal={J. R. Stat. Soc. Ser. B Stat. Methodol.},
  volume={85},
  number={2},
  pages={179--211},
  year={2023},
  publisher={Oxford University Press US}
}

@article{munda2012parfm,
  title={parfm: Parametric frailty models in {R}},
  author={Munda, Marco and Rotolo, Federico and Legrand, Catherine},
  journal={J. Stat. Softw.},
  volume={51},
  pages={1--20},
  year={2012}
}

@Manual{ishwaran2023random,
    title = {randomForestSRC: Fast Unified Random Forests for Survival, Regression, and Classification (RF-SRC)},
    author = {Hemant Ishwaran and Michael S. Lauer and Eugene H. Blackstone and Min Lu and Udaya B. Kogalur},
    year = {2023},
    note = {R package version 3.2.3. https://CRAN.R-project.org/package=randomForestSRC}
  }

@article{ali2005herd,
  title={Herd immunity conferred by killed oral cholera vaccines in Bangladesh: a reanalysis},
  author={Ali, Mohammad and Emch, Michael and Von Seidlein, Lorenz and Yunus, Mohammad and Sack, David A and Rao, Malla and Holmgren, Jan and Clemens, John D},
  journal={The Lancet.},
  volume={366},
  number={9479},
  pages={44--49},
  year={2005},
  publisher={Elsevier}
}

@article{root2011role,
  title={The role of vaccine coverage within social networks in cholera vaccine efficacy},
  author={Root, Elisabeth D and Giebultowicz, Sophia and Ali, Mohammad and Yunus, Mohammad and Emch, Michael},
  journal={PLoS One},
  volume={6},
  number={7},
  pages={e22971},
  year={2011},
  publisher={Public Library of Science San Francisco, USA}
}

@article{clemens1988field,
  title={Field trial of oral cholera vaccines in Bangladesh: results of one year of follow-up},
  author={Clemens, John D and Harris, Jeffrey R and Sack, David A and Chakraborty, J and Ahmed, Faruque and Stanton, Bonita F and Khan, Moslem Uddin and Kay, Bradford A and Huda, N and Khan, MR and others},
  journal={J. Infect. Dis.},
  volume={158},
  number={1},
  pages={60--69},
  year={1988},
  publisher={The University Chicago Press}
}

@article{robins2000correcting,
  title={Correcting for noncompliance and dependent censoring in an AIDS clinical trial with inverse probability of censoring weighted (IPCW) log-rank tests},
  author={Robins, James M and Finkelstein, Dianne M},
  journal={Biometrics.},
  volume={56},
  number={3},
  pages={779--788},
  year={2000},
  publisher={Wiley Online Library}
}

@article{ozenne2020estimation,
  title={On the estimation of average treatment effects with right-censored time to event outcome and competing risks},
  author={Ozenne, Brice Maxime Hugues and Scheike, Thomas Harder and St{\ae}rk, Laila and Gerds, Thomas Alexander},
  journal={Biometrical J.},
  volume={62},
  number={3},
  pages={751--763},
  year={2020},
  publisher={Wiley Online Library}
}

@article{vaida2000proportional,
  title={Proportional hazards model with random effects},
  author={Vaida, Florin and Xu, Ronghui},
  journal={Stat. Med.},
  volume={19},
  number={24},
  pages={3309--3324},
  year={2000},
  publisher={Wiley Online Library}
}

@article{fan2006trees,
  title={Trees for correlated survival data by goodness of split, with applications to tooth prognosis},
  author={Fan, Juanjuan and Su, Xiao-Gang and Levine, Richard A and Nunn, Martha E and LeBlanc, Michael},
  journal={J. Am. Statist. Assoc.},
  volume={101},
  number={475},
  pages={959--967},
  year={2006},
  publisher={Taylor \& Francis}
}

@article{hong2006evaluating,
author = {Guanglei Hong and Stephen W Raudenbush},
title = {Evaluating Kindergarten Retention Policy: A case study of causal inference for multilevel observational data},
journal = {J. Am. Statist. Assoc.},
volume = {101},
number = {475},
pages = {901-910},
year = {2006},
publisher = {Taylor & Francis},
doi = {10.1198/016214506000000447},
eprint = {https://doi.org/10.1198/016214506000000447}
}

@article{biau2012analysis,
  title={Analysis of a random forests model},
  author={Biau, G{\'e}rard},
  journal={J. Mach. Learn. Res.},
  volume={13},
  pages={1063--1095},
  year={2012},
  publisher={JMLR. org}
}

@article{cui2022consistency,
  title={Consistency of survival tree and forest models: Splitting bias and correction},
  author={Cui, Yifan and Zhu, Ruoqing and Zhou, Mai and Kosorok, Michael},
  journal={Statist. Sinica.},
  volume={32},
  number={3},
  pages={1245--1267},
  year={2022}
}

@article{sun2019counting,
  title={Counting process-based dimension reduction methods for censored outcomes},
  author={Sun, Qiang and Zhu, Ruoqing and Wang, Tao and Zeng, Donglin},
  journal={Biometrika.},
  volume={106},
  number={1},
  pages={181--196},
  year={2019},
  publisher={Oxford University Press}
}

@article{aronow2017estimating,
  title={Estimating average causal effects under general interference, with application to a social network experiment},
  author={Aronow, Peter M and Samii, Cyrus},
  journal={Ann. Appl. Statist.},
  volume={11},
  number={4},
  pages={1912--1947},
  year={2017},
  publisher={Institute of Mathematical Statistics}
}

@article{salditt2023parametric,
  title={Parametric and nonparametric propensity score estimation in multilevel observational studies},
  author={Salditt, Marie and Nestler, Steffen},
  journal={Stat. Med.},
  volume={42},
  number={23},
  pages={4147--4176},
  year={2023},
  publisher={Wiley Online Library}
}

@article{chang2022flexible,
  title={Flexible propensity score estimation strategies for clustered data in observational studies},
  author={Chang, Ting-Hsuan and Nguyen, Trang Quynh and Lee, Youjin and Jackson, John W and Stuart, Elizabeth A},
  journal={Stat. Med.},
  volume={41},
  number={25},
  pages={5016--5032},
  year={2022},
  publisher={Wiley Online Library}
}

@Manual{dorie2022dbarts,
    title = {dbarts: discrete Bayesian additive regression trees sampler},
    author = {Dorie, V},
    year = {2022},
    note = {R package version 0.9-26. https://cran.r-project.org/web/packages/dbarts/dbarts.pdf}
  }

@article{hajjem2017generalized,
  title={Generalized mixed effects regression trees},
  author={Hajjem, Ahlem and Larocque, Denis and Bellavance, Fran{\c{c}}ois},
  journal={Statist. Probab. Lett.},
  volume={126},
  pages={114--118},
  year={2017},
  publisher={Elsevier}
}

@article{rotnitzky2005inverse,
  title={Inverse probability weighted estimation in survival analysis},
  author={Rotnitzky, Andrea and Robins, James},
  journal={Encycl. Biostat.},
  volume={4},
  pages={2619--2625},
  year={2005},
  publisher={Citeseer}
}

@article{chernozhukov2014gaussian,
  title={Gaussian approximation of suprema of empirical processes},
  author={Chernozhukov, Victor and Chetverikov, Denis and Kato, Kengo},
  journal={Ann. Statist.},
  volume={42},
  number={4},
  pages={1564--1597},
  year={2014}
}

@article{belloni2015uniform,
  title={Uniform post-selection inference for least absolute deviation regression and other Z-estimation problems},
  author={Belloni, Alexandre and Chernozhukov, Victor and Kato, Kengo},
  journal={Biometrika.},
  volume={102},
  number={1},
  pages={77--94},
  year={2015},
  publisher={Oxford University Press}
}

@article{chernozhukov2024conditional,
  title={Conditional Influence Functions},
  author={Chernozhukov, Victor and Newey, Whitney K and Syrgkanis, Vasilis},
  journal={arXiv:2412.18080},
  year={2024}
}


\spacingset{1.0} 
\newgeometry{margin = 0.7in}
\pagebreak
\begin{center}
    \textbf{\LARGE Supplementary Materials}
\end{center}
\setcounter{section}{0}
\setcounter{equation}{0}
\setcounter{figure}{0}
\setcounter{table}{0}
\setcounter{page}{1}
\makeatletter
\renewcommand{\theequation}{S\arabic{equation}}
\renewcommand{\thefigure}{S\arabic{figure}}
\renewcommand{\thetable}{S\arabic{table}}
\renewcommand{\bibnumfmt}[1]{[S#1]}
\renewcommand{\citenumfont}[1]{S#1}
\renewcommand{\thesection}{\Alph{section}}

\fontsize{10}{12}\selectfont

\section{Theoretical properties and proofs}

In this section, the proofs of Lemma 1, Proposition 1, Theorems 1 -- 4, 
and the large sample properties of the proposed estimators under the subsampling approximation and bounding modifications are presented.


\subsection{Proof of Lemma 1}

First, assume \textup{(I1) -- (I3)} in the main text. The identifiability of the target estimands from the full data follows from the results in \cite{lee2025efficient}.
Specifically,
\begin{align*}
	\Psi(\cR; \Bw)
	=&
	\textstyle
	E \left\{ 
		\suma
		\Bw(\ai, \BX_i, N_i)^\top
		\cR(\Ti(\ai))
	\right\}
	\\
	=&
	\textstyle
	E \left[ 
		\suma
		\Bw(\ai, \BX_i, N_i)^\top
		E \left\{
			\cR(\Ti(\ai)) | \BX_i, N_i
		\right\}
	\right]
	\\
	=&
	\textstyle
	E \left[ 
		\suma
		\Bw(\ai, \BX_i, N_i)^\top
		E \left\{
			\cR(\Ti) | \Ai = \ai, \BX_i, N_i
		\right\}
	\right]
\end{align*}
where the second equality follows from the law of iterated expectation and the third equality follows from causal consistency (I1) and conditional exchangeability (I2).
Note that due to weight-conditional positivity \textup{(I3)},
if $\Bw(\ai, \BX_i, N_i) \ne \textbf{0}_{N_i}$, then $\Pr(\Ai = \ai \mid \BX_i, N_i) > 0$,
and thus
$E \left\{
    \cR(\Ti) \mid \Ai = \ai, \BX_i, N_i
\right\}$ is identifiable.
Otherwise, if $\Bw(\ai, \BX_i, N_i) = \textbf{0}_{N_i}$, 
identifiability of
$E \left\{
    \cR(\Ti) \mid \Ai = \ai, \BX_i, N_i
\right\}$
does not matter.
Since $\Bw(\ai, \BX_i, N_i)$ is a functional of $P_{\AXN}$ and thus identifiable, 
the target causal estimand $\Psi(\cR;\Bw)$ is identifiable from the full data $\Zi = (\Ti, \Ai, \BX_i, N_i), i = 1,\dots,m$.

Next, assume (I1) -- (I5). 
Note that
\begin{align*}
    E \left\{
        \frac
            {\Deltaij \cR(\Yij) }
            { S_{ij}^C(\Yij | \AXNi) }
        \middle| \Ai, \BX_i, N_i 
    \right\}
    =&
    E \left[
        E \left\{
            \frac
                {\indicator(\Tij \le \Cij) \cR(\Tij) }
                { S_{ij}^C(\Tij | \AXNi) }
            \middle| \Tij, \AXNi 
        \right\}
        \middle| \AXNi 
    \right]
    \\
    =&
    E \left[
        \frac
            {E \{ \indicator(\Tij \le \Cij) | \Tij, \AXNi \}
			\cR(\Tij) }
            { S_{ij}^C(\Tij | \AXNi) }
        \middle| \AXNi 
    \right]
    \\
    =&
    E \left\{
        \cR(\Tij)
        \middle| \AXNi 
    \right\}
\end{align*}
where the last equality holds from 
\begin{align*}
    E \{ \indicator(\Tij \le \Cij) | \Tij, \AXNi \}
    =&
    \int_{0}^{\infty} 
        \indicator(\Tij \le u)
        f_{\Cij | \Tij, \AXNi} (u |  \Tij, \AXNi) 
    du
    \\
    =&
    \int_{\Tij}^{\infty} 
        f_{\Cij | \AXNi} (u | \AXNi) 
    du
    \\
    =&
    S_{ij}^C(\Tij | \AXNi)
\end{align*}
using (I4) $C_{ij} \indep T_{ij} | \Ai, \BX_i, N_i$,
where
$f_{\Cij | \Tij, \AXNi} (u |  \Tij, \AXNi)$ and
$f_{\Cij | \AXNi} (u | \AXNi)$ are the conditional density function of $\Cij$ given $\Tij, \AXNi$ and $\Cij$ given $\AXNi$, respectively,
which concludes the proof.
\hfill $\blacksquare$


\subsection{Construction of the proposed estimating function}

Motivated by \cite{tsiatis06} Chapter 10.4,
the proposed estimating equation is given by
\[\scalebox{1}{$
	\displaystyle
	0 =
	\sum_{i=1}^{m}
	\sumj
		\Bigg[
			\frac{\Delta_{ij}}{\SCij{\Yij}}
			\phiFs_{ij}(\cR; \Zi)
			+
			\int_{0}^{\infty}
				\frac
					{E \big\{ 
						\phiFs_{ij}(\cR; \Zi)
						\mid \Tij \ge r, \AXNi
					\big\}}
					{\SCij{r}}
				d\MijC
		\Bigg],
$}\]
where
$\varphi_{ij}^{F,*}(\cR; \mathbf{Z}_i) 
:= \phiF_{ij}(\cR; \Zi) - \Psi(\cR;\Bw)
= \suma \OR_{ij}(\cR; \Zi, \ai) + \text{BC}_{ij}(\cR; \Zi) - \Psi(\cR; \Bw)$
is the individual-level EIF using full data $\Zi$ in the main text.
Below it is shown that
\begin{align*}    
    \frac
        {\Delta_{ij}  \phiFs_{ij}(\cR; \Zi)}
        {\SCij{\Yij}}
    +
    \int_{0}^{\infty}
        \frac
            {E \big\{ 
                \phiFs_{ij}(\cR; \Zi)
                | \Tij \ge r, \AXNi
            \big\} }
            {\SCij{r}}
        d\MijC
    =
    \phiP_{ij}(\cR; \mathbf{O}_i)
    -
    \Psi(\cR;\Bw)
\end{align*}   
where 
$\phiP_{ij}(\cR; \mathbf{O}_i) 
= 
\suma \OR_{ij}(\cR; \Oi, \ai)
\allowbreak
+
\allowbreak
\IPCWBC_{ij}(\cR; \Oi)
\allowbreak
+
\allowbreak
\AUG_{ij}(\cR; \Oi)$
is the proposed estimating function in the main text.

First, let
\begin{align*}
        \xi_{ij}^{F}(\cR; \AXNi) 
        :=& 
        \sumaline
            \big\{ w_j(\aXNi) + \phi_j(\AXNi; \ai) \big\}
            E\{\cR(\Tij) | \Ai = \aXNi\}
        \\
        &-
        \frac
            {w_j(\AXNi) }
            {\textup{pr}(\AbarXNi)}
		E\{\cR(\Tij) | \AXNi\}
        - 
        \Psi(\cR;\Bw)
    \end{align*}
which is a function of $(\AXNi)$. 
Then,
\begin{align*}
    \varphi_{ij}^{F,*}(\cR; \Zi) 
    =
    \xi_{ij}^{F}(\cR; \AXNi) 
    +
    \frac
        {w_j(\AXNi) }
        {\textup{pr}(\AbarXNi)}
    \cR(\Tij) 
    .
\end{align*}
Since $\xi_{ij}^{F}(\cR; \AXNi)$ is a function of $(\AXNi)$ only (but not $T_{ij}$),
we have
\begin{align*}
    E \big\{ 
        \phiFs_{ij}(\cR; \Zi)
        \big| \Tij \ge r, \AXNi
    \big\}
    &=
    \xi_{ij}^{F}(\cR; \AXNi) 
    +
	\frac
		{w_j(\AXNi)}
		{\textup{pr}(\AbarXNi)}
    E \left\{ 
        \cR(\Tij) 
        \middle| \Tij \ge r, \AXNi
    \right\}
    \\
    &=
    \xi_{ij}^{F}(\cR; \AXNi) 
    + 
    \frac
        {w_j(\AXNi)}
        {\textup{pr}(\AbarXNi)}
    \frac
        {E[\cR(\Tij) \indicator(\Tij \ge r) | \AXNi]}
        {\STij{r}}
\end{align*}
where the last equality follows from the fact that
\begin{align*}
	E[\cR(\Tij) L_i | \AXNi]
	=&
	E[E\{\cR(\Tij) L_i | L_i, \AXNi\} | \AXNi]
	\\
	=&
	E\{\cR(\Tij)| L_i=1, \AXNi\} pr(L_i=1 | \AXNi)
\end{align*}
where $L_i = \indicator(\Tij \ge r)$,
and thus
$
	E\{\cR(\Tij) | \Tij \ge r, \AXNi\}
	=
	E[\cR(\Tij) \indicator(\Tij \ge r) | \AXNi] 
    \allowbreak
    / 
    pr(\Tij \ge r | \AXNi)
$
.
Therefore,
\begin{align*}
    &
    \int_{0}^{\infty}
        \frac
            {E \big\{ 
                \phiFs_{ij}(\cR; \Zi)
                | \Tij \ge r, \AXNi
            \big\} }
            {\SCij{r}}
    d\MijC
    \\
    &=
    \int_{0}^{\infty}
        \frac
            {\xi_{ij}^{F}(\cR; \AXNi)}
            {\SCij{r}}
    d\MijC
    +
    \frac
        {w_j(\AXNi)}
        {\textup{pr}(\AbarXNi)}
    \int_{0}^{\infty}
        \frac
        {E[\cR(\Tij) \indicator(\Tij \ge r) | \AXNi]}
        {\SCij{r} \STij{r}}
    d\MijC
    \\
    &=
    \left\{
        1 - \frac{\Deltaij}{\SCij{\Yij}}
    \right\}
    \xi_{ij}^{F}(\cR; \AXNi)
    +
    \AUG_{ij}(\cR; \Oi)
\end{align*}
where the last line follows because
\begin{align*}
    \int_{0}^{\infty}
        \frac
            {d\MijC}
            {\SCij{r}}
    =
    1 - \frac{\Deltaij}{\SCij{\Yij}}
    .
\end{align*}
Thus,
\begin{align*}
    &
    \frac
        {\Delta_{ij}}
        {\SCij{\Yij}}
        \phiFs_{ij}(\cR; \Zi)
    +
    \int_{0}^{\infty}
        \frac
            {E \big\{ 
                \phiFs_{ij}(\cR; \Zi)
                | \Tij \ge r, \AXNi
            \big\} }
            {\SCij{r}}
        d\MijC
    \\
    &=
    \frac
        {\Delta_{ij}}
        {\SCij{\Yij}}
    \left\{
        \xi_{ij}^{F}(\cR; \AXNi) 
        +
        \frac
            {w_j(\AXNi) }
            {\textup{pr}(\AbarXNi)}
        \cR(\Tij)
    \right\}
    \\
    & \quad +
    \left\{
        1 - \frac{\Deltaij}{\SCij{\Yij}}
    \right\}
    \xi_{ij}^{F}(\cR; \AXNi)
    +
    \AUG_{ij}(\cR; \Oi)
    \\
    &=
    \phiP_{ij}(\cR; \mathbf{O}_i)
    -
    \Psi(\cR;\Bw)
    .
\end{align*}
\hfill $\blacksquare$


\subsection{Proof of Proposition 1}

For simplicity, the subscript expressing cluster index is omitted in the following.
Also, we slightly abuse notation by suppressing the dependency on random variables if there is no ambiguity,
for example, 
$\BF^T \big(r | \aXN \big)
=
\BF^T \big(r | \BA = \aXN \big)$.

First, note that the proposed estimating function can be expressed as follows:
\begin{align*}
    \phiP(\cR; \BO, \Beta)
    =
    \sumaN
        \OR(\cR; \BO, \Ba) 
    +
    \IPCWBC(\cR; \BO)
    +
    \AUG(\cR; \BO)
\end{align*}
where
\begin{align*}
    \OR(\cR; \BO, \Ba) 
    =&
    \big\{ \Bw(\aXN) + \Bphi(\AXN; \Ba) \big\}^\top
    E\{\cR(\BT) | \aXN\} 
    \\
    \IPCWBC(\cR; \BO)
	=&
    \frac{1}{H(\AXN)}
    \Bw(\AXN)^\top 
    \left[ 
        \frac
            {\Delta_j  \cR(Y_j)}
            {\SCj{Y_j}}
        - 
        E\{\cR(T_j) | \AXN\} 
    \right]_{j=1}^N
    \\
	\AUG(\cR; \BO)
    =&
    \frac{1}{H(\AXN)}
    \Bw(\AXN)^\top 
    \left[ 
        \int_{0}^{\infty}
            \frac
                {E\{\cR(T_j) \indicator(T_j \ge r) | \AXN\}}
                {\SCj{r} \STj{r}}
        d\MjC
    \right]_{j=1}^N
    .
\end{align*}
From Lemma 1,
$
    E \left\{
        \sumaN
            \Bw(\aXN)^\top
            E\{\cR(\BT) | \aXN\} 
    \right\}
    =
    \Psi(\cR;\Bw)
    .
$
Next, since $\Bphi$, the CIF of $\Bw$,
satisfies
$E \left\{
    \Bphi(\AXN; \Ba)
    \middle| \XxNn
\right\}
=0$ for all $\axn$,
it follows that
$$    
E \Bigg\{
    \sumaN
        \Bphi(\AXN; \Ba)^\top
        E\{\cR(\BT) | \aXN\} 
\Bigg\}
=
E \Bigg[
    \sumaN
        E \big\{
            \Bphi(\AXN; \Ba)
            \big| \XN
        \big\}^\top
        \allowbreak
        E\{\cR(\BT) | \aXN\} 
\Bigg]
=
0
.
$$
Third, from the proof of Lemma 1, it follows that
\begin{align*}
    E \left\{
        \frac{1}{H(\AXN)}
        \Bw(\AXN)^\top 
        \left[ 
            \frac
                {\Delta_j  \cR(Y_j)}
                {\SCj{Y_j}}
        \right]_{j=1}^N
    \right\}
    =&
    E \left\{
        \frac{1}{H(\AXN)}
        \Bw(\AXN)^\top 
        \cR(\BT)
    \right\}
	\\
	=&
	E \left[
        \frac{1}{H(\AXN)}
        \Bw(\AXN)^\top 
        E\{\cR(\BT) | \AXN\}
    \right]
	.
\end{align*}
Finally,
\begin{align*}
    E \left[ 
        \int_{0}^{\infty}
            \frac
                {E\{\cR(T_j) \indicator(T_j \ge r) | \AXN\}}
                {\SCj{r} \STj{r}}
        d\MjC
        \middle| \AXN
    \right]
    =
    0
\end{align*}
since 
$\STj{r}, \SCj{r}$, and 
$E\{\cR(T_j) \indicator(T_j \ge r) | \AXN\} 
=
\int_r^\infty
	\cR(t) 
dF_{j}^T(t|\AXN)$ 
are predictable with respect to $\AXN$, and thus the integral with respect to 
$\MjC
= 
\indicator(Y_j \le r, \Delta_j = 0)
+
\int_{0}^{\min\{Y_j, r\}}
d \{\log{\SCj{u}}\}$
yields a mean zero martingale \citep{ozenne2020estimation},
which concludes the proof of $E\{\phiP(\cR; \BO)\} = \Psi(\cR;\Bw)$.
\hfill $\blacksquare$

\subsection{Proof of Theorem 1}
\label{proof:thm1}

To establish the asymptotic properties of the proposed estimators, we follow the approach of \citet{lee2025efficient}. Assume sample splitting is approximately uniform, such that $m_k/m = K^{-1} + O(m^{-1})$. Let $\mathbb{P}f(\mathbf{O}) = \int f(\mathbf{o}) dP(\mathbf{o})$ denote the expectation of $f(\mathbf{O})$, treating $f$ as fixed even when it is estimated from the sample. For example, if $\widehat{f}$ is a prediction model trained on data $D$ independent of $\mathbf{O}$, then $\mathbb{P}\widehat{f}(\mathbf{O}) = E\big\{\widehat{f}(\mathbf{O}) \big| D \big\}$. 

For the estimand $\Psi(\cR;\Bw) = \mathbb{P}\big\{\varphi(\cR; \BO, \Beta)\big\}$ and its estimator $\widehat{\Psi}(\cR;\Bw) = \sumkline \bPmk \big\{\phiP(\cR; \BO, \Betahatk)\big\}$, we have the decomposition:
\begin{align}
    \widehat{\Psi}(\cR;\Bw) - \Psi(\cR;\Bw)
    =
    (\bPm - \bP) \varphi(\cR; \BO, \Beta) 
    +
    B_{m,1}
    +
    B_{m,2}
    +
    B_{m,3},
    \label{pf:thm1:decomp}
\end{align}
where
\begin{align*}
    B_{m,1} 
    =& 
    \sumk
        \bPmk \big\{\varphi(\cR; \BO, \Beta)\big\}
        \left( 1 - \frac{K m_k}{m} \right),
    \\
    B_{m,2} 
    =&
    \sumk
        (\bPmk - \bP) \big\{\phiP(\cR; \BO, \Betahatk) - \phiP(\cR; \BO, \Beta)\big\},
    \\
    B_{m,3}
    =&
    \sumk
        \bP \big\{\phiP(\cR; \BO, \Betahatk) - \phiP(\cR; \BO, \Beta)\big\}.
\end{align*}

The first term, $(\bPm - \bP) \varphi(\cR; \BO, \Beta)$, is $O_{P}(m^{-1/2})$ by the central limit theorem. The second term, $B_{m,1}$, is bounded as:
\begin{align*}
    \bPmk \big\{\varphi(\cR; \BO, \Beta)\big\} \left(1 - \frac{K m_k}{m}\right)
    =
    \{\Psi(\cR;\Bw) + \oP{1}\}O(m^{-1})
    =
    \OP{m^{-1}}.
\end{align*}

The third term, $B_{m,2}$, is bounded as:
\begin{align*}
    (\bPmk - \bP) \big\{\phiP(\cR; \BO, \Betahatk) - \phiP(\cR; \BO, \Beta)\big\}
    =
    O_{P} \left( 
        m_k^{-1/2}
        \normp{\phiP(\cR; \BO, \Betahatk) - \phiP(\cR; \BO, \Beta)}
    \right),
\end{align*}
from Lemma 1 in \citet{kennedy22}. It remains to bound $\normp{\varphidiffk}$ and $\mathbb{P} \big\{\varphidiffk\big\}$.

For simplicity, omit subscript $(k)$ in $\Betahatk$, and let $D$ denote the data used to train $\Betahatk$. Let $\BO = (\BY, \mathbf{\Delta}, \AXN)$ denote a new cluster-level observation independent of $D$. 

For simplicity, use shorthand notation by suppressing the dependency on random variables if there is no ambiguity as follows:
$\BFThat(r|\Ba) = \BFThat(r|\aXN)$,
$\BFThat(r) = \BFThat(r|\AXN)$,
$\BSChat(r|\Ba) = \BSChat(r|\aXN)$,
$\BSChat(r) = \BSChat(r|\AXN)$,
$\Hhat(\Ba) = \Hhat(\aXN)$,
$\Bwhat(\Ba) = \Bwhat(\aXN)$,
$\Bphihat(\Ba) = \Bphihat(\AXN; \Ba)$,
and define 
$\BF^T(r|\Ba)$,
$\BS^C(r|\Ba)$,
$H(\Ba)$,
$\Bw(\Ba)$,
$\Bphi(\Ba)$
similarly.

Then,
\begin{align*}
    \phiP(\cR; \BO, \Betahat)
    =
    \sumaN
        \ORhat(\cR; \BO, \Ba) 
        +
        \IPCWBChat(\cR; \BO)
        +
        \AUGhat(\cR; \BO),
\end{align*}
where
\begin{align*}
    \ORhat(\cR; \BO, \Ba) 
    =&
    \big\{\Bwhat(\Ba) + \Bphihat(\Ba)\big\}^\top
    \widehat{\BG}^{\cR}(0|\Ba), 
    \\
    \IPCWBChat(\cR; \BO)
    =&
    \frac{1}{\Hhat(\BA)}
    \Bwhat(\BA)^\top 
    \left[ 
        \frac{\Delta_j  \cR(Y_j)}{\SCjhat(Y_j)}
        - 
        \widehat{G}_j^{\cR}(0)
    \right]_{j=1}^N, 
    \\
    \AUGhat(\cR; \BO)
    =&
    \frac{1}{\Hhat(\BA)}
    \Bwhat(\BA)^\top 
    \left[ 
        \int_{0}^{\infty}
            \frac{\widehat{G}_j^{\cR}(r)}{\SCjhat(r) \STjhat(r)}
        d\MCjhat
    \right]_{j=1}^N.
\end{align*}
Here, 
${G}_j^{\cR}(r|\Ba) 
= 
E\{\cR(T_j) \indicator(T_j \ge r) | \aXN\} 
=
\int_r^\infty \cR(t) dF_{j}^T(t|\aXN)$
and
${G}_j^{\cR}(r) 
= 
E\{\cR(T_j) \indicator(T_j \ge r) | \AXN\}$.
Their estimators are:
$\widehat{G}_j^{\cR}(r|\Ba) 
= 
\int_r^\infty \cR(t) d\widehat{F}_{j}^T(t|\aXN)$
and
$\widehat{G}_j^{\cR}(r) 
= 
\int_r^\infty \cR(t) d\widehat{F}_{j}^T(t|\AXN)$.
The vectorized versions are:
$\widehat{\BG}^{\cR}(r|\Ba) 
= 
\left[\widehat{G}_j^{\cR}(r|\Ba)\right]_{j=1}^N$
and
$\widehat{\BG}^{\cR}(r)
=
\left[\widehat{G}_j^{\cR}(r)\right]_{j=1}^N$.

First, consider 
$
\mathbb{P} \big\{ \varphidiff \big\}
$.
We have the following decomposition from iterated expectation:
\begin{align}
    &
    \mathbb{P} \big\{ \varphidiff \big\}
    \nonumber
    \\
    &=
    \ED{[}{]}{
        \sumaN 
            \big\{ \Bwhat(\Ba) + \Bphihat(\Ba) \big\}^\top 
            \widehat{\BG}^{\cR}(0|\Ba)
    }
    \nonumber
    \\
    & \quad +
    \ED{[}{]}{
        \frac{1}{\Hhat(\BA)}
        \Bwhat(\BA)^\top 
        \left[ 
            \frac
                {S_j^C(T_j)  \cR(T_j)}
                {\SCjhat(T_j)}
        \right]_{j=1}^N
    }
    -
    \ED{[}{]}{
        \frac{1}{\Hhat(\BA)}
        \Bwhat(\BA)^\top 
        \widehat{\BG}^{\cR}(0)
    }
    \nonumber
    \\
    & \quad+
    \ED{[}{]}{
        \frac{1}{\Hhat(\BA)}
        \Bwhat(\BA)^\top 
        \left[ 
            \int_{0}^{\infty}
                \frac
                    {\widehat{G}_j^{\cR}(r)}
                    {\SCjhat(r) \STjhat(r)}
            d\MCjhat
        \right]_{j=1}^N
    }
    \nonumber
    \\
    & \quad -
    \ED{[}{]}{
        \sumaN 
            \Bw(\Ba)^\top 
            {\BG}^{\cR}(0|\Ba)
    }
    \nonumber
    \\
    & =
    \ED{(}{)}{
        \sumaN 
            \left[ 
                \Bwhat(\Ba) - \Bw(\Ba)
                + \E{\{}{\}}{\Bphihat(\Ba) \middle| D, \mathbf{X}, N}
            \right]^\top 
            {\BG}^{\cR}(0|\Ba)
    }
    \label{pf:thm1:wdiff2}
    \\
    & \quad +
    \ED{[}{]}{
        \sumaN 
            \big\{ \Bphihat(\Ba) - \Bphi(\Ba)  \big\}^\top 
            \big\{\widehat{\BG}^{\cR}(0|\Ba) - {\BG}^{\cR}(0|\Ba)\big\}
    }
    \label{pf:thm1:phidiffFdif}
    \\
    & \quad +
    \ED{[}{]}{
        \sumaN 
            \Bwhat(\Ba)^\top 
            \big\{\widehat{\BG}^{\cR}(0|\Ba) - {\BG}^{\cR}(0|\Ba)\big\}
    }
    \label{pf:thm1:Fdiff}
    \\
    & \quad +
    \ED{[}{]}{
        \frac{1}{\Hhat(\BA)}
        \Bwhat(\BA)^\top 
        \left[ 
            \frac
                {S_j^C(T_j)}
                {\SCjhat(T_j)}
            \cR(T_j)
        \right]_{j=1}^N
    }
    \label{pf:thm1:SCI}
    \\
    & \quad -
    \ED{[}{]}{
        \frac{1}{\Hhat(\BA)}
        \Bwhat(\BA)^\top 
        \widehat{\BG}^{\cR}(0)
    }
    \label{pf:thm1:F}
    \\
    & \quad +
    \ED{[}{]}{
        \frac{1}{\Hhat(\BA)}
        \Bwhat(\BA)^\top 
        \left[ 
            \int_{0}^{\infty}
                \frac
                    {\widehat{G}_j^{\cR}(r)}
                    {\SCjhat(r) \STjhat(r)}
            d\MCjhat
        \right]_{j=1}^N
    }
    \label{pf:thm1:stieltjes}
\end{align}
where the last equality follows from
$
E \left\{ 
    \Bphi(\mathbf{A}, \mathbf{X}, N; \mathbf{a})
    \middle| 
    \mathbf{X}, N
\right\}  = 0
$.
To analyze each term, we use the following inequalities:

\vspace{-0.2cm}
\begin{enumerate}[label=(\roman*)]
    \item 
    $\abs{x_1y_1 - x_2y_2} 
    \lesssim
    \abs{x_1-x_2} + \abs{y_1-y_2}$ 
    for bounded $x_1, x_2, y_1, y_2$;
    
    \item
    $\ED{[}{]}{\sum_{\Ba \in \cAN} f(\aXN)} 
    \lesssim
    \ED{[}{]}{f(\AXN)} 
    \lesssim
    \ED{[}{]}{\sum_{\Ba \in \cAN} f(\aXN)}$ 
    for a function $f$;
    
    \item
    $\ED{[}{]}{f(\BO)} 
    \lesssim 
    \normp{f}$ 
    for a function $f$ (Cauchy-Schwarz inequality);
    
    \item
    $\ED{[}{]}{f(\BO)^2}
    \lesssim
    \ED{[}{]}{\abs{f(\BO)}}$ 
    for bounded $f$;
    
    \item
    $\ED{[}{]}{(Z_1^2 + \dots + Z_n^2)^{1/2}} 
    \lesssim
    \ED{[}{]}{\abs{Z_1} + \dots + \abs{Z_n}}
    \lesssim 
    \ED{[}{]}{(Z_1^2 + \dots + Z_n^2)^{1/2}}$ 
    for finite $n$.
\end{enumerate}
Here, $\alpha \lesssim \beta$ means $\alpha \le C \beta$ for some constant $C$. Inequality (ii) follows from the iterated expectation with respect to $\XN$ and the boundedness of $H(\AXN)$, and (v) follows from bounding sums of squares and cross terms for finite $n$.

First, (\ref{pf:thm1:stieltjes}) is computed from the Stieltjes integration.
Note that
\begin{align*}
    \EDA{\{}{\}}{
        d\MCjhat
    }
    &=
    E \left\{
        d\MjC
        | \AXN
    \right\}
    +
    E \left\{
        \indicator(Y_j \ge r)
        | \AXN
    \right\}
    \left(
        \frac
            {d\SCjhat}
            {\SCjhat}
        -
        \frac
            {dS_j^C}
            {S_j^C}
    \right)
    (r|\AXN)
    \\
    &=
    -
    \STj{r}
    \SCjhat(r|\AXN)
    d
    \left(
        \frac
            {S_j^C}
            {\SCjhat}
    \right)
    (r|\AXN)
    ,
\end{align*}
since
$d\MjC$ is mean zero martingale with respect to $(\AXN)$,
$\indicator(Y_j \ge r) = \indicator(T_j \ge r)\indicator(C_j \ge r)$,
and
\begin{align*}
    d
    \left(
        \frac
            {S_j^C}
            {\SCjhat}
    \right)
    (r|\AXN)
    =
    -
    \left(
        \frac
            {S_j^C}
            {\SCjhat}
    \right)
    \left( 
        \frac
            {d\SCjhat}
            {\SCjhat}
        -
        \frac
            {dS_j^C}
            {S_j^C}
    \right)
    (r|\AXN)
    .
\end{align*}
It follows that,
\begin{align*}
    &
    \EDA{\{}{\}}{
        \int_{0}^{\infty}
            \frac
                {\widehat{G}_j^{\cR}(r)}
                {\SCjhat(r) \STjhat(r)}
        d\MCjhat
    }
    \\
    &=
    \int_{0}^{\infty}
        \frac
            {\widehat{G}_j^{\cR}(r)}
            {\SCjhat(r) \STjhat(r)}
    E \left\{
        d\MCjhat
        | D, \AXN
    \right\}
    \\
    &=
    -
    \int_{0}^{\infty}
    \left\{
        \frac
            {S^T_j}
            {\STjhat}
        (r)
        \widehat{G}_j^{\cR}(r)
    \right\}
    d
    \left(
        \frac
            {S_j^C}
            {\SCjhat}
    \right)
    (r)
    \\
    &=
    -
    \left[
        \left\{
            \frac
                {S^T_j}
                {\STjhat}
            (r)
            \widehat{G}_j^{\cR}(r)
        \right\}
        \frac
            {S_j^C}
            {\SCjhat}
        (r)
    \right]_{0}^{\infty}
    +
    \int_{0}^{\infty}
    \left\{
        d\left(
            \frac
                {S^T_j}
                {\STjhat}
        \right)
        (r)
        \widehat{G}_j^{\cR}(r)
        +
        \frac
            {S^T_j}
            {\STjhat}
        (r)
        d\widehat{G}_j^{\cR}(r)
    \right\}
    \frac
        {S_j^C}
        {\SCjhat}
    (r)
    \\
    &=
    \widehat{G}_j^{\cR}(0)
    -
    \int_{0}^{\infty}
        \frac
            {S^C_j}
            {\SCjhat}
        (r)
        \cR(r)
        S^T_j(r)
        \lambdaTjhat(r)
    dr
    +
    \int_{0}^{\infty}
        \frac
            {S^C_j}
            {\SCjhat}
        (r)
        \frac
            {S^T_j}
            {\STjhat}    
        (r)
        \widehat{G}_j^{\cR}(r)
        \left(
            \lambdaTjhat
            -
            \lambda^T_j
        \right)
        (r)
    dr
    \\
    &=
    \widehat{G}_j^{\cR}(0)
    -
    \int_{0}^{\infty}
        \frac
            {S^C_j}
            {\SCjhat}
        (r)
        \cR(r)
    dF^T_j(r)
    +
    \int_{0}^{\infty}
        \frac
            {S^T_j}
            {\STjhat}    
        (r)
        \frac
            {S^C_j}
            {\SCjhat}
        (r)
        \left(
            \lambdaTjhat
            -
            \lambda^T_j
        \right)
        (r)
        \left\{
            \int_{r}^{\infty}
                \cR(t)
                \STjhat(t)\lambdaTjhat(t)
            dt
            -
            \STjhat(r)
            \cR(r)
        \right\}
    dr
\end{align*}
where the third equality is from integration by parts,
the fourth is from
\begin{align*}
    d
    \left(
        \frac
            {S_j^T}
            {\STjhat}
    \right)
    (r)
    =
    -
    \left(
        \frac
            {S_j^T}
            {\STjhat}
    \right)
    (r)
    \left( 
        \frac
            {d\STjhat}
            {\STjhat}
        -
        \frac
            {dS_j^T}
            {S_j^T}
    \right)
    (r)
    =
    \left(
        \frac
            {S_j^T}
            {\STjhat}
    \right)
    (r)
    \left( 
        \lambdaTjhat
        -
        \lambda^T_j
    \right)
    (r)
    dr
    ,
\end{align*}
and
$d\widehat{G}_j^{\cR}(r) = \cR(r)d\STjhat(r) = -\cR(r)\STjhat(r)\lambdaTjhat(r)dr$,
and the fifth is from 
$dF^T_j(r)
=
S^T_j(r)
\lambda^T_j(r)
dr$.
Thus,
\begin{align}
    (\ref{pf:thm1:stieltjes}) 
    =&
    \underbrace{
    \ED{[}{]}{
        \frac{1}{\Hhat(\BA)}
        \Bwhat(\BA)^\top 
        \widehat{\BG}^{\cR}(0)
    }
    }_{(\ref{pf:thm1:F})}
    -
    \underbrace{
    \ED{[}{]}{
        \frac{1}{\Hhat(\BA)}
        \Bwhat(\BA)^\top 
        \left[
            \frac
                {S_j^C}
                {\SCjhat}
            (T_j)
            \cR(T_j)
        \right]_{j=1}^N
    }
    }_{(\ref{pf:thm1:SCI})}
    \nonumber
    \\
    &+
    \ED{[}{]}{
        \frac{1}{\Hhat(\BA)}
        \Bwhat(\BA)^\top 
        \left[
            \int_{0}^{\infty}
                \frac
                    {S^T_j}
                    {\STjhat}    
                (r)
                \frac
                    {S^C_j}
                    {\SCjhat}
                (r)
                \left(
                    \lambdaTjhat
                    -
                    \lambda^T_j
                \right)
                (r)
                \left\{
                    \int_{r}^{\infty}
                        \cR(t)
                        \STjhat(t)\lambdaTjhat(t)
                    dt
                    -
                    \STjhat(r)
                    \cR(r)
                \right\}
            dr
        \right]_{j=1}^N
    }
    \label{pf:thm1:lambdadiff}
\end{align}

Next, from 
\begin{align*}
    (\STjhat-S^T_j)(t)
    =
    \STjhat(t)
    \left\{
        1 - \exp((\LambdaTjhat - \Lambda_j^T)(t))
    \right\}
    =
    -
    \STjhat(t)  
    \int_{0}^{t}
        \frac
            {S^T_j}
            {\STjhat}    
        (r)
        (\lambdaTjhat- \lambda^T_j)(r)
    dr
    ,
\end{align*}
it follows that
\begin{align*}
    &
    \widehat{G}_j^{\cR}(0|\AXN) - {G}_j^{\cR}(0|\AXN)
    \\
    =&
    \int_{0}^{\infty}
        \cR(t)
    d(\widehat{F}^T_j - F_j^T)(t|\AXN)
    \\
    =&
    \int_{0}^{\infty}
        \cR(t)
        S^T_j(t)
        (\lambdaTjhat- \lambda^T_j)(t)
    dt
    -
    \int_{0}^{\infty}
        \cR(t)
        \lambdaTjhat(t)
        \STjhat(t)  
        \int_{0}^{t}
            \frac
                {S^T_j}
                {\STjhat}    
            (r)
            (\lambdaTjhat- \lambda^T_j)(r)
        dr
    dt
    .
\end{align*}
Thus, from
\begin{align*}
    (\ref{pf:thm1:Fdiff})
    =&
    \ED{[}{]}{
        \sumaN 
            \Bwhat(\Ba)^\top 
            \big\{\widehat{\BG}^{\cR}(0|\Ba) - {\BG}^{\cR}(0|\Ba)\big\}
    }
    \\
    =&
    \ED{(}{)}{
        \E{[}{]}{
            \frac{1}{H(\AXN)}
            \Bwhat(\AXN)^\top 
            \left\{
                \widehat{\BG}^{\cR}(0|\AXN) - {\BG}^{\cR}(0|\AXN)
            \right\}
        \middle| D, \XN
        }
    },
\end{align*}
it follows that
\begin{align}
    &
    (\ref{pf:thm1:Fdiff}) + (\ref{pf:thm1:lambdadiff})
    \nonumber
    \\
    =&
    \ED{[}{]}{
        \frac{1}{H(\BA)}
        \Bwhat(\BA)^\top 
        \left[
            \int_{0}^{\infty}
                \cR(r)
                S^T_j(r)
                \left\{
                    1
                    -
                    \frac{H}{\Hhat}(\BA)
                    \frac
                        {S^C_j}
                        {\SCjhat}
                    (r)
                \right\}
                \left(
                    \lambdaTjhat
                    -
                    \lambda^T_j
                \right)
                (r)
            dr
        \right]_{j=1}^N
    }
    \nonumber
    \\
    &+
    \ED{[}{]}{
        \frac{1}{H(\BA)}
        \Bwhat(\BA)^\top 
        \left[
            \frac{H}{\Hhat}(\BA)
            \int_{0}^{\infty}
                \frac
                    {S^T_j}
                    {\STjhat}    
                (r)
                \frac
                    {S^C_j}
                    {\SCjhat}
                (r)
                \left(
                    \lambdaTjhat
                    -
                    \lambda^T_j
                \right)
                (r)
                \int_{r}^{\infty}
                    \cR(t)
                    \STjhat(t)\lambdaTjhat(t)
                dt
            dr
        \right]_{j=1}^N
    }
    \nonumber
    \\
    &+
    \ED{[}{]}{
        \frac{1}{H(\BA)}
        \Bwhat(\BA)^\top 
        \left[
            -
            \int_{0}^{\infty}
                \cR(t)
                \lambdaTjhat(t)
                \STjhat(t)
                \int_{0}^{t}
                    \frac
                        {S^T_j}
                        {\STjhat}    
                    (r)
                    (\lambdaTjhat- \lambda^T_j)(r)
                dr
            dt
        \right]_{j=1}^N
    }
    \nonumber
    \\
    =&
    \ED{[}{]}{
        \frac{1}{H(\BA)}
        \Bwhat(\BA)^\top 
        \left[
            \int_{0}^{\infty}
                \left\{
                    1
                    -
                    \frac{H}{\Hhat}(\BA)
                    \frac
                        {S^C_j}
                        {\SCjhat}
                    (r)
                \right\}
                \left(
                    \lambdaTjhat
                    -
                    \lambda^T_j
                \right)
                (r)
                S^T_j(r)
                \left\{
                    \cR(r)
                    -
                    \frac
                        {\widehat{G}_j^{\cR}(r)}
                        {\STjhat(r)}
                \right\}
            dr
        \right]_{j=1}^N
    }
    \label{pf:thm1:lambdadiffHSdiff}
\end{align}
where the last equality follows from the change of integral region 
from  $(0 \le r \le t, 0 \le t \le \infty)$
to $(r \le t \le \infty, 0 \le r \le \infty)$. 
In conclusion,
\begin{align*}
    &
    \mathbb{P} \big\{ \varphidiff \big\}
    \nonumber
    \\
    &=
    (\ref{pf:thm1:wdiff2}) 
    + 
    (\ref{pf:thm1:phidiffFdif}) 
    +
    (\ref{pf:thm1:lambdadiffHSdiff})
    \\
    &= 
    \ED{[}{]}{
        \sumaN 
            \left\{
                \Bwhat(\aXN) - \Bw(\aXN)
                + 
                \sum_{\Ba' \in \cAN}
                    \Bphihat(\Ba', \XN;\Ba)
                    H(\Ba',\XN)
            \right\}^\top 
            \BG^{\cR}(0 | \aXN)
    }
    \\
    & \quad +
    \ED{[}{]}{
        \sumaN 
            \big\{ \Bphihat(\AXN;\Ba) - \Bphi(\AXN;\Ba) \big\}^\top 
            \big\{ \widehat{\BG}^{\cR}(0 | \aXN) - \BG^{\cR}(0 | \aXN)\big\}
    }
    \\
    & \quad +
    \ED{[}{]}{
        \frac{1}{H(\BA)}
        \Bwhat(\BA)^\top 
        \left[
            \STjhat(\cR)
            \int_{0}^{\infty}
                \left\{
                    1
                    -
                    \frac{H}{\Hhat}(\BA)
                    \frac
                        {S^C_j}
                        {\SCjhat}
                    (r)
                \right\}
                \left(
                    \lambdaTjhat
                    -
                    \lambda^T_j
                \right)
                (r)
                S^T_j(r)
                \left\{
                    \cR(r)
                    -
                    \frac
                        {\widehat{G}_j^{\cR}(r)}
                        {\STjhat(r)}
                \right\}
            dr
        \right]_{j=1}^N
    }
    ,
\end{align*}
and each term in the above decomposition is bounded by
\begin{align*}
    (\ref{pf:thm1:wdiff2}) 
    &\lesssim
    \ED{[}{]}{
        \sumaN
        \bigg|\bigg|
            \Bwhat(\aXN) - \Bw(\aXN)
            + 
            \sum_{\Ba' \in \cAN}
                \Bphihat(\Ba', \XN;\Ba)
                H(\Ba',\XN)
        \bigg|\bigg|_2^2
    }^{1/2}
    =
    \OP{r_{\Bw}^2}
    \\
    (\ref{pf:thm1:phidiffFdif}) 
    &\lesssim
    \ED{[}{]}{
        \sumaN 
            \normt{\Bphihat(\Ba) - \Bphi(\Ba)}^2
    }^{1/2}
    \ED{[}{]}{
        \int_{0}^{\infty}
            \normt{
                (\BlambdaThat - \Blambda^T) (r|\AXN)
            }^2
        dr
    }^{1/2}
    =
    O_{P}(r_{\Bphi} r_{\Blambda^T})
    \\
    (\ref{pf:thm1:lambdadiffHSdiff}) 
    &\lesssim
    \ED{[}{]}{
        \sum_{j=1}^{N}
                \left\{
                    \int_{0}^{\infty}
                        (\lambdaTjhat - \lambda^T_j) (r|\AXN)^2
                    dr
                \right\}^{1/2}
                \left\{
                    \int_{0}^{\infty}
                        \left(
                            1 
                            - 
                            \frac{H}{\Hhat}(\AXN)
                            \frac
                                {S_j^C}
                                {\SCjhat}
                            (r|\AXN)
                        \right)^2
                    dr
                \right\}^{1/2}
    }
    \\
    &\lesssim
    \ED{[}{]}{
        \int_{0}^{\infty}
            \normt{
                (\BlambdaThat - \Blambda^T) (r|\AXN)
            }^2
        dr
    }^{1/2}
    \\
    & \ \ \cdot
    \left\{
        \ED{[}{]}{
            (\Hhat-H)(\AXN)^2
        }^{1/2}
        +
        \ED{[}{]}{
            \int_{0}^{\infty}
                \normt{
                    (\BSChat - \BS^C)
                    (r|\AXN)
                }^2
            dr
        }^{1/2}
    \right\}
    \\
    &=
    O_P(r_{\Blambda^T} (r_H + r_{\BS^C}))
\end{align*}
from Cauchy-Schwarz inequality and boundedness of nuisance functions,
and the fact that
\begin{align*}
    |\widehat{G}_j^{\cR}(0|\AXN) - {G}_j^{\cR}(0|\AXN)|
    \lesssim
    \int_{0}^{\infty}
        |\lambdaTjhat- \lambda^T_j|(r)
    dr
    .
\end{align*}

Therefore,
\begin{align}
    \mathbb{P} \big\{ \varphidiff \big\}
    =
    \OP{
        r_{\Bw}^2 
        + 
        r_{\Blambda^T} (r_H + r_{\Bphi} + r_{\BS^C})
    }
    \label{pf:thm1:Pdiff}
\end{align}
and thus 
$B_{m,3} 
=
\OP{
    r_{\Bw}^2 
    + 
    r_{\Blambda^T} (r_H + r_{\Bphi} + r_{\BS^C})
}$.

Next, consider $||\varphidiff||$.
We first prove that $\varphidiff$ is bounded.
From the boundedness of nuisance functions and their estimators,
it is easy to show that $\OR$ and $\IPCWBC$ are bounded.
Also, from 
$\MjC
= 
\indicator(Y_j \le r, \Delta_j = 0)
- 
\int_{0}^{r}
    \indicator(u \le Y_j)
    \lambdaCj{u}
    \allowbreak
du$,
we have
\begin{align}
    \int_{0}^{\infty}
        L_j(r)
    dM^C_j(r)
    =
    \indicator(\Delta_j = 0)
    \cR(Y_j)
    L_j(Y_j)
    -
    \int_{0}^{\infty}
        L_j(r)
        \indicator(r \le Y_j)
        \lambda_j^C(r)
    dr
    \label{pf:thm1:stieltjesexpansion}
\end{align}
by the computation of Stieltjes integration,
where
$L_j(r) = G_j^{\cR}(r) \{S^C_j(r) S^T_j(r)\}^{-1}$.
Since $L_j(r)$ and $\lambda_j^C(r)$ in the above equation are bounded, it follows that $\AUG$ is also bounded.
Therefore, $\phiP(\cR; \BO, \Beta)$ is bounded,
and similarly, $\phiP(\cR; \BO, \Betahat)$ is also bounded.
Thus,
$
\normp{\varphidiff}^2
\allowbreak
=
\ED{[}{]}{ \{\varphidiff\}^2 }
\lesssim
E \big[
    \abs{\varphidiff}
    \big| D
\big]
.$
From
\begin{align*}
    \abs{\varphidiff}
    \le &
    \sumaN
    |{(\ORhat -\OR)(\cR; \BO, \Ba)}|
    +
    |(\IPCWBChat -\IPCWBC)(\cR; \BO)|
    \\
    &+
    |(\AUGhat -\AUG)(\cR; \BO)|
,
\end{align*}
it remains to bound each difference.

First, from
\begin{align*}
    \abs{(\ORhat -\OR)(\cR; \BO, \Ba)}
    &=
    \abs{
        \big\{ 
            \Bwhat(\Ba) + \Bphihat(\Ba) 
        \big\}^\top
        \widehat{\BG}^{\cR}(0|\Ba)
        -
        \big\{ 
            \Bw(\Ba) + \Bphi(\Ba) 
        \big\}^\top
        \BG^{\cR}(0|\Ba)
    }
    \\
    &\lesssim
    \normt{(\Bwhat-\Bw)(\Ba)}
    +
    \normt{(\Bphihat-\Bphi)(\Ba)}
    +
    \normt{(\widehat{\BG}^{\cR}-{\BG}^{\cR})(0|\Ba)}
    ,
\end{align*}
we obtain
\begin{align*}
    &
    \ED{[}{]}{
        \sumaN
            \abs{(\ORhat -\OR)(\cR; \BO, \Ba)}
    }
    \\
    &\lesssim
    \ED{[}{]}{
        \sumaN
            \normt{(\Bwhat-\Bw)(\aXN)}^2
    }^{1/2}
    +
    \ED{[}{]}{
        \sumaN
            \normt{(\Bphihat-\Bphi)(\AXN;\Ba)}^2
    }^{1/2}
    \\
    & \quad +
    \ED{[}{]}{
        \int_{0}^{\infty}
            \normt{
                (\BlambdaThat - \Blambda^T) (r|\AXN)
            }^2
        dr
    }^{1/2}
    \\
    &=
    \OP{r_{\Bw}^2 + r_{\Bphi}}
    +
    \OP{r_{\Bphi}}
    +
    \OP{r_{\Blambda^T}}
    .
\end{align*}
The bound for $\Bwhat - \Bw$ is derived from
\begin{align*}
    &
    \ED{[}{]}{
        \sumaN \normt{(\Bwhat - \Bw)(\aXN)}^2
    }
    \\
    \lesssim &
    \ED{[}{]}{
        \sumaN \normt{\Bwhat(\Ba) - \Bw(\Ba) + 
        E\{ \Bphihat | D, \mathbf{X}, N \}}^2
    }
    +
    \ED{[}{]}{
        \sumaN \normt{E\{ \Bphihat(\Ba) | D, \mathbf{X}, N \}}^2
    }
    \\
    = &
    \ED{[}{]}{
        \sumaN \normt{\Bwhat(\Ba) - \Bw(\Ba) + 
        E\{ \Bphihat(\Ba) | D, \mathbf{X}, N \}}^2
    }
    +
    \ED{[}{]}{
        \sumaN \normt{E\{ (\Bphihat - \Bphi)(\Ba) | D, \mathbf{X}, N \}}^2
    }
    \\
    \lesssim &
    \ED{[}{]}{
        \sumaN 
        \norm{
            (\Bwhat - \Bw)(\aXN) 
            + 
            \sum_{\Ba' \in \cAN}
                \Bphihat(\Ba',\XN;\Ba) H(\Ba',\XN)
        }_2^2
    }
    \\ 
    &\quad +
    \ED{[}{]}{
        \sumaN \normt{ 
            (\Bphihat - \Bphi)(\AXN;\Ba)
        }^2
    }
    \\
    = &
    \OP{r_{\Bw}^4 + r_{\Bphi}^2}
\end{align*}
using
$\E{\{}{\}}{
    \normt{
        \E{(}{)}{
            V | W
        }
    }^2
}
\le
\E{(}{)}{
    \normt{V}^2
}
$
and
$E\{\phit | D, \mathbf{X}, N \} 
= 
E \left\{ 
    \phi(\mathbf{A}, \mathbf{X}, N; \mathbf{a})
    \middle| 
    \mathbf{X}, N
\right\}
=
0$.
Similarly, we have
\begin{align*}
    &
    \abs{(\IPCWBChat -\IPCWBC)(\cR; \BO)}
    \\
    &\lesssim
    \normt{(\Bwhat-\Bw)(\BA)}
    +
    \abs{(\Hhat-H)(\BA)}
    +
    \sum_{j}
        \cR(T_j)
        \abs{\SCjhat(T_j) - S^C_j(T_j)}
    +
    \normt{(\widehat{\BG}^{\cR}-{\BG}^{\cR})(0)}
\end{align*}
from
\begin{align*}
    \abs{    
        \frac
            {\Delta_j  \cR(Y_j)}
            {\SCjhat(Y_j)}
        -
        \frac
            {\Delta_j  \cR(Y_j)}
            {S^C_j(Y_j)}
    }
    =
    \frac
        {\Delta_j  \cR(T_j) 
        \abs{\SCjhat(T_j) - S^C_j(T_j)}}
        {\SCjhat(T_j) S^C_j(T_j)}
    \lesssim
    \cR(T_j)
    \abs{
        \SCjhat(T_j) - S^C_j(T_j)
    }
    .
\end{align*}
Thus, we obtain the following bound for $\IPCWBChat-\IPCWBC$:
\begin{align*}
    &
    \ED{[}{]}{
        \abs{(\IPCWBChat -\IPCWBC)(\cR; \BO)}
    }
    \\
    &\lesssim
    \ED{[}{]}{
        \normt{(\Bwhat-\Bw)(\AXN)}^2
    }^{1/2}
    +
    \ED{[}{]}{
        (\Hhat - H)(\AXN)^2
    }^{1/2}
    \\
    & \quad +
    \ED{[}{]}{
        \left\{
            \int_{0}^{\infty}
                \normt{(\BSChat-\BS^C)(u|\AXN)}
            du
        \right\}^2
    }^{1/2}
    +
    \ED{[}{]}{
        \int_{0}^{\infty}
            \normt{
                (\BlambdaThat - \Blambda^T) (r|\AXN)
            }^2
        dr
    }^{1/2}
    \\
    &=
    \OP{r_{\Bw}^2 + r_{\Bphi}}
    +
    \OP{r_{H}}
    +
    \OP{r_{\BS^C}}
    +
    \OP{r_{\Blambda^T}}
    ,
\end{align*}
where the second inequality follows from
\begin{align*}
    \ED{[}{]}{
        \normt{\vecj{
            \cR(T_j)
            \abs{\SCjhat(T_j) - S^C_j(T_j)}
        }}
    }
    & \lesssim
    \ED{[}{]}{
        \sum_{j=1}^N
            \cR(T_j)
            \abs{\SCjhat(T_j) - S^C_j(T_j)}
    }
    \\
    & =
    \ED{[}{]}{
        \sum_{j=1}^N
        \E{\{}{\}}{
                \cR(T_j)
                \abs{\SCjhat(T_j) - S^C_j(T_j)}
            \middle| D,\AXN        
        }
    }
    \\
    & \lesssim
    \ED{[}{]}{
        \sum_{j=1}^N
        \int_{0}^{\infty}
            \abs{\SCjhat(u) - S^C_j(u)}
        dF_j^T(u|\AXN)
    }
    \\
    & \lesssim
    \ED{[}{]}{
        \int_{0}^{\infty}
            \sum_{j=1}^N
            \abs{\SCjhat(u) - S^C_j(u)}
        du
    }
    (\because) \ dF_j^T(u) = S_j^T(u) \lambda_j^T(u) du \lesssim du
    \\
    & \lesssim
    \ED{[}{]}{
            \int_{0}^{\infty}
                \normt{(\BSChat-\BS^C)(u|\AXN)}
            du
    }
    \\
    & \lesssim
    \normp{
        \int_{0}^{\infty}
            \normt{(\BSChat-\BS^C)(u|\AXN)}
        du
    }
    \\
    & = 
    \OP{r_{\BS^C}}
\end{align*}

Finally, from (\ref{pf:thm1:stieltjesexpansion}),
we obtain
\begin{align*}
    \abs{(\AUGhat -\AUG)(\cR; \BO)}
    &\lesssim
    \normt{(\Bwhat-\Bw)(\BA)}
    +
    \abs{(\Hhat-H)(\BA)}
    +
    \sum_j
        \cR(C_j)
        \abs{(\widehat{L}_j-L_j)(C_j)}
    \\
    & \quad +
    \sum_{j}
        \int_{0}^{\infty}
            \abs{(\widehat{L}_j-L_j)(r)}
        dr
    +
    \sum_{j}
        \int_{0}^{\infty}
            \abs{(\lambdaCjhat-\lambda_j^C)(r)}
        dr
    \\
    & \lesssim
    \normt{(\Bwhat-\Bw)(\BA)}
    +
    \abs{(\Hhat-H)(\BA)}
    +
    \normt{(\BSThat-\BS^T)(\cR|\AXN)}
    \\
    & \quad + 
    \sum_j
        \cR(C_j)
        \abs{(\STjhat-S^T_j)(C_j)}
    +
    \sum_j
        \cR(C_j)
        \abs{(\SCjhat-S^C_j)(C_j)}
    \\
    & \quad +
    \sum_{j}
        \int_{0}^{\infty}
            \abs{(\SCjhat-S^C_j)(r)}
        dr
    +
    \sum_{j}
        \int_{0}^{\infty}
            \abs{(\lambdaTjhat-\lambda^T_j)(r)}
        dr
    +
    \sum_{j}
        \int_{0}^{\infty}
            \abs{(\lambdaCjhat-\lambda_j^C)(r)}
        dr
    ,
\end{align*}
where the last inequality follows from
\begin{align*}
    \abs{(\widehat{L}_j-L_j)(r)}
    =
    \abs{
        \frac
            {\widehat{G}_j^{\cR}(r)}
            {\SCjhat(r) \STjhat(r)}
        -
        \frac
            {G_j^{\cR}(r)}
            {S^C_j(r) S^T_j(r)}
    }
    \lesssim
    \abs{(\SCjhat-S^C_j)(r)}
    +
    \abs{(\STjhat-S^T_j)(r)}
    +
    \int_{0}^{\infty}
        |\lambdaTjhat- \lambda^T_j|(r)
    dr
    .
\end{align*}
Similar to the derivation of bound for $\IPCWBChat-\IPCWBC$,
we obtain
\begin{align*}
    &
    \ED{[}{]}{
        \abs{(\AUGhat -\AUG)(\cR; \BO)}
    }
    \\
    &\lesssim
    \ED{[}{]}{
        \normt{(\Bwhat-\Bw)(\AXN)}^2
    }^{1/2}
    +
    \ED{[}{]}{
        (\Hhat - H)(\AXN)^2
    }^{1/2}
    +
    \ED{[}{]}{
        \normt{(\BFThat-\BF^T)(\cR|\AXN)}^2
    }^{1/2}
    \\
    & \quad +
    \ED{[}{]}{
        \left\{
            \int_{0}^{\infty}
                \normt{(\BFThat-\BF^T)(u|\AXN)}
            du
        \right\}^2
    }^{1/2}
    +
    \ED{[}{]}{
        \left\{
            \int_{0}^{\infty}
                \normt{(\BSChat-\BS^C)(u|\AXN)}
            du
        \right\}^2
    }^{1/2}
    \\
    &=
    \OP{r_{\Bw}^2 + r_{\Bphi}}
    +
    \OP{r_{H}}
    +
    \OP{r_{\Blambda^T}}
    +
    \OP{r_{\BS^C}}
    ,
\end{align*}
where the following relationship is employed:
\begin{align*}
    \int_{0}^{\infty}
        (\FTjhat - F_j^T)(u|\AXN)
    du
    =&
    \int_{0}^{\infty}
            \exp (-\LambdaTjhat(u|\AXN))
            - 
            \exp \Big(-\Lambda_j^T(u|\AXN)\Big)
    du
    \\
    \lesssim &
    \int_{0}^{\infty}
        \abs{
            (\LambdaTjhat - \Lambda_j^T)(u|\AXN)
        }
    du
    \\
    =&
    \int_{0}^{\infty}
        \abs{
            \int_{0}^{u}
                (\lambdaTjhat - \lambda_j^T)(r|\AXN)
            dr
        }
    du
    \\
    \lesssim &
    \int_{0}^{\infty}
        \int_{0}^{\infty}
            \abs{
                (\lambdaTjhat - \lambda_j^T)(r|\AXN)
            }
        dr
    du
    \\
    \lesssim &
    \int_{0}^{\infty}
        \abs{
            (\lambdaTjhat - \lambda_j^T)(r|\AXN)
        }
    dr
\end{align*}
and therefore
\begin{align*}
    \ED{[}{]}{
        \left\{
            \int_{0}^{\infty}
                \normt{(\BFThat-\BF^T)(u|\AXN)}
            du
        \right\}^2
    }^{1/2}
    & \lesssim
    \ED{[}{]}{
        \left\{
            \int_{0}^{\infty}
                \sum_j
                    \abs{(\FTjhat - F_j^T)(u|\AXN)}
            du
        \right\}^2
    }^{1/2}
    \\
    & \lesssim
    \ED{[}{]}{
        \left\{
        \int_{0}^{\infty}
            \sum_j
            \abs{
                (\lambdaTjhat - \lambda_j^T)(r|\AXN)
            }
        dr
        \right\}^2
    }^{1/2}
    \\
    & \lesssim 
    \ED{[}{]}{
        \left\{
        \int_{0}^{\infty}
            \normt{(\BlambdaThat-\Blambda^T)(u|\AXN)}
        dr
        \right\}^2
    }^{1/2}
    \\
    & =
    \OP{r_{\Blambda^T}}
    .
\end{align*}
Combining the bounds for each difference, we have
\begin{align}
    \normp{\varphidiff}
    =
    \OP{
        r_{\Bw}^2 + r_{\Bphi} + r_{\Blambda^T} 
        + r_{H} + r_{\BS^C}
    }
    \label{pf:thm1:normdiff}
\end{align}
and thus
$B_{m,2} 
=
m^{-1/2}
\OP{
    r_{\Bw}^2 + r_{\Bphi} + r_{\Blambda^T} 
    + r_{H} + r_{\BS^C}
}$.

In conclusion, 
from (\ref{pf:thm1:Pdiff}) and (\ref{pf:thm1:normdiff}), 
the decomposition (\ref{pf:thm1:decomp}) is given by
\begin{align*}
    &
    \widehat{\Psi}(\cR;\Bw) - \Psi(\cR;\Bw)
    \\
    &=
    (\bPm - \bP)
            \varphi(\cR; \BO, \Beta) 
    +
    \OP{m^{-1}}
    +
    m^{-1/2}
    \OP{
        r_{\Bw}^2 + r_{\Bphi} + r_{\Blambda^T} 
        + r_{H} + r_{\BS^C}
    }
    +
    \OP{
        r_{\Bw}^2 
        + 
        r_{\Blambda^T} (r_H + r_{\Bphi} + r_{\BS^C})
    }
    .
\end{align*}
Therefore, we have the following results:
\\
\textup{(i) \underline{(Consistency)}}
If $\rw = o(1)$ 
and 
$\rT(\rH + \rphi + \rC) = o(1)$ 
as $m \to \infty$,
then 
\begin{align*}
    \widehat{\Psi}(\cR;\Bw) - \Psi(\cR;\Bw)
    =
    (\mathbb{P}_m - \mathbb{P})
    \varphi(\cR; \BO, \Beta) 
    +
    \oP{1}
    =
    \oP{1}
\end{align*}
from the law of large numbers, 
and thus
$\widehat{\Psi}(\cR;\Bw) \overset{p}{\to} \Psi(\cR;\Bw)$.
\\
(ii) \underline{(Asymptotic Normality)}
If 
$\rphi + \rH + \rT + \rC = o(1)$, 
$\rw = o(m^{-1/4})$, 
and $\rT(\rH + \rphi + \rC) = o(m^{-1/2})$ 
as $m \to \infty$,
then
\begin{align*}
    \sqrt{m}\{\widehat{\Psi}(\cR;\Bw) - \Psi(\cR;\Bw)\}
    =
    \sqrt{m}(\mathbb{P}_m - \mathbb{P})
    \varphi(\cR; \BO, \Beta) 
    +
    \oP{1}
    ,
\end{align*}
and thus
$
\sqrt{m}\{\widehat{\Psi}(\cR;\Bw) - \Psi(\cR;\Bw)\}
\overset{d}{\to}
N(0,\sigma(\cR; \Bw)^2)
$
from the central limit theorem,
where 
$
\sigma(\cR; \Bw)^2
=
\textup{Var}
    \big\{ 
        \varphi(\cR; \mathbf{O}, \boldsymbol{\eta}) 
    \big\}
$.
\\
(iii) \underline{(Consistent Variance Estimator)}
Assume $\rw + \rphi + \rH + \rT + \rC = o(1)$ as $m \to \infty$.
To prove $\widehat{\sigma}(\cR;\Bw) \overset{p}{\to} \sigma(\cR; \Bw)$,
note that
$$
    \widehat{\sigma}(\cR;\Bw)^2
    = 
    \sumk
    \bPmk
    \Big[ 
        \big\{ \allowbreak
            \varphi(\cR; \mathbf{O}, \Betahatk) 
        \big\}^2
    \Big]
    - 
    \widehat{\Psi}(\cR;\Bw)^2
$$
and
$$
    \sigma(\cR; \Bw)^2 
    = 
    \mathbb{P} \allowbreak
    \Big[ 
        \big\{
            \varphi(\cR; \mathbf{O}, \Beta) 
        \big\}^2
    \Big]
    - 
    \Psi(\cR;\Bw)^2
    .
$$
Since 
$\widehat{\Psi}(\cR;\Bw) \overset{p}{\to} \Psi(\cR;\Bw)$ from (i),
it suffices to show
$$
\mathbb{P}_m^k \allowbreak
    \Big[ 
        \big\{ \allowbreak
            \varphi(\cR; \mathbf{O}, \Betahatk) \allowbreak
        \big\}^2
    \Big]
\overset{p}{\to}
\mathbb{P} \allowbreak
    \Big[ 
        \big\{ \allowbreak
            \varphi(\cR; \mathbf{O}, \Beta) \allowbreak
        \big\}^2
    \Big]
.
$$
From the conditional law of large numbers 
(Theorem 4.2. in \cite{majerek2005conditional}),
$$
\mathbb{P}_m^k \allowbreak
    \Big[ 
        \big\{ \allowbreak
            \varphi(\cR; \mathbf{O}, \Betahatk) \allowbreak
        \big\}^2
    \Big]
-
\mathbb{P} \allowbreak
    \Big[ 
        \big\{ \allowbreak
            \varphi(\cR; \mathbf{O}, \Betahatk) \allowbreak
        \big\}^2
    \Big]
=\oP{1}
.
$$
Thus, it suffices to show
$$
\mathbb{P} \allowbreak
    \Big[ 
        \big\{ \allowbreak
            \varphi(\cR; \mathbf{O}, \Betahatk) \allowbreak
        \big\}^2
        -
        \big\{ \allowbreak
            \varphi(\cR; \mathbf{O}, \Beta) \allowbreak
        \big\}^2
    \Big]
=\oP{1}
.
$$
Using some algebra and from Cauchy-Schwarz inequality,
\begin{align*}
    &
    \mathbb{P} \Big[ 
        \big\{
            \varphi(\cR; \mathbf{O}, \Betahatk) 
        \big\}^2
        -
        \big\{
            \varphi(\cR; \mathbf{O}, \Beta)
        \big\}^2
    \Big]
    \\
    = &
    \mathbb{P} \Big[ 
        \big\{
            \varphidiffk
        \big\}^2
    \Big]
    +
    2
    \mathbb{P} \Big[ 
        \big\{
            \varphidiffk
        \big\}
        \varphi(\cR; \mathbf{O}, \Beta)
    \Big]
    \\
    \le &
    \mathbb{P} \Big[ 
        \big\{
            \varphidiffk
        \big\}^2
    \Big]
    +
    2
    \left(
        \mathbb{P} \Big[ 
            \big\{
                \varphidiffk
            \big\}^2
        \Big]
    \right)^{\frac{1}{2}}
    \left(
        \mathbb{P} \Big[ 
            \big\{
                \varphi(\cR; \mathbf{O}, \Beta)
            \big\}^2
        \Big]
    \right)^{\frac{1}{2}}
    .
\end{align*}
Here,
\begin{align*}
\left(
    \mathbb{P} \Big[ 
        \big\{
            \varphidiffk
        \big\}^2
    \Big]
\right)^{\frac{1}{2}}
    =
    \normp{ \varphidiffk }
    =
    \OP{\rw^2 + \rphi + \rH + \rT + \rC}
    =
    \oP{1}
\end{align*}
under $\rw + \rphi + \rH + \rT + \rC = o(1)$
and
\begin{align*}
    \left(
        \mathbb{P} \Big[ 
            \big\{
                \varphi(\cR; \mathbf{O}, \Beta)
            \big\}^2
        \Big]
    \right)^{\frac{1}{2}}
    =
    \sigma(\cR; \Bw)
    =\OP{1}
\end{align*}
which finishes the proof.
Thus, if all conditions hold,
$
\sqrt{m}\{\widehat{\Psi}(\cR;\Bw) - \Psi(\cR;\Bw)\}/\widehat{\sigma}(\cR;\Bw)
\overset{d}{\to}
N(0,1)
$.
\hfill $\blacksquare$

\subsection{Proof of Theorem 2}

To prove the weak convergence of the proposed estimator 
$m^{1/2}\big\{
    \widehat{\Psi}(\cR, \cdot)
    - 
    \Psi(\cR, \cdot)
\big\}
\rightsquigarrow
\mathbb{G}(\cdot)$
to a Gaussian process in $\ell^{\infty} (\Theta)$
as $m \to \infty$,
begin with a decomposition of the bias:
\begin{align*}
    \widehat{\Psi}(\cR, \cdot)
    - 
    \Psi(\cR, \cdot)
    =
    (\bPm - \bP) \varphi(\cR, \theta; \BO, \Beta) 
    +
    B_{m,1}(\theta)
    +
    B_{m,2}(\theta)
    +
    B_{m,3}(\theta)
\end{align*}
where 
$\theta \in \Theta$
and
\begin{align*}
    B_{m,1}(\theta) 
    =& 
    \sumk
        \bPmk \{ \phiG{\cR}{\theta} \}
        \left( 1 - \frac{K m_k}{m} \right)
    ,
    \\
    B_{m,2}(\theta) 
    =&
    \sumk
        (\bPmk - \bP) \big\{ 
            \phiGhatk{\cR}{\theta} - \phiG{\cR}{\theta}
        \big\}
    ,
    \\
    B_{m,3}(\theta)
    =&
    \sumk
        \bP \big\{ 
            \phiGhatk{\cR}{\theta} - \phiG{\cR}{\theta}
        \big\}
    .
\end{align*}
In the following, it is shown that
(i) the process 
$\sqrt{m}(\bPm - \bP)\phiG{\cR}{\cdot}$ 
weakly converges to the Gaussian process $\mathbb{G}(\cdot)$ in $\ell^{\infty} (\Theta)$,
and
(ii)
the remainder processes $B_{m,1}(\cdot)$, $B_{m,2}(\cdot)$, and $B_{m,3}(\cdot)$ are $\oP{m^{-1/2}}$ in $\ell^{\infty} (\Theta)$,
which together establish the desired result.

First, (i) holds from the fact that the function class 
$\cF_{{\varphi}} 
= 
\{ 
    \varphi(\cR, \theta; \BO, \Beta)
    : \theta \in \Theta
\}$
is a Donsker class for any fixed ${\boldsymbol{\eta}}$.
To show that 
$\mathcal{F}_{\varphi}$
is Donsker, 
first note that
\begin{align*}
    \phiP(\cR, \theta; \BO, \Beta)
    =
    \sumaN
        \OR(\cR, \theta; \BO, \Ba) 
        +
        \IPCWBC(\cR, \theta; \BO)
        +
        \AUG(\cR, \theta; \BO),
\end{align*}
where
\begin{align*}
        \OR(\cR, \theta; \BO, \Ba) 
        =&
        \big\{ \Bw(\aXN;\theta) + \Bphi(\AXN; \Ba; \theta) \big\}^\top
        \BG^{\cR} \big(0 | \aXN \big) 
        , 
        \\
        \IPCWBC(\cR, \theta; \BO)
        =&
        \frac{1}{H(\AXN)}
        \Bw(\AXN; \theta)^\top 
        \left[ 
            \frac
                {\Delta_j  \cR(Y_j)}
                {\SCj{Y_j}}
            - 
            G_j^{\cR}(0|\AXN)
        \right]_{j=1}^N
        , 
        \\
        \AUG(\cR, \theta; \BO)
        =&
        \frac{1}{H(\AXN)}
        \Bw(\AXN;\theta)^\top 
        \left[ 
            \int_{0}^{\infty}
                \frac
                    {G_j^{\cR}(r|\AXN)}
                    {\SCj{r} \STj{r}}
            dM^C_j(r)
        \right]_{j=1}^N
        .
\end{align*}
From the assumption that the function classes
$\cF_{\Bw} = \{\Bw(\axn;\theta): \theta \in \Theta\}$
and
$\cF_{\Bphi} = \{\Bphi(\Ba',\xn;\Ba, \theta): \theta \in \Theta\}$
are Donsker classes,
and the fact that the class of products and sums of functions in Donsker classes is Donsker \citep{kennedy16}, it follows that $\cF_{{\varphi}}$ is also a Donsker class.

Next, (ii) is shown. 
From the proof of Theorem 1,
\begin{align*}
    \sup_{\theta \in \Theta}
    \abs{
        \bPmk
            \left\{
                \varphi(\cR, \theta; \BO, \Beta) 
            \right\}
            \left(
            1
            -
            \frac{K m_k}{m}
            \right)
    }
    =
    \sup_{\theta \in \Theta}
    \abs{
        \{\Psi(\cR;\theta) + \oP{1}\}O(m^{-1})
    }
    =
    \OP{m^{-1}}
    ,
\end{align*}
and thus 
$\norm{B_{m,1}}_{\ell^{\infty} (\Theta)} = \oP{m^{-1/2}}$.
Additionally, it follows that
\begin{align*}
    &
    \normp{\phiP(\cR, \theta; \BO, \Betahatk) - \phiP(\cR, \theta; \BO, \Beta)}
    \\
    & \lesssim
    \normp{
        \int_{0}^{\infty}
            \normt{
                \big(
                    \BlambdaThatk
                    -
                    \Blambda^T
                \big)(r|\AXN)
            }
        dr
    }
    +
    \normp{
        \int_{0}^{\infty}
            \normt{
                \big(
                    \BSChatk
                    -
                    \BS^C
                \big)(r|\AXN)
            }
        dr
    }
    \\
    & \quad +
    \normp{
        \abs{
            \big(
                \Hhatk
                -
                H
            \big)(\AXN)
        }
    }
    +
    \sup_{\theta \in \Theta}
    \normp{
        \textstyle \sumaN
            \normt{
                \big(
                    \Bphihatk - \Bphi
                \big)
                (\mathbf{A}, \mathbf{X}, N; \mathbf{a}, 
                \theta)
            }
    }
    \\
    & \quad + 
    \sup_{\theta \in \Theta}
    \normp{
            \normt{
                \big(
                \Bwhatk
                -
                \Bw
                \big)
                (\mathbf{a}, \mathbf{X}, N; \theta)
                +
                \textstyle \sum_{\mathbf{a}' \in \mathcal{A}(N)}
                    \Bphihatk(\mathbf{a}', \mathbf{X}, N; \mathbf{a}, \theta)
                    H(\mathbf{a}', \mathbf{X}, N)
            }
    }
    \\
    & =  
    \OP{
        r_{\Bw}^2 + r_{\Bphi} + r_{\Blambda^T} 
        + r_{H} + r_{\BS^C}
    }
    \\
    & = 
    \oP{1}
\end{align*}
under the conditions given in Theorem 2.
Therefore,
\begin{align*}
    \sup_{\theta \in \Theta}
    \abs{
        (\bPmk - \bP) \big\{ 
            \phiP(\cR, \theta; \BO, \Betahatk) - \phiP(\cR, \theta; \BO, \Beta) 
        \big\}
    }
    &=
    O_{P} \left( 
        m_k^{-1/2}
        \sup_{\theta \in \Theta}
            \normp{\phiP(\cR, \theta; \BO, \Betahatk) - \phiP(\cR, \theta; \BO, \Beta)}
    \right)
    \\
    &=
    \oP{m^{-1/2}}
\end{align*}
and thus
$\norm{B_{m,2}}_{\ell^{\infty} (\Theta)} = \oP{m^{-1/2}}$.
Finally, from
\begin{align*}
    &
    \sup_{\theta \in \Theta}
    \mathbb{P} \big\{ 
        \phiP(\cR, \theta; \BO, \Betahatk) - \phiP(\cR, \theta; \BO, \Beta) 
    \big\}
    \\
    & \lesssim
    \sup_{\theta \in \Theta}
    \normp{
            \normt{
                \big(
                \Bwhatk
                -
                \Bw
                \big)
                (\mathbf{a}, \mathbf{X}, N; \theta)
                +
                \textstyle \sum_{\mathbf{a}' \in \mathcal{A}(N)}
                    \Bphihatk(\mathbf{a}', \mathbf{X}, N; \mathbf{a}, \theta)
                    H(\mathbf{a}', \mathbf{X}, N)
            }
    }
    \\
    & \quad +
    \normp{
        \int_{0}^{\infty}
            \normt{
                \big(
                    \BlambdaThatk
                    -
                    \Blambda^T
                \big)(r|\AXN)
            }
        dr
    }
    \left(
        \normp{
            \int_{0}^{\infty}
                \normt{
                    \big(
                        \BSChatk
                        -
                        \BS^C
                    \big)(r|\AXN)
                }
            dr
        }
        +
        \normp{
            \abs{
                \big(
                    \Hhatk
                    -
                    H
                \big)(\AXN)
            }
        }
    \right)
    \\
    & \quad +
    \normp{
        \int_{0}^{\infty}
            \normt{
                \big(
                    \BlambdaThatk
                    -
                    \Blambda^T
                \big)(r|\AXN)
            }
        dr
    }
    \sup_{\theta \in \Theta}
    \normp{
        \textstyle \sumaN
            \normt{
                \big(
                    \Bphihatk - \Bphi
                \big)
                (\mathbf{A}, \mathbf{X}, N; \mathbf{a}, 
                \theta)
            }
    }
    \\
    & =  
    \OP{
        r_{\Bw}^2 
        + 
        r_{\Blambda^T}(r_{\Bphi} + r_{H} + r_{\BS^C})
    }
    ,
\end{align*}
it follows that
$\norm{B_{m,3}}_{\ell^{\infty} (\Theta)} = \oP{m^{-1/2}}$,
under the conditions given in Theorem 2,
which completes the proof. 
\hfill $\blacksquare$


\subsection{Proof of Theorem 3}

The proof follows the approach of \cite{kennedy19} to establish the approximation using multiplier bootstrap process. Define the following processes:
\begin{align*}
    &
    \psi_{m}(\theta) 
    =
    m^{1/2}
    (\bPm - \bP) \varphi(\cR, \theta; \BO, \Beta) 
    =
    \sumi
        \left\{
        \frac
            {\varphi(\cR, \theta; \Oi, \Beta) - \Psi(\cR, \theta)}
            {\sigma(\cR, \theta) / m^{1/2}}
        \right\},
    \\
    &
    \widehat{\psi}_m(\theta) 
    = 
    \frac
        {\widehat{\Psi}(\cR, \theta)-\Psi(\cR, \theta)}
        {\widehat{\sigma}(\cR, \theta) / m^{1/2}},
    \\
    &
    \psi_{m}^{*}(\theta) 
    =
    \sumi
        \left[
            \xi_i
            \left\{
            \frac
                {\varphi(\cR, \theta; \Oi, \Beta) - \Psi(\cR, \theta)}
                {\sigma(\cR, \theta) / m^{1/2}}
            \right\}
        \right],
    \\
    &
    \widehat{\psi}_{m}^{*}(\theta) 
    =
    \sumk
    \sumik
        \left[
            \xi_i
            \left\{
            \frac
                {\varphi(\cR, \theta; \Oi, \Betahatk) - \widehat{\Psi}(\cR, \theta)}
                {\widehat{\sigma}(\cR, \theta) / m^{1/2}}
            \right\}
        \right],
\end{align*}
where the star superscripts denote multiplier bootstrap processes. Let $\mathbb{G}$ denote a mean-zero Gaussian process with covariance function
\[
E \{ 
    \mathbb{G}(\theta_1) 
    \mathbb{G}(\theta_2) 
\} 
= 
\textup{Cov} \{ 
    \tilde{\varphi}(\cR, \theta_1; \mathbf{O}, \boldsymbol{\eta}),
    \tilde{\varphi}(\cR, \theta_2; \mathbf{O}, \boldsymbol{\eta}) 
\},
\]
where 
\[
\tilde{\varphi}(\cR, \theta; \mathbf{O}, \boldsymbol{\eta})
=
\frac{\varphi(\cR, \theta; \mathbf{O}, \boldsymbol{\eta}) - \Psi(\cR, \theta)}{\sigma(\cR, \theta)}
\]
is the standardized version of $\varphi(\cR, \theta; \mathbf{O}, \boldsymbol{\eta})$.

The critical value $\widehat{c}_{\alpha}$ is defined as the $(1-\alpha)$ quantile of the bootstrap process $\widehat{\psi}_{m}^{*}(\cdot)$, i.e.,
\[
\textup{pr}\left\{
    ||\widehat{\psi}_{m}^{*}||_{\ell^{\infty} (\Theta)}
    \le 
    \widehat{c}_{\alpha}
\right\}
= 1-\alpha+o(1).
\]
Since
\begin{align*}
    &
    \textup{pr}\left\{
        \widehat{\Psi}(\cR, \theta)
        -
        \widehat{c}_{\alpha}
        \frac{\widehat{\sigma}(\cR, \theta)}{m^{1/2}}
        \le
        \Psi(\cR, \theta)
        \le
        \widehat{\Psi}(\cR, \theta)
        +
        \widehat{c}_{\alpha}
        \frac{\widehat{\sigma}(\cR, \theta)}{m^{1/2}}
        \textup{, for all } \theta \in \Theta
    \right\}
    \\
    &=
    \text{pr}\left\{
        \sup_{\theta \in \Theta}
        \left|
            \frac
                {\widehat{\Psi}(\cR, \theta)
                - 
                \Psi(\cR, \theta)}
                {\widehat{\sigma}(\cR, \theta) / m^{1/2}}
        \right| 
        \le 
        \widehat{c}_{\alpha} 
    \right\}
    =
    \textup{pr}\left\{
        ||\widehat{\psi}_{m}||_{\ell^{\infty} (\Theta)}
        \le 
        \widehat{c}_{\alpha}
    \right\},
\end{align*}
it remains to show that
\[
\left|
    \textup{pr}\left\{
        ||\widehat{\psi}_{m}^{*}||_{\ell^{\infty} (\Theta)}
        \le 
        \widehat{c}_{\alpha}
    \right\}
    -
    \textup{pr}\left\{
        ||\widehat{\psi}_{m}||_{\ell^{\infty} (\Theta)}
        \le 
        \widehat{c}_{\alpha}
    \right\}
\right|
= o(1).
\]

In the proof of Theorem 2, it was shown that 
$||\widehat{\psi}_{m} - \psi_m||_{\ell^{\infty} (\Theta)} = o_P(1)$,
which implies 
$\big| ||\widehat{\psi}_{m}||_{\ell^{\infty} (\Theta)} - ||\psi_m||_{\ell^{\infty} (\Theta)} \big| = o_P(1)$.
Additionally, by Corollary 2.2 in \cite{chernozhukov2014gaussian} (Gaussian approximation to the supremum of empirical processes),
$\big| ||{\psi}_{m}||_{\ell^{\infty} (\Theta)} - ||\mathbb{G}||_{\ell^{\infty} (\Theta)} \big| = o_P(1)$,
and thus 
$\big| ||\widehat{\psi}_{m}||_{\ell^{\infty} (\Theta)} - ||\mathbb{G}||_{\ell^{\infty} (\Theta)} \big| = o_P(1)$.
Therefore, by Lemma 2.3 in \cite{chernozhukov2014gaussian} (Gaussian approximation in distribution functions), it follows that
\[
\sup_{c \in \bR}
\Big| 
    \textup{pr}\left\{
        ||\widehat{\psi}_{m}||_{\ell^{\infty} (\Theta)}
        \le c
    \right\}
-
    \textup{pr}\left\{
        ||\mathbb{G}||_{\ell^{\infty} (\Theta)}
        \le c
    \right\}
\Big|
= 
o(1).
\]

Next, by Corollary 4 of \cite{belloni2015uniform} (uniform convergence of the bootstrap process), we have
$\big| ||\widehat{\psi}_{m}^{*}||_{\ell^{\infty} (\Theta)} - ||\mathbb{G}||_{\ell^{\infty} (\Theta)} \big| = o_P(1)$.
Again, by Lemma 2.3 of \cite{chernozhukov2014gaussian}, it follows that 
\[
\sup_{c \in \bR}
\Big| 
    \textup{pr}\left\{
        ||\widehat{\psi}_{m}^{*}||_{\ell^{\infty} (\Theta)}
        \le c
    \right\}
-
    \textup{pr}\left\{
        ||\mathbb{G}||_{\ell^{\infty} (\Theta)}
        \le c
    \right\}
\Big|
= 
o(1).
\]

In conclusion, combining these two results, we have
\[
\sup_{c \in \bR}
\Big| 
    \textup{pr}\left\{
        ||\widehat{\psi}_{m}^{*}||_{\ell^{\infty} (\Theta)}
        \le c
    \right\}
-
    \textup{pr}\left\{
        ||\widehat{\psi}_{m}||_{\ell^{\infty} (\Theta)}
        \le c
    \right\}
\Big|
= 
o(1),
\]
which yields the desired result.
\hfill $\blacksquare$


\subsection{Proof of Theorem 4}

The proof of the weak convergence of the proposed estimator 
$m^{1/2}\big\{
    \widehat{\Psi}(\cdot, \cdot)
    - 
    \Psi(\cdot, \cdot)
\big\}
\rightsquigarrow
\mathbb{G}(\cdot, \cdot)$
in $\ell^{\infty} ([0,\infty) \times \Theta)$
as $m \to \infty$
when $\cR(T; \tau) = \indicator(T \le \tau)$
follows a similar approach to that of Theorem 2. 

The decomposition of the bias is expressed as:
\[
\widehat{\Psi}(\tau, \theta) - \Psi(\tau, \theta)
=
(\bPm - \bP) \varphi(\tau, \theta; \BO, \Beta) 
+
B_{m,1}(\tau, \theta)
+
B_{m,2}(\tau, \theta)
+
B_{m,3}(\tau, \theta),
\]
where $(\tau, \theta) \in [0,\infty) \times \Theta$, and the terms are defined as:
\begin{align*}
    B_{m,1}(\tau, \theta) 
    =& 
    \sumk
        \bPmk \{ \phiG{\tau}{\theta} \}
        \left( 1 - \frac{K m_k}{m} \right),
    \\
    B_{m,2}(\tau, \theta) 
    =&
    \sumk
        (\bPmk - \bP) \big\{ 
            \phiGhatk{\tau}{\theta} - \phiG{\tau}{\theta}
        \big\},
    \\
    B_{m,3}(\tau, \theta)
    =&
    \sumk
        \bP \big\{ 
            \phiGhatk{\tau}{\theta} - \phiG{\tau}{\theta}
        \big\}.
\end{align*}

To establish the weak convergence, it is shown that:
(i) the process 
$\sqrt{m}(\bPm - \bP)\phiG{\cdot}{\cdot}$ 
weakly converges to the Gaussian process $\mathbb{G}(\cdot, \cdot)$ in $\ell^{\infty} ([0,\infty) \times \Theta)$,
and
(ii) the remainder processes $B_{m,1}(\cdot,\cdot)$, $B_{m,2}(\cdot,\cdot)$, and $B_{m,3}(\cdot,\cdot)$ are $\oP{m^{-1/2}}$ in $\ell^{\infty} ([0,\infty) \times \Theta)$.

The first result follows from the fact that the function class 
$\cF_{{\varphi}} 
= 
\{ 
    \varphi(\tau, \theta; \BO, \Beta)
    : (\tau, \theta) \in [0,\infty) \times \Theta
\}$
is Donsker for any fixed ${\boldsymbol{\eta}}$.
To show that 
$\mathcal{F}_{\varphi}$
is Donsker, note that
\begin{align*}
    \phiP(\tau, \theta; \BO, \Beta)
    =
    \sumaN
        \OR(\tau, \theta; \BO, \Ba) 
        +
        \IPCWBC(\tau, \theta; \BO)
        +
        \AUG(\tau, \theta; \BO),
\end{align*}
\begin{align*}
    \OR(\tau, \theta; \BO, \Ba) 
    =&
    \big\{ \Bw(\aXN; \theta) + \Bphi(\AXN; \Ba; \theta) \big\}^\top
    \BF^T \big(\tau | \aXN \big), 
    \\
    \IPCWBC(\tau, \theta; \BO)
    =&
    \frac{1}{H(\AXN)}
    \Bw(\AXN; \theta)^\top 
    \left[ 
        \frac
            {\Delta_j  \indicator(Y_j \leq \tau)}
            {\SCj{Y_j}}
        - 
        \FTj{\tau}
    \right]_{j=1}^N, 
    \\
    \AUG(\tau, \theta; \BO)
    =&
    \frac{1}{H(\AXN)}
    \Bw(\AXN; \theta)^\top 
    \left[ 
        \int_{0}^{\infty}
            \frac
                {\STj{r} - \STj{\tau}}
                {\SCj{r} \STj{r}}
            \indicator(r \leq \tau)
        dM^C_j(r)
    \right]_{j=1}^N.
\end{align*}
Here, the class of functions 
$L_j(r) 
= 
\indicator(r \leq \tau)\{S^T_j(r) - S^T_j(\tau)\} 
\big/ 
\{S^C_j(r) S^T_j(r)\}$ 
is Donsker because classes of indicator functions, bounded monotone functions, and smooth functions are Donsker, and the class of products and sums of functions in Donsker classes is also Donsker \citep{kennedy16}. 
Furthermore, from (\ref{pf:thm1:stieltjesexpansion}), the class of functions
\begin{align*}
    \int_{0}^{\infty}
        L_j(r)
    dM^C_j(r)
    =
    \indicator(\Delta_j = 0)
    \indicator(Y_j \le \tau)
    L_j(Y_j)
    -
    \int_{0}^{\tau}
        L_j(r)
        \indicator(r \le Y_j)
        \lambda_j^C(r)
    dr
\end{align*}
is also Donsker because the function
$
    I_j(\tau)
    =
    \int_{0}^{\tau}
        L_j(r)
        \indicator(r \le Y_j)
        \lambda_j^C(r)
    dr
$
is differentiable with respect to $\tau$. 
Therefore, $\varphi(\tau, \theta; \BO, \Beta)$ is a sum of products of functions in Donsker classes, and thus (i) holds.

The second result follows by bounding the remainder terms using similar arguments as in Theorem 2.
\hfill $\blacksquare$


\subsection{Algorithm for the construction of the SBS-NCF estimator}

\begin{algorithm}
    \caption{Construction of the SBS-NCF estimator.} \label{alg:estimator2}
        \begin{spacing}{1.1}
        \begin{algorithmic}
            
            \Require
            Data $(\mathbf{O}_1, \dots, \mathbf{O}_m)$; number of folds $K$; split repetitions $S$; subsampling degree $r$ 
            
            \Ensure 
            SBS-NCF Estimator $\widehat{\Psi}^{\textup{SBS}} (\cR; \Bw)$ and variance estimator $\widehat{\sigma}^{\textup{SBS}} (\cR; \Bw)^2$

            \For{$s = 1, \dots, S$}
                
                \State 
                Randomly partition the data into $K$ disjoint folds; $G_i \in \{1, \dots, K\}$: fold assignment
                
                \For{$k = 1, \dots, K$}
                    
                    \State 
                    Set $m_k = \sum_{i=1}^{m} \indicator(G_i = k)$
                    
                    \State
                    Train $\Beta$ on $\{\Oi: G_i \neq k\}$, 
                    and denote it as
                    $\Betahatk
                    =
                    \big(
                        \BFThatk,
                        \BSChatk,
                        \Hhatk, 
                        \Bwhatk, 
                        \Bphihatk
                    \big)$
                    
                    \State
                    Compute mean IPW weight:
                    $\BV_{(k)} = \bPmk \{ \Bwhatk (\AXN)^\top \textbf{1} / \Hhatk (\AXN) \}$
                                        
                    \For{$i: G_i = k$}
                        
                        \State
                        Compute the following 

                        \State
                        \quad
                        \underline{\textit{Subsampling}}:
                        $
                        \ORhat_{i,(k)}^{\textup{Sub}} 
                        = 
                        r^{-1}
                        \sum_{q=1}^{r}
                        \ORhatk\big(\ai^{(q)}\big)
                        /\Hhatk\big(\ai^{(q)}\big)$,
                        for
                        $\ai^{(q)} \overset{iid}{\sim} \Hhatk (\cdot)$ 
                                                
                        \State
                        \quad
                        \underline{\textit{Bounding}}:
                        $
                        \IPCWBChat_{i,(k)}^{\textup{Bdd}} 
                        = 
                        \BV_{(k)}^{-1} \ 
                        \IPCWBChat_{i,(k)}$,
                        $\AUGhat_{i,(k)}^{\textup{Bdd}} 
                        = 
                        \BV_{(k)}^{-1} \ 
                        \AUGhat_{i,(k)}$

                        \State
                        Compute
                        $\phiP^{\textup{BS}}(\cR; \Oi, \Betahatk) 
                        =
                        \ORhat_{i,(k)}^{\textup{Sub}} 
                        +
                        \IPCWBChat_{i,(k)}^{\textup{Bdd}}
                        +
                        \AUGhat_{i,(k)}^{\textup{Bdd}}$
                        
                    \EndFor

                    \State
                    Average to yield
                    $\bPmk \big\{ 
                        \phiP^{\textup{BS}}(\cR; \BO, \Betahatk) 
                    \big\}$
                \EndFor
                \State
                Compute
                $\widehat{\Psi}_{s}^{\textup{BS}} (\cR; \Bw)
                =
                \sumkline
                \bPmk \big\{ 
                    \phiP^{\textup{BS}}(\cR; \BO, \Betahatk) 
                \big\}$
                \State
                Compute
                $\widehat{\sigma}_{s}^{\textup{BS}} (\cR; \Bw)^2
                =
                \sumkline
                \mathbb{P}_m^k 
                \Big[ 
                    \big\{ 
                        \phiP^{\textup{BS}}(\cR; \BO, \Betahatk) 
                        - 
                        \widehat{\Psi}_{s}^{\textup{BS}} (\cR; \Bw)
                    \big\}^2
                \Big]$
            \EndFor
            
            \State
            \underline{\textit{Split-robust}} estimator:
            $\widehat{\Psi}^{\textup{SBS}} (\cR; \Bw)
            =
            \textup{median}\{
                \widehat{\Psi}_{s}^{\textup{BS}} (\cR; \Bw)
                : s = 1, \dots, S
            \}$

            \State
            \underline{\textit{Split-robust}} variance estimator:
            $\widehat{\sigma}^{\textup{SBS}} (\cR; \Bw)^2
            =
            \textup{median}\{
                \widehat{\sigma}_{s}^{\textup{BS}} (\cR; \Bw)^2
                : s = 1, \dots, S
            \}$
            
        \end{algorithmic}
        \end{spacing}
\end{algorithm}


\subsection{Large sample property under subsampling approximation}

Here, the large sample properties of the proposed NCF estimators are presented when the subsampling approximation is applied.
First, from the proof of Theorem 1 (supplementary material Section \ref{proof:thm1}), the NCF estimator is given by
$
\widehat{\Psi}(\cR;\Bw)
=
\sumkline
    \bPmk \big\{ 
        \phiP(\cR; \BO, \Betahatk) 
    \big\}
$,
where
\begin{align*}
    \phiP(\cR; \BO, \Betahat)
    =
    \textstyle \sumaN
        \ORhat(\cR; \BO, \Ba) 
        +
        \IPCWBChat(\cR; \BO)
        +
        \AUGhat(\cR; \BO)
    ,
\end{align*}
and here the subscript ${(k)}$ is sometimes omitted for simplicity.

The subsampling approximation is utilized since the summation 
$\sumaN \ORhat(\cR; \BO, \Ba)$ can be computationally intensive.
Now consider a probability distribution $f$ on $\mathcal{A}(N)$
such that
$\sumaN f(\Ba) = 1$ 
and 
$f(\mathbf{a}) > 0$ for $\mathbf{a} \in \mathcal{A}(N)$ such that $\Bw(\aXN) > 0$.
Let
$\mathbf{a}^{*} 
= 
\big(\mathbf{a}^{(1)}, \dots,  \mathbf{a}^{(r)}\big)$
denote a random sample from $f$ such that
$\mathbf{a}^{(q)} \overset{\mathrm{iid}}{\sim} f(\cdot), q=1,\dots,r$,
which is conditionally independent of the observed treatment vector $\BA$ given $\XN$.
Then, 
\begin{align*}
    \sumaN  \ORhat(\cR; \BO, \Ba)
    =
    \sumaN 
        \frac
            {\ORhat(\cR; \BO, \Ba)}
            {f(\mathbf{a})}
        \times
        f(\mathbf{a})
    =
    E_{\mathbf{a}^{(\cdot)}} \left\{
        \frac
            {\ORhat(\cR; \BO, \Ba^{(\cdot)})}
            {f(\mathbf{a}^{(\cdot)})}
        \middle| D, \mathbf{O}
    \right\}
\end{align*}
which can be approximated by 
\begin{align*}
    \frac{1}{r}
    \sum_{q=1}^{r}
        \frac
            {\ORhat(\cR; \BO, \Ba^{(q)})}
            {f(\mathbf{a}^{(q)})}
    ,
\end{align*}
where $D$ denotes the data independent of $\mathbf{O}$ that $\widehat{\boldsymbol{\eta}}$ was trained on, as in the proof of Theorem 1.
In the main text,
for cluster $i$ in split group $G_i = k$,
the training data is $D = \{\BO_i: G_i \ne k \}$,
and $f$ was chosen to be $\Hhatk\big(\cdot, \XNi \big)$.

Now, let 
$\mathbf{O}^{*} 
= (\mathbf{O}, \mathbf{a}^{*}) 
= (\mathbf{Y}, \mathbf{\Delta}, \mathbf{A}, \mathbf{X}, N, \mathbf{a}^{*})$.
Also, define the approximated $\phiP(\cR; \BO, \Beta)$ by
\begin{align*}
    \varphi^{\textup{Sub}}(\cR; \BO^*, \Beta) 
    =
    \sumaq 
        \frac{\OR(\cR; \BO, \aq)}{f(\aq)}
    +
    \IPCWBC(\cR; \BO)
    +
    \AUG(\cR; \BO)
\end{align*}
and the approximated estimator $\phiP(\cR; \BO, \Betahat)$ by
\begin{align*}
    \varphi^{\textup{Sub}}(\cR; \BO^*, \Betahat) 
    =
    \sumaq 
        \frac{\ORhat(\cR; \BO, \aq)}{f(\aq)}
    +
    \IPCWBChat(\cR; \BO)
    +
    \AUGhat(\cR; \BO)
    .
\end{align*}
Note that
$E\{
    \varphi^{\textup{Sub}}(\cR; \BO^*, \Beta) 
    | \mathbf{O}
\}
=
\phiP(\cR; \BO, \Beta)$
and 
$E\{
    \varphi^{\textup{Sub}}(\cR; \BO^*, \Betahat)  
    | D, \mathbf{O}
\}
=
\phiP(\cR; \BO, \Betahat)$.
Then, the proposed estimator under subsampling approximation is
\begin{align*}
    \widehat{\Psi}^{\textup{Sub}}(\cR; \Bw) 
    =
    \frac{1}{K} 
    \sum_{k=1}^{K} 
    \mathbb{P}_m^k 
    \big\{ 
        \varphi^{\textup{Sub}}(\cR; \BO^*, \Betahatk) 
    \big\}
\end{align*}
and the estimand is 
$\Psi(\cR;\Bw)
=
\mathbb{P}\{ 
    \varphi^{\textup{Sub}}(\cR; \BO^*, \Beta) 
\}$,
which follows from the iterated expectation.
Also, the variance estimator of $\widehat{\Psi}^{\textup{Sub}}(\cR; \Bw)$ is given by
$
    \widehat{\sigma}^{\textup{Sub}}(\cR; \Bw)^2
    = 
    K^{-1}
    \sum_{k=1}^{K} 
    \mathbb{P}_m^k 
    \Big[ 
        \big\{ 
            \varphi^{\textup{Sub}}(\cR; \BO^*, \Betahatk)  
            - 
            \widehat{\Psi}^{\textup{Sub}}(\cR; \Bw)
        \big\}^2
    \Big]
$.
The following theorem gives the large sample properties of the proposed estimator under subsampling approximation.

\begin{theorem}
\label{thm:app}
Under the same conditions stated in Theorem 1 in the main text, 
\begin{quote}
\setlength{\leftskip}{-0.5cm}
\textup{(i) \underline{(Consistency)}}
$\widehat{\Psi}^{\textup{Sub}}(\cR; \Bw) \overset{p}{\to} \Psi(\cR;\Bw)$

\textup{(ii) \underline{(Asymptotic Normality)}}
$
\sqrt{m}\{\widehat{\Psi}^{\textup{Sub}}(\cR; \Bw) - \Psi(\cR;\Bw)\}
\overset{d}{\to}
N(0,\sigma^{\textup{Sub}}(\cR; \Bw)^{2})
$,
where 
\begin{align*}
    \sigma^{\textup{Sub}}(\cR; \Bw)^{2}
    =
    \textup{Var}
        \big\{ 
            \varphi^{\textup{Sub}}(\cR; \BO^*, \Beta) 
        \big\}
    =
    \sigma(\cR; \Bw)^2
    +
    \frac{1}{r}
    E\left[
        \textup{Var}\left\{
            \frac
                {\textup{OR}(\cR; \BO, \Ba^{(\cdot)})}
                {f(\mathbf{a}^{(\cdot)})}
            \middle| \mathbf{O}
        \right\}
    \right]
\end{align*}

\textup{(iii) \underline{(Consistent Variance Estimator)}}
$\widehat{\sigma}^{\textup{Sub}}(\cR; \Bw)^2$ is a consistent estimator of the asymptotic variance of $\widehat{\Psi}^{\textup{Sub}}(\cR; \Bw)$.
Therefore,
$
\sqrt{m}\{
    \widehat{\Psi}^{\textup{Sub}}(\cR; \Bw) 
    - 
    \Psi(\cR;\Bw)
\}
/
\widehat{\sigma}^{\textup{Sub}}
\overset{d}{\to}
N(0,1)
$
\end{quote}

\end{theorem}

In conclusion, under mild conditions, 
the proposed estimator under the subsampling approximation is consistent and asymptotically normal, 
with asymptotic variance approaching $\sigma(\cR; \Bw)^2$ as $r \to \infty$.
The decreasing finite sample variance of $\widehat{\Psi}^{\textup{Sub}}(\cR; \Bw)$ is numerically illustrated in Section \ref{simul:r}.
The proof of the above Theorem is as follows.\\

\noindent
(\textit{Proof of Theorem \ref{thm:app}})

We can derive the decomposition of 
$\widehat{\Psi}^{\textup{Sub}}(\cR;\Bw) - \Psi(\cR;\Bw)$
similar to the proof of Theorem 1 (supplementary material Section \ref{proof:thm1}), given by
\begin{align*}
    \widehat{\Psi}^{\textup{Sub}}(\cR;\Bw) - \Psi(\cR;\Bw)
    =
    \frac{1}{K}
    \sum_{k=1}^{K}
    \Big[
        &
        (\bPmk - \bP) \phiPsub
        +
        (\bPmk - \bP) \big\{ 
            \phiPsubhatk - \phiPsub
        \big\}
        \\
        & +
        \bP \big\{ \phiPsubhatk - \phiPsub \big\}
    \Big]
    .
\end{align*}
The first term $(\mathbb{P}_m^k - \mathbb{P}) \phiPsub$ is $O_{P}(m^{-1/2})$ from the central limit theorem, 
and the second term is
\begin{align*}
(\mathbb{P}_m^k - \mathbb{P})
\big\{ \phiPsubhatk - \phiPsub \big\}
=
O_{P} \left( 
    \frac
    {|| \phiPsubhatk - \phiPsub ||}
    {m_k^{1/2} }
\right)
\end{align*}
from Lemma 1 in \citet{kennedy22}.
It remains to analyze 
$||\phiPsubhatk - \phiPsub||$
and
$\mathbb{P} \big\{ \phiPsubhatk - \phiPsub \big\}$.
As previous, omit subscript $(k)$ in $\widehat{\boldsymbol{\eta}}_{(k)}$ for notational convenience.
First, we have
\begin{align*}
    \mathbb{P} \big\{ \phiPsubhat - \phiPsub \big\}
    = &
    \ED{[}{]}{ 
        \E{\{}{\}}{
            \phiPsubhat - \phiPsub |D, \mathbf{O}
        }
    }
    \\
    = &
    \ED{\{}{\}}{ \varphidiff }
    \\
    = & 
    \mathbb{P} \big\{ \varphidiff \big\}
    \\
    = &
    \OP{
        r_{\Bw}^2 
        + 
        r_{\Blambda^T} (r_H + r_{\Bphi} + r_{\BS^C})
    }
\end{align*}
from the iterated expectations, and the last equality follows from the proof of Theorem 1.
Next, we have 
\begin{align*}
    &
    \normp{\phiPsubhat - \phiPsub}^2
    \\
    & =
    \ED{[}{]}{ \{ \phiPsubhat - \phiPsub \}^2 }
    \\
    & \lesssim
    E \big[
        \abs{ \phiPsubhat - \phiPsub }
        \big| D
    \big]
    \\
    & \le
    \ED{[}{]}{ 
        \sumaq 
            \frac
            {|(\ORhat - \OR)(\cR; \BO, \aq)|}
            {f(\aq)}
    }
    +
    \ED{[}{]}{ |(\IPCWBChat -\IPCWBC)(\cR; \BO)| }
    +
    \ED{[}{]}{ |(\AUGhat -\AUG)(\cR; \BO)| }
    \\
    & \lesssim
    \ED{[}{]}{ 
        \sumaN 
            {|(\ORhat - \OR)(\cR; \BO, \aq)|}
    }
    +
    \ED{[}{]}{ |(\IPCWBChat -\IPCWBC)(\cR; \BO)| }
    +
    \ED{[}{]}{ |(\AUGhat -\AUG)(\cR; \BO)| }
    \\
    & = 
    \OP{
        r_{\Bw}^2 + r_{\Blambda^T} + r_H + r_{\Bphi} + r_{\BS^C}
    }
\end{align*}
from the boundedness assumption of quantities,
and the last equality follows from the convergence rate bounds from the proof of Theorem 1.
In conclusion, 
following the similar approach in the proof of Theorem 1,
we have
\begin{align*}
    &
    \widehat{\Psi}^{\textup{Sub}}(\cR;\Bw) - \Psi(\cR;\Bw)
    \\
    &=
    (\bPm - \bP) \phiPsub
    +
    m^{-1/2}
    \OP{
        r_{\Bw}^2 + r_{\Bphi} + r_{\Blambda^T} 
        + r_{H} + r_{\BS^C}
    }
    +
    \OP{
        r_{\Bw}^2 
        + 
        r_{\Blambda^T} (r_H + r_{\Bphi} + r_{\BS^C})
    }
    ,
\end{align*}
and thus the consistency and asymptotic normality of $\widehat{\Psi}^{\textup{Sub}}(\cR;\Bw)$ holds under the same condition for $\widehat{\Psi}(\cR;\Bw)$ in the main text.
Furthermore,
\begin{align*}
    \sigmawsub
    = &
    \textup{Var}
        \big\{ 
            \phiPsub
        \big\}
    \\
    =&
    \V{[}{]}{
        \E{\{}{\}}{
            \phiPsub
            | \mathbf{O}
        }
    }
    +
    \E{[}{]}{ 
        \V{\{}{\}}{ 
            \phiPsub
            | \mathbf{O}
        } 
    }
    \\
    =&
    \V{\{}{\}}{
        \varphi(\cR; \BO, \Beta) 
    }
    \\
    & +
    \E{[}{]}{ 
        \V{\{}{\}}{ 
            \sumaq 
                \frac{\OR(\cR; \BO, \aq)}{f(\aq)}
            +
            \IPCWBC(\cR; \BO)
            +
            \AUG(\cR; \BO)
            \middle| \BO
        } 
    }
    \\
    = &
    \sigma(\cR; \Bw)^2
    +
    \frac{1}{r}
    E\left[
        \textup{Var}\left\{
            \frac
                {\OR(\cR; \BO, \Ba^{(\cdot)})}
                {f(\mathbf{a}^{(\cdot)})}
            \middle| \mathbf{O}
        \right\}
    \right]
    .
\end{align*}

It remains to show the consistency of $\widehat{\sigma}^{\textup{Sub}}(\cR; \Bw)^2$.
First, the variance estimator of $\widehat{\Psi}^{\textup{Sub}}(\cR;\Bw)$
is
\begin{align*}
    \widehat{\sigma}^{\textup{Sub}}(\cR; \Bw)^2 
    = 
    \sumk
    \mathbb{P}_m^k 
    \Big[ 
        \big\{ 
            \varphi^{\textup{Sub}}(\cR; \BO^*, \Betahatk)  
        \big\}^2
    \Big]
    -
    \widehat{\Psi}^{\textup{Sub}}(\cR; \Bw)^2
\end{align*}
and the asymptotic variance of $\widehat{\Psi}^{\textup{Sub}}(\cR;\Bw)$ is
\begin{align*}
    \sigmawsub
    =
    \textup{Var}
        \big\{ 
            \phiPsub
        \big\}
    =
    \bP \Big[ 
        \big\{
            \phiPsub
        \big\}^2
    \Big]
    -
    \Psi(\cR;\Bw)^2
    .
\end{align*}
Since $\widehat{\Psi}^{\textup{Sub}}(\cR;\Bw) \overset{p}{\to} \Psi(\cR;\Bw)$,
it suffices to show
$$
\mathbb{P}_m^k
    \Big[ 
        \big\{
            \varphi^{\textup{Sub}}(\cR; \BO^*, \Betahatk)  
        \big\}^2
    \Big]
\overset{p}{\to}
\mathbb{P}
    \Big[ 
        \big\{
            \phiPsub
        \big\}^2
    \Big]
.
$$
This can be shown exactly the same as in the proof of Theorem 1, using the conditional law of large numbers and Cauchy-Schwarz inequality.
\hfill $\blacksquare$

\subsection{Bounded Estimators}
\label{sec:bounded}

This section describes the construction and large sample properties of the bounded estimator.
Here, assume that the outcome transformation is bounded, such as $\cR(T) = \min\{T, h\}$, where $h$ is a fixed time point, or $\cR(T; \tau) = \indicator(T \le \tau)$.

We first present results for the bounded estimators of the estimand $\mu(\cR; Q)$, with results for other estimands to follow.
Note that the weight function $\Bw$ for $\mu(\cR; Q)$ is $\Bw(\AXN) = N^{-1} [Q(\BA | \XN)]_{j=1}^N$,
and
\begin{align*}
    E \left\{
        \frac
            {\Bw(\AXN)^\top \textbf{1} }
            {H(\AXN)}
    \right\}
    = &
    E \left[
        E \left\{
            \frac
                {\Bw(\AXN)^\top \textbf{1} }
                {H(\AXN)}
            \middle|
            \XN
        \right\}
    \right]
    \\
    = &
    E \Bigg\{
        {\textstyle \sumaN}
            \frac
                {\Bw(\aXN)^\top \textbf{1} }
                {H(\aXN)}
            \times
            \underbrace{\bP(\BA = \Ba | \XN)}_{=H(\aXN)}
    \Bigg\}
    =
    E \left\{
        \textstyle
        \sumaN
            Q(\Ba | \XN)
    \right\}
    =
    1
    .
\end{align*}
This implies that $m_k$ in the proposed estimator can be replaced by 
$\sum_{i: G_i = k} 
    \Bwhatk (\AXNi) ^\top \textbf{1} 
    \allowbreak
    / 
    \allowbreak
    \Hhatk (\AXNi)$.
Informed by this, the bounded NCF estimator is given by
\begin{align*}
        \widehat{\Psi}^{\textup{Bdd}} (\cR; \Bw)
        =&
        \sumk
        \Bigg[
            \sumik
            \Bigg\{
                \suma
                    \ORhatk(\cR; \Oi, \ai) 
            \Bigg\}
            \\
            & \quad \quad \quad \quad 
            +
            \left\{
                \sum_{i: G_i = k} 
                    \frac
                        {\Bwhatk (\AXNi) ^\top \textbf{1} }
                        {\Hhatk (\AXNi)}
            \right\}^{-1}
            \sum_{i: G_i = k}
            \Bigg\{
                \IPCWBChatk(\cR; \Oi)
                +
                \AUGhatk(\cR; \Oi)
            \Bigg\}
        \Bigg]
        \\
        =&
        \sumk
        \sumik
        \Bigg[
            \suma
                \ORhatk(\cR; \Oi, \ai) 
            +
            \IPCWBChatk^{\textup{Bdd}}(\cR; \Oi)
            +
            \AUGhatk^{\textup{Bdd}}(\cR; \Oi)
        \Bigg]
\end{align*}
where
$
\IPCWBChat_{(k)}^{\textup{Bdd}} (\cR; \Oi)
= 
\BV_{(k)}^{-1} \ 
\IPCWBChat_{(k)} (\cR; \Oi)
$,
$
\AUGhat_{(k)}^{\textup{Bdd}} (\cR; \Oi)
= 
\BV_{(k)}^{-1} \ 
\AUGhat_{(k)} (\cR; \Oi)
$,
and                     
$\BV_{(k)} 
\allowbreak
= 
\allowbreak
m_k^{-1}
\allowbreak
\sum_{i: G_i = k} 
    \Bwhatk (\AXNi) ^\top \textbf{1} / \Hhatk (\AXNi)$
is the mean IPW weight estimate in group $k$.

\begin{align*}
    \ORhat(\cR; \BO, \Ba) 
    =&
    \big\{\Bwhat(\Ba) + \Bphihat(\Ba)\big\}^\top
    \widehat{\BG}^{\cR}(0|\Ba), 
    \\
    \IPCWBChat(\cR; \BO)
    =&
    \frac{1}{\Hhat(\BA)}
    \Bwhat(\BA)^\top 
    \left[ 
        \frac{\Delta_j  \cR(Y_j)}{\SCjhat(Y_j)}
        - 
        \widehat{G}_j^{\cR}(0)
    \right]_{j=1}^N, 
    \\
    \AUGhat(\cR; \BO)
    =&
    \frac{1}{\Hhat(\BA)}
    \Bwhat(\BA)^\top 
    \left[ 
        \int_{0}^{\infty}
            \frac{\widehat{G}_j^{\cR}(r)}{\SCjhat(r) \STjhat(r)}
        d\MCjhat
    \right]_{j=1}^N.
\end{align*}

Then, the boundedness of the proposed bounded NCF estimator can be shown as follows.
First, note that
$
    \textstyle
    \sumaN \Bw(\mathbf{a}, \mathbf{X}, N)
    =
    \sumaN N^{-1} [Q(\Ba | \XN)]_{j=1}^{N} 
    =
    N^{-1} \left[ \sumaN Q(\Ba | \XN) \right]_{j=1}^{N} 
    =
    N^{-1} \mathbf{1} 
    .
$
Thus, the estimator of the weight function $\Bwhatk$ should satisfy this condition, ensuring that
$\suma \Bwhatk(\ai) = N_i^{-1} \mathbf{1}$.
Next, 
since 
$\sum_{\Ba \in \cAn} \allowbreak
\Bw(\mathbf{a}, \mathbf{x}, n) 
= 
n^{-1} \mathbf{1}$ 
is constant,
it follows that its CIF $\sum_{\Ba \in \cAn} \Bphi(\AXN; \Ba)$ is Therefore, the estimator $\Bphihatk$ should also satisfy this condition, ensuring that
$\suma \Bphihatk(\ai) = \mathbf{0}$.
From these, the outcome regression term
\begin{align*}
    \textstyle
    \suma 
        \ORhatk(\cR; \Oi, \ai)
    =&
    \textstyle
    \suma 
        \big\{ \Bwhatk(\ai) + \Bphihatk(\ai) \big\}^\top
        \widehat{\BG}_i^{\cR}(0|\ai)
    \\
    =&
    \textstyle
    \sumjline
    \textstyle
    \suma
        \big\{ \widehat{w}_{ij,(k)}(\ai) + \widehat{\phi}_{ij,(k)}(\ai) \big\}
        \widehat{G}_{ij,(k)}^{\cR}(0|\ai)
\end{align*}
is the weighted average of $\widehat{G}_{ij,(k)}^{\cR}(0|\ai)$ over $j$ and $\ai \in \cANi$,
and thus bounded.

Secondly, consider 
$\IPCWBChat_{(k)}^{\textup{Bdd}} (\cR; \Oi)
= 
\BV_{(k)}^{-1} \ 
\IPCWBChat_{(k)} (\cR; \Oi)$.
Note that
\begin{align*}
    &
    \sumik \Bigg\{
        \IPCWBChatk^{\textup{Bdd}}(\cR; \Oi)
    \Bigg\}
    \\
    =&
    \left\{
        \sum_{i: G_i = k} 
            \frac
                {\Bwhatk (\Ai) ^\top \textbf{1} }
                {\Hhatk (\Ai)}
    \right\}^{-1}
    \sum_{i: G_i = k} \Bigg\{
        \IPCWBChatk(\cR; \Oi)
    \Bigg\}
    \\
    =&
    \left\{
        \sum_{i: G_i = k} 
            \frac
                {\Bwhatk (\Ai) ^\top \textbf{1} }
                {\Hhatk (\Ai)}
    \right\}^{-1}
    \sum_{i: G_i = k} \Bigg\{
        \frac{1}{\Hhatk(\Ai)}
        \Bwhatk(\Ai)^\top 
        \left[ 
            \frac
                {\Delta_{ij}  \cR(Y_{ij})}
                {\widehat{S}_{ij,(k)}^C(Y_{ij})}
            - 
            \widehat{G}_{ij,(k)}^{\cR}(0|\ai)
        \right]_{j=1}^N
    \Bigg\}
\end{align*}
and thus this term is bounded by 
\begin{align*}
    \max_{i: G_i = k, 1 \le j \le N_i}
    \left\{
        \left| 
            \frac
                {\Delta_{ij}  \cR(Y_{ij})}
                {\widehat{S}_{ij,(k)}^C(Y_{ij})}
            - 
            \widehat{G}_{ij,(k)}^{\cR}(0|\ai)
        \right|
    \right\}
    \le
    \max_{i: G_i = k, 1 \le j \le N_i} |\cR(\Yij)|
    (1+d^{-1})
\end{align*}
where $\widehat{S}_{ij,(k)}^C(r) > d$ for some $d \in (0,1)$ from the Assumption (L1).
Similar logic applies to the augmentation term AUG.
Therefore, the proposed bounded NCF estimator is bounded.

The following theorem gives the large sample properties of the proposed estimator under bounding modification.

\begin{theorem}
\label{thm:bounding}
Under the same conditions stated in Theorem 1 in the main text, the proposed bounded estimator has the same large sample properties as the original (unbounded) estimator, i.e.,
\begin{quote}
\setlength{\leftskip}{-0.5cm}
\textup{(i) \underline{(Consistency)}}
$\widehat{\Psi}^{\textup{Bdd}}(\cR; \Bw) \overset{p}{\to} \Psi(\cR;\Bw)$

\textup{(ii) \underline{(Asymptotic Normality)}}
$
\sqrt{m}\{\widehat{\Psi}^{\textup{Bdd}}(\cR; \Bw) - \Psi(\cR;\Bw)\}
\overset{d}{\to}
N(0,\sigma(\cR; \Bw)^{2})
$

\textup{(iii) \underline{(Consistent Variance Estimator)}}
$\widehat{\sigma}^{\textup{Bdd}}(\cR; \Bw)^2$ is a consistent estimator of the asymptotic variance of $\widehat{\Psi}^{\textup{Bdd}}(\cR; \Bw)$.
Therefore,
$
\sqrt{m}\{
    \widehat{\Psi}^{\textup{Bdd}}(\cR; \Bw) 
    - 
    \Psi(\cR;\Bw)
\}
/
\widehat{\sigma}^{\textup{Bdd}}
\overset{d}{\to}
N(0,1)
$
\end{quote}

\end{theorem}

The finite sample performance of $\widehat{\Psi}^{\textup{Bdd}}(\cR; \Bw)$ is numerically illustrated in Section \ref{simul:bounded},
such that bounding improves finite sample performance without affecting asymptotic properties.

The proof of the above Theorem is as follows.\\

\noindent
(\textit{Proof of Theorem \ref{thm:bounding}})

We first present results for the bounded estimators of the estimand $\mu(\cR; Q)$, with results for other estimands to follow.
Begin with the following decomposition:
\begin{align*}
        &
        \widehat{\Psi}^{\textup{Bdd}} (\cR; \Bw) - \widehat{\Psi}(\cR;\Bw) 
        \\
        =&
        \sumk
        \Bigg(
            \Bigg[
                \left\{
                    \sum_{i: G_i = k} 
                        \frac
                            {\Bwhatk (\AXNi) ^\top \textbf{1} }
                            {\Hhatk (\AXNi)}
                \right\}^{-1}
                -
                \frac{1}{m_k}
            \Bigg]
            \sum_{i: G_i = k}
            \Bigg\{
                \IPCWBChatk(\cR; \Oi)
                +
                \AUGhatk(\cR; \Oi)
            \Bigg\}
        \Bigg)
        \\
        =&
        \sumk
            \Bigg(
                \BV_{(k)}^{-1}
                -
                1
            \Bigg)
        \Bigg[
            \sumik
            \Bigg\{
                \IPCWBChatk(\cR; \Oi)
                +
                \AUGhatk(\cR; \Oi)
            \Bigg\}
        \Bigg]
\end{align*}
Now, consider the asymptotics of $\BV_{(k)}$ as follows:
\begin{align*}
    \BV_{(k)}
    = &
    \sumik
        \frac
            {\Bwhatk (\AXNi) ^\top \textbf{1} }
            {\Hhatk (\AXNi)}
    =
    \bPmk \left\{
        \frac
            {\Bwhatk (\AXN) ^\top \textbf{1} }
            {\Hhatk (\AXN)}
    \right\}
    \\
    = &
    \underbrace{
        (\bPmk - \bP) \left\{
            \frac
                {\Bw (\AXN) ^\top \textbf{1} }
                {H (\AXN)}
        \right\}    
    }_{\OP{m_k^{-1/2}} \ (\because) \textup{ CLT}}
    +
    \underbrace{
        (\bPmk - \bP) \left\{
            \frac
                {\Bwhatk (\AXN) ^\top \textbf{1} }
                {\Hhatk (\AXN)}
            -
            \frac
                {\Bw (\AXN) ^\top \textbf{1} }
                {H (\AXN)}
        \right\}
    }_{\OP{m_k^{-1/2}} \ (\because) \textup{ Lemma 1 in \citet{kennedy22}}}
    +
    \underbrace{
        \bP \left\{
            \frac
                {\Bwhatk (\AXN) ^\top \textbf{1} }
                {\Hhatk (\AXN)}
        \right\}
    }_{1 + \OP{r_H}}
\end{align*}
where the first term in the last equation is $\OP{m_k^{-1/2}}$ from the central limit theorem, 
the second term is also $\OP{m_k^{-1/2}}$ from Lemma 1 in \citet{kennedy22} and boundedness of nuisance functions,
and the third term is bounded as follows:
\begin{align*}
    \bP \left\{
        \frac
            {\Bwhatk (\AXN) ^\top \textbf{1} }
            {\Hhatk (\AXN)}
    \right\}
    - 
    1
    =&
    \bP \left\{
        \sumaN
            \frac
                {\Bwhatk (\aXN) ^\top \textbf{1} }
                {\Hhatk (\aXN)}
            H(\aXN)
        -
        \sumaN
            \Bwhatk (\aXN) ^\top \textbf{1}
    \right\}
    \\
    =&
    \bP \left\{
        \sumaN
            \frac
                {1}
                {\Hhatk (\aXN)}
            \left\{ H(\aXN) - \Hhatk (\aXN) \right\}
            \Bwhatk (\aXN) ^\top \textbf{1} 
    \right\}
    \\
    =&
    o_{P} \left(
        \normp{
            \big(\Hhatk-H\big)(\AXN)
        }
    \right)
    =
    O_{P}(r_H)
\end{align*}
where the last equality follows from the boundedness of nuisance function estimators and H\"older's inequality.
Therefore, we have
$\BV_{(k)}^{-1} - 1
=
O_{P}(r_H)$.
It remains to analyze 
\begin{align*}
    \sumik
        \Bigg\{
            \IPCWBChatk(\cR; \Oi)
            +
            \AUGhatk(\cR; \Oi)
        \Bigg\}
    =
    \bPmk \left\{
        \IPCWBChatk(\cR; \BO)
        +
        \AUGhatk(\cR; \BO)
    \right\}
    .
\end{align*}
Similar to the proof of Theorem 1 (supplementary material Section \ref{proof:thm1}), we have
\begin{align*}
    \bPmk \left\{
        \IPCWBChatk(\cR; \BO)
        +
        \AUGhatk(\cR; \BO)
    \right\}
    =&
    \underbrace{
        (\bPmk - \bP) \left\{
            \IPCWBC(\cR; \BO)
            +
            \AUG(\cR; \BO)
        \right\}
    }_{\OP{m_k^{-1/2}} \ (\because) \textup{ CLT}}
    \\
    &+
    \underbrace{
        (\bPmk - \bP) 
        \left\{
            (\IPCWBChatk
            +
            \AUGhatk
            -
            \IPCWBC
            -
            \AUG)
            (\cR; \BO)
        \right\}
    }_{\OP{m_k^{-1/2}} \ (\because) \textup{ Lemma 1 in \citet{kennedy22}}}
    \\
    &+
    \underbrace{
        \bP
        \left\{
            \IPCWBChatk(\cR; \BO)
            +
            \AUGhatk(\cR; \BO)
        \right\}
    }_{\OP{\rT}}
\end{align*}
where
the first term is $\OP{m_k^{-1/2}}$ from the central limit theorem,
the second term is also $\OP{m_k^{-1/2}}$ from Lemma 1 in \citet{kennedy22} and boundedness of nuisance functions,
and the third term is $\OP{\rT}$ from the proof of Theorem 1, i.e.,
\begin{align*}
    \bP
    \left\{
        \IPCWBChatk(\cR; \BO)
        +
        \AUGhatk(\cR; \BO)
    \right\}
    =
    &
    \underbrace{
        \ED{[}{]}{
            \frac{1}{\Hhat(\BA)}
            \Bwhat(\BA)^\top 
            \left[ 
                \frac
                    {S_j^C(T_j)}
                    {\SCjhat(T_j)}
                \cR(T_j)
            \right]_{j=1}^N
        }
    }_{(\ref{pf:thm1:SCI})}
    -
    \underbrace{
        \ED{[}{]}{
            \frac{1}{\Hhat(\BA)}
            \Bwhat(\BA)^\top 
            \widehat{\BG}^{\cR}(0)
        }
    }_{(\ref{pf:thm1:F})}
    \\
    & \quad +
    \underbrace{
        \ED{[}{]}{
            \frac{1}{\Hhat(\BA)}
            \Bwhat(\BA)^\top 
            \left[ 
                \int_{0}^{\infty}
                    \frac
                        {\widehat{G}_j^{\cR}(r)}
                        {\SCjhat(r) \STjhat(r)}
                d\MCjhat
            \right]_{j=1}^N
        }
    }_{(\ref{pf:thm1:stieltjes})}
    \\
    = &
    -
    \underbrace{
        \ED{[}{]}{
            \frac{1}{\Hhat(\BA)}
            \Bwhat(\BA)^\top 
            \left[
                \STjhat(\cR)
                \int_{0}^{\infty}
                    \frac
                        {S_j^T}
                        {\STjhat}
                    (r)
                    (\lambdaTjhat - \lambda^T_j) (r)
                    \frac
                        {S_j^C}
                        {\SCjhat}
                    (r)
                dr
            \right]_{j=1}^N
        }
    }_{(\ref{pf:thm1:lambdadiff})}
    \\
    = &
    \OP{\rT}
\end{align*}
where the last equality follows from the boundedness of nuisance function estimators and H\"older's inequality.
Therefore,
\begin{align*}
        \widehat{\Psi}^{\textup{Bdd}} (\cR; \Bw) - \widehat{\Psi}(\cR;\Bw) 
        =
        \sumk
            \big(
                \BV_{(k)}^{-1}
                -
                1
            \big)
        \bPmk \left\{
            \IPCWBChatk(\cR; \BO)
            +
            \AUGhatk(\cR; \BO)
        \right\}
        = \OP{\rH \rT}
        .
\end{align*}
From
\begin{align*}
    &
    \widehat{\Psi}^{\textup{Bdd}} (\cR; \Bw) - \Psi(\cR;\Bw)
    =
    \underbrace{
        \{ \widehat{\Psi}^{\textup{Bdd}} (\cR; \Bw) - \widehat{\Psi}(\cR;\Bw) \}
    }_{\OP{\rH \rT}}
    +
    \{ \widehat{\Psi}(\cR;\Bw) - \Psi(\cR;\Bw) \}
\end{align*}
and
\begin{align*}
    &
    \widehat{\Psi}(\cR;\Bw) - \Psi(\cR;\Bw)
    \\
    &=
    (\bPm - \bP)
            \varphi(\cR; \BO, \Beta) 
    +
    m^{-1/2}
    \OP{
        r_{\Bw}^2 + r_{\Bphi} + r_{\Blambda^T} 
        + r_{H} + r_{\BS^C}
    }
    +
    O_{P}(
        r_{\Bw}^2 
        + 
        r_{\Blambda^T}r_H 
        +
        r_{\Blambda^T}r_{\Bphi} 
        + 
        r_{\Blambda^T}r_{\BS^C}
    )
    ,
\end{align*}
we have
\begin{align*}
    &
    \widehat{\Psi}^{\textup{Bdd}} (\cR; \Bw) - \Psi(\cR;\Bw)
    \\
    &=
    (\bPm - \bP)
            \varphi(\cR; \BO, \Beta) 
    +
    m^{-1/2}
    \OP{
        r_{\Bw}^2 + r_{\Bphi} + r_{\Blambda^T} 
        + r_{H} + r_{\BS^C}
    }
    +
    \OP{
        r_{\Bw}^2 
        + 
        r_{\Blambda^T} r_H 
        + 
        r_{\Blambda^T} r_{\Bphi} 
        + 
        r_{\Blambda^T} r_{\BS^C}
    }
    ,
\end{align*}
which implies that $\widehat{\Psi}^{\textup{Bdd}} (\cR; \Bw)$ has the same large sample properties as $\widehat{\Psi}(\cR;\Bw)$.

Now, it remains to prove that the variance estimator under the bounded estimator is consistent. 
The variance estimator of $\widehat{\Psi}^{\textup{Bdd}} (\cR; \Bw)$ is given as follows:
\begin{align*}
    \widehat{\sigma}^{\textup{Bdd}}(\cR; \Bw)^2
    = 
    \sumk
    \bPmk
    \Big[ 
        \big\{
            \varphi^{\textup{Bdd}}(\cR; \mathbf{O}, \Betahatk) 
        \big\}^2
    \Big]
    - 
    \widehat{\Psi}^{\textup{Bdd}} (\cR; \Bw)^2
    .
\end{align*}
Note that the variance estimator of the original (unbounded) estimator is 
$$
    \widehat{\sigma}(\cR;\Bw)^2
    = 
    \sumk
    \bPmk
    \Big[ 
        \big\{ \allowbreak
            \varphi(\cR; \mathbf{O}, \Betahatk) 
        \big\}^2
    \Big]
    - 
    \widehat{\Psi}(\cR;\Bw)^2
$$
which is a consistent estimator of 
$$
    \sigma(\cR; \Bw)^2 
    = 
    \mathbb{P} \allowbreak
    \Big[ 
        \big\{
            \varphi(\cR; \mathbf{O}, \Beta) 
        \big\}^2
    \Big]
    - 
    \Psi(\cR;\Bw)^2
$$
under the given conditions.
Since
$\widehat{\Psi}^{\textup{Bdd}} (\cR; \Bw) \overset{p}{\to} \Psi(\cR;\Bw)$
under the same condition of
$\widehat{\Psi}(\cR;\Bw) \overset{p}{\to} \Psi(\cR;\Bw)$,
it suffices to consider
\begin{align*}
    \bPmk
    \Big[ 
        \big\{
            \varphi^{\textup{Bdd}}(\cR; \mathbf{O}, \Betahatk) 
        \big\}^2
    \Big]
    -
    \mathbb{P}_m^k \allowbreak
    \Big[ 
        \big\{ \allowbreak
            \varphi(\cR; \mathbf{O}, \Betahatk) \allowbreak
        \big\}^2
    \Big]
    .
\end{align*}
Indeed,
\begin{align*}
    &
    \bPmk
    \Big[ 
        \big\{
            \varphi^{\textup{Bdd}}(\cR; \BO, \Betahatk) 
        \big\}^2
    \Big]
    -
    \mathbb{P}_m^k
    \Big[ 
        \big\{ 
            \varphi(\cR; \BO, \Betahatk)
        \big\}^2
    \Big]
    \\
    =&
    \frac{1}{m_k}
    \sum_{i:G_i = k}
        \left\{ 
            \varphi^{\textup{Bdd}}(\cR; \Oi, \Betahatk) 
            -
            \varphi(\cR; \Oi, \Betahatk)
        \right\}
        \left\{ 
            \varphi^{\textup{Bdd}}(\cR; \Oi, \Betahatk) 
            +
            \varphi(\cR; \Oi, \Betahatk)
        \right\}
    \\
    =&
    \underbrace{
        \left(
            \BV_{(k)}^{-1}
            -
            1
        \right)
    }_{\OP{r_H}}
    \underbrace{
        \frac{1}{m_k}
        \sum_{i:G_i = k}
            \left\{
                \IPCWBChatk(\cR; \Oi)
                +
                \AUGhatk(\cR; \Oi)
            \right\}
            \left\{ 
                \varphi^{\textup{Bdd}}(\cR; \Oi, \Betahatk) 
                +
                \varphi(\cR; \Oi, \Betahatk)
            \right\}
    }_{\OP{1}}
    \\
    =&
    \OP{r_H}
    ,
\end{align*}
and thus
\begin{align*}
    \widehat{\sigma}^{\textup{Bdd}}(\cR; \Bw)^2
    -
    \widehat{\sigma}(\cR;\Bw)^2
    =
    \OP{r_H}
    .
\end{align*}
Combined with this, 
from the proof of Theorem 1, 
\begin{align*}
    \widehat{\sigma}(\cR;\Bw)^2
    -
    \sigma(\cR; \Bw)^2
    =
    \OP{
        r_{\Bw} + r_{\Bphi} + r_{\Blambda^T} 
        + 
        r_H 
        + r_{\BS^C}
    }
\end{align*}
and thus
\begin{align*}
    \widehat{\sigma}^{\textup{Bdd}}(\cR; \Bw)^2
    -
    \sigma(\cR; \Bw)^2
    =
    \OP{
        r_{\Bw} + r_{\Bphi} + r_{\Blambda^T} 
        + r_{H} + r_{\BS^C}
    }.
\end{align*}
In conclusion, 
$\widehat{\sigma}^{\textup{Bdd}}(\cR; \Bw)^2$ 
is a consistent estimator of the asymptotic variance of 
$\widehat{\Psi}^{\textup{Bdd}} (\cR; \Bw)$, 
i.e., $\sigma(\cR; \Bw)^2$, 
under the same conditions $\widehat{\sigma}(\cR;\Bw)^2$ is a consistent estimator of $\sigma(\cR; \Bw)^2$.

The above results are about the estimand 
$\mu(\cR; Q)$, and the same logic applies to $\mu_a(\cR; Q), a \in \{0,1\}$.
The weight function $\Bw$ for $\mu_a(\cR; Q)$ is 
$\Bw(\AXN) = N^{-1} [\indicator(A_{j} = a) Q(\BA_{(-j)}| \XN)]_{j=1}^N$,
and thus
\begin{align*}
    E \left\{
        \frac
            {\Bw(\AXN)^\top \textbf{1} }
            {H(\AXN)}
    \right\}
    = &
    E \left\{
        \sumaN
            \Bw(\aXN)^\top \textbf{1}
    \right\}
    =
    E \left\{
        \sumaN
        \frac{1}{N}
        \sum_{j=1}^N 
            \indicator(a_{j} = a) Q(\Ba_{(-j)}| \XN)
    \right\}
    \\
    = &
    E \left\{
        \frac{1}{N}
        \sum_{j=1}^N 
        \sum_{\Ba_{(-j)} \in \mathcal{A}(N-1)}
            Q(\Ba_{(-j)}| \XN)
    \right\}
    =
    E \left\{
        \frac{1}{N}
        \sum_{j=1}^N 
        1
    \right\}
    =
    1
    ,
\end{align*}
and also 
\begin{align*}
    \sumaN 
        \Bw(\mathbf{a}, \mathbf{X}, N)
    =
    \sumaN 
        \frac{1}{N} [\indicator(a_{j} = a) Q(\Ba_{(-j)}| \XN)]_{j=1}^N
    =
    \frac{1}{N} 
    \left[         
        \sum_{\Ba_{(-j)} \in \mathcal{A}(N-1)}
            Q(\Ba_{(-j)}| \XN)
    \right]_{j=1}^{N} 
    =
    \frac{1}{N} \mathbf{1} 
    .
\end{align*}
Therefore, the bounded estimator for $\mu_a(\cR; Q)$ can be constructed similarly to the case of $\mu(\cR; Q)$, and the large sample properties can be shown similarly.

For the causal effects, such as $DE(\cR; Q) = \mu_1(\cR; Q) - \mu_0(\cR; Q)$, the bounded estimator is constructed differently, since the weight function $\Bw$ for these causal effects does not satisfy the condition 
$E \left\{
        \Bw(\AXN)^\top \textbf{1} / H(\AXN)
\right\}
\allowbreak
=
1$.
Rather, the bounded estimator for $DE(\cR; Q)$, for example, is given as follows:
\begin{align*}
    \widehat{DE}^{\textup{Bdd}} (\cR; Q) 
    =
    \widehat{\mu}_1^{\textup{Bdd}} (\cR; Q) 
    - 
    \widehat{\mu}_0^{\textup{Bdd}} (\cR; Q) 
\end{align*}
and the variance estimator is given by
\begin{align*}
    \widehat{\sigma}_{DE}^{\textup{Bdd}}(\cR; Q)^2
    = 
    \sumk
    \bPmk
    \Big[ 
        \big\{
            \varphi_{DE}^{\textup{Bdd}}(\cR; \mathbf{O}, \Betahatk) 
            -
            \widehat{DE}^{\textup{Bdd}} (\cR; Q)
        \big\}^2
    \Big]
    =
    \sumk
    \bPmk
    \Big[ 
        \big\{
            \varphi_{DE}^{\textup{Bdd}}(\cR; \mathbf{O}, \Betahatk) 
        \big\}^2
    \Big]
    - 
    \widehat{DE}^{\textup{Bdd}} (\cR; Q)^2
\end{align*}
where
\begin{align*}
    \varphi_{DE}^{\textup{Bdd}}(\cR; \Oi, \Betahatk) 
    =
    \varphi_{\mu_1}^{\textup{Bdd}}(\cR; \Oi, \Betahatk) 
    -
    \varphi_{\mu_0}^{\textup{Bdd}}(\cR; \Oi, \Betahatk) 
\end{align*}
and
$\varphi_{\mu_1}^{\textup{Bdd}}$ and $\varphi_{\mu_0}^{\textup{Bdd}}$ are 
the proposed estimating equations with bounding modifications for 
$\mu_1(\cR; Q)$ and $\mu_0(\cR; Q)$,
respectively.

The boundedness of $\widehat{DE}^{\textup{Bdd}} (\cR; Q)$ is given from the boundedness of $\widehat{\mu}_1^{\textup{Bdd}} (\cR; Q)$ and
$\widehat{\mu}_0^{\textup{Bdd}} (\cR; Q)$.
The large sample properties of $\widehat{DE}^{\textup{Bdd}} (\cR; Q)$ are the same as those of $\widehat{DE} (\cR; Q)$,
which can be derived from the following decomposition:
\begin{align*}
    \widehat{DE}^{\textup{Bdd}} (\cR; Q) 
    -
    \widehat{DE} (\cR; Q) 
    =
    \underbrace{
        \big\{
            \widehat{\mu}_1^{\textup{Bdd}} (\cR; Q) 
            - 
            \widehat{\mu}_1 (\cR; Q) 
        \big\}
    }_{\OP{\rH \rT}}
    -
    \underbrace{
        \left\{
            \widehat{\mu}_0^{\textup{Bdd}} (\cR; Q) 
            -
            \widehat{\mu}_0 (\cR; Q) 
        \right\}
    }_{\OP{\rH \rT}}
    =
    \OP{\rH \rT}
    .
\end{align*}
Finally, the consistency of the variance estimator is given by
\begin{align*}
    &
    \bPmk
    \Big[ 
        \big\{
            \varphi_{DE}^{\textup{Bdd}}(\cR; \BO, \Betahatk) 
        \big\}^2
    \Big]
    -
    \mathbb{P}_m^k
    \Big[ 
        \big\{ 
            \varphi_{DE}(\cR; \BO, \Betahatk)
        \big\}^2
    \Big]
    \\
    =&
    \frac{1}{m_k}
    \sum_{i:G_i = k}
        \underbrace{
            \left\{ 
                \varphi_{DE}^{\textup{Bdd}}(\cR; \Oi, \Betahatk) 
                -
                \varphi_{DE}(\cR; \Oi, \Betahatk)
            \right\}        
        }_{\OP{r_H}}
        \left\{ 
            \varphi_{DE}^{\textup{Bdd}}(\cR; \Oi, \Betahatk) 
            +
            \varphi_{DE}(\cR; \Oi, \Betahatk)
        \right\}
    \\
    =&
    \OP{r_H}
    ,
\end{align*}
using the fact that
\begin{align*}
    &
    \varphi_{DE}^{\textup{Bdd}}(\cR; \Oi, \Betahatk) 
    -
    \varphi_{DE}(\cR; \Oi, \Betahatk)
    \\
    =&
        \left\{ 
            \varphi_{\mu_1}^{\textup{Bdd}}(\cR; \Oi, \Betahatk) 
            -
            \varphi_{\mu_1}(\cR; \Oi, \Betahatk) 
        \right\}
    -
        \left\{ 
            \varphi_{\mu_0}^{\textup{Bdd}}(\cR; \Oi, \Betahatk) 
            -
            \varphi_{\mu_0}(\cR; \Oi, \Betahatk) 
        \right\}
    \\
    =&
    \OP{\BV_{\mu_1, (k)}^{-1}-1}
    -
    \OP{\BV_{\mu_0, (k)}^{-1}-1}
    =
    \OP{r_H}
\end{align*}
where
\begin{align*}
    \BV_{\mu_1, (k)}
    = &
    \sumik
        \frac
            {\widehat{\Bw}_{\mu_1, (k)} (\AXNi) ^\top \textbf{1} }
            {\Hhatk (\AXNi)}
    =
    1 + \OP{r_H}
\end{align*}
is the mean IPW weight estimate in group $k$ 
and
$\widehat{\Bw}_{\mu_1, (k)}$ is the estimated weight function in group $k$ for the estimand $\mu_1(\cR;Q)$,
respectively,
and $\BV_{\mu_0, (k)}$ is defined similarly.

The same logic applies to the bounded estimator for the other causal effects.

\hfill $\blacksquare$ \\

\newpage

\section{Details on example policies}

This section presents details on the large sample properties of the proposed estimators under the example policies.
Assume the CIF of 
$Q(\mathbf{a} | \mathbf{x}, n)$
is
$\phi_Q(\mathbf{A}_i, \mathbf{X}_i, N_i; \mathbf{a})$
for fixed 
$(\mathbf{a}, \mathbf{x}, n) 
\in 
\mathcal{A}(n) \times \mathcal{X}(n) \times \mathbb{N}$.
For notational convenience, the cluster index $i$ is omitted sometimes, while $j$ indicates the unit index.
Let
$\Bw_o(\aXN) 
= 
N^{-1} \big[ Q(\Ba|\XN) \big]_{j=1}^{N}$
and 
$\Bw_a(\aXN)
= 
N^{-1} \big[ \indicator(a_{j} = a) Q(\Ba_{(-j)}| \XN) \big]_{j=1}^{N}, 
\allowbreak a \in \{0,1\}$
denote the weight function $\Bw$ for $\mu(\cR; Q)$ and $\mu_a(\cR; Q)$, respectively.
Similarly, let
$\Bphi_o(\mathbf{A}, \mathbf{X}, N; \mathbf{a})$ and
$\Bphi_a(\mathbf{A}, \mathbf{X}, N; \mathbf{a})$
denote the CIF of
$\Bw_o(\mathbf{a}, \mathbf{x}, n)$ and
$\Bw_a(\mathbf{a}, \mathbf{x}, n)$, respectively,
which are given by
$\Bphi_o(\mathbf{A}, \mathbf{X}, N; \mathbf{a})
= 
N^{-1} \big[ \phi_{Q}(\mathbf{A}, \mathbf{X}, N; \mathbf{a}) \big]_{j=1}^{N}$
and 
$\Bphi_a(\mathbf{A}, \mathbf{X}, N; \mathbf{a})
= 
N^{-1} \big[ \indicator(a_{j} = a) \phi_{Q}(\mathbf{A}, \mathbf{X}, N; \mathbf{a}_{(-j)}) \big]_{j=1}^{N}, 
\allowbreak a \in \{0,1\}$,
where
$\phi_{Q}(\mathbf{A}, \mathbf{X}, N; \mathbf{a}_{(-j)})
=
\allowbreak
\phi_{Q}(\mathbf{A}, \mathbf{X}, N; (1,\mathbf{a}_{(-j)})) 
+
\allowbreak
\phi_{Q}(\mathbf{A}, \mathbf{X}, N; (0,\mathbf{a}_{(-j)}))$
is the CIF of $Q(\mathbf{a}_{(-j)} | \mathbf{x}, n)$.

Under this specification, the proposed estimating functions of $\mu(\cR;Q)$ and $\mu_a(\cR;Q)$, 
namely $\varphi_{\mu(\cR;Q)}(\cR; \BO, \Beta)$ and $\varphi_{\mu_a(\cR;Q)}(\cR; \BO, \Beta)$, 
are given by
\begin{align*}
    \varphi_{\mu(\cR;Q)}(\cR; \BO, \Beta)
    = &
    \frac{1}{N}
    \sum_{j=1}^{N}
    \begin{aligned}[t]
    \Bigg[  
        &
        \sum_{\mathbf{a} \in \mathcal{A}(N)}
            \big\{ 
                    Q(\mathbf{a} | \mathbf{X}, N) 
                    + 
                    \phi_Q(\mathbf{A}, \mathbf{X}, N;\mathbf{a}) 
                \big\}
            G_{j}^{\cR}(0 | \aXN )
        \\
        & + 
        \frac
            {Q(\mathbf{A} | \mathbf{X}, N) }
            {H(\mathbf{A},\mathbf{X}, N)}
        \left\{ 
            \frac
                {\Delta_j \cR(Y_j)}
                {S^C_j(Y_j|\AXN)}
            - 
            G_{j}^{\cR}(0 | \aXN )
        \right\}
        \\
        & +
        \frac
            {Q(\mathbf{A} | \mathbf{X}, N) }
            {H(\mathbf{A},\mathbf{X}, N)}
        \int_{0}^{\infty}
            \frac
                {G_{j}^{\cR}(r | \aXN )}
                {S^C_j( r | \AXN ) S^T_j( r | \AXN )}
        dM_j^C(r)
    \Bigg]
    \end{aligned}
    \\
    \varphi_{\mu_a(\cR;Q)}(\cR; \BO, \Beta)
    = &
    \frac{1}{N}
    \sum_{j=1}^{N}
    \begin{aligned}[t]
    \Bigg[  
        &
        \sum_{\mathbf{a} \in \mathcal{A}(N)}
            \mathbbm{1}(a_{j} = a) 
            \big\{ 
                    Q(\mathbf{a}_{(-j)} | \mathbf{X}, N) 
                    + 
                    \phi_Q(\mathbf{A}, \mathbf{X}, N;\mathbf{a}_{(-j)}) 
                \big\}
            G_{j}^{\cR}(0 | \aXN )
        \\
        & + 
        \frac
            {\mathbbm{1}(A_{j} = a) 
            Q(\mathbf{A}_{(-j)} | \mathbf{X}, N) }
            {H(\mathbf{A},\mathbf{X}, N)}
        \left\{ 
            \frac
                {\Delta_j \cR(Y_j)}
                {S^C_j(Y_j|\AXN)}
            - 
            G_{j}^{\cR}(0 | \aXN )
        \right\}
        \\
        & + 
        \frac
            {\mathbbm{1}(A_{j} = a) 
            Q(\mathbf{A}_{(-j)} | \mathbf{X}, N) }
            {H(\mathbf{A},\mathbf{X}, N)}
        \int_{0}^{\infty}
            \frac
                {G_{j}^{\cR}(r | \aXN )}
                {S^C_j( r | \AXN ) S^T_j( r | \AXN )}
        dM_j^C(r)
    \Bigg]
    .
    \end{aligned}
\end{align*}
Based on these, the inference procedure for $\mu(\cR;Q)$, $\mu_a(\cR;Q)$, and causal effects can be described as follows:
\begin{enumerate}[leftmargin = 50pt, itemsep = 0pt, parsep = 0pt, label=\textbf{Step \arabic*}]

    \item Determine the policy distribution $Q$ of interest
    
    \item Compute $\phi_Q$
    
    \item Construct $\Bw_o(\mathbf{a}, \mathbf{x}, n)$ and $\Bw_a(\mathbf{a}, \mathbf{x}, n)$ for $a \in \{0,1\}$ from $Q$
    
    \item Construct $\Bphi_o(\mathbf{A}, \mathbf{X}, N; \mathbf{a})$ and $\Bphi_a(\mathbf{A}, \mathbf{X}, N; \mathbf{a})$ from $\phi_Q$

    \item Obtain the proposed estimating function of $\mu(\cR;Q)$ and $\mu_a(\cR;Q)$, namely, $\varphi_{\mu(\cR;Q)}(\cR; \mathbf{O}, \boldsymbol{\eta})$ and $\varphi_{\mu_a(\cR;Q)}(\cR;\mathbf{O}, \boldsymbol{\eta})$
    
    \item Compute the proposed estimator
    $\widehat{\mu}(\cR;Q) 
    =
    K^{-1} 
    \sum_{k=1}^{K} 
    \mathbb{P}_m^k 
    \big\{ 
        \varphi_{\mu(\cR;Q)}(\cR;\mathbf{O}, \widehat{\boldsymbol{\eta}}_{(k)}) 
    \big\}$
    and
    $\widehat{\mu}_a(\cR;Q) 
    =
    K^{-1} 
    \sum_{k=1}^{K} 
    \allowbreak
    \mathbb{P}_m^k 
    \big\{ 
        \varphi_{\mu_a(\cR;Q)}(\cR; \mathbf{O}, \widehat{\boldsymbol{\eta}}_{(k)}) 
    \big\}$
    and their variance estimators. 
    Estimators of causal effects (e.g., direct effect) are obtained by the difference between the estimators $\widehat{\mu}(\cR;Q)$, $\widehat{\mu}_a(\cR;Q)$.

    \item Check sufficient conditions for consistency and asymptotic normality of the proposed estimators according to Theorem 1.

    \item Perform inference on target causal estimands based on the large sample properties of the proposed estimators.

\end{enumerate}

\noindent
Regarding the large sample properties conditions in step 7, 
it can be shown that 
\begin{align*}    
    \textstyle
     \norm\Big{
        \sum_{\mathbf{a}\in \mathcal{A}(N)}
            \norm\big{
                \big(
                    \widehat{\Bphi}_o - \Bphi_o
                \big)
                (\mathbf{A}, \mathbf{X}, N; \mathbf{a})
            }_2
    }_{L_2(\mathbb{P})}
    =
    O \left(
        \norm\Big{
            \sum_{\mathbf{a}\in \mathcal{A}(N)}
                \big|
                    \big(
                        \widehat{\phi}_Q - \phi_Q
                    \big)
                    (\mathbf{A}, \mathbf{X}, N; \mathbf{a})
                \big|
        }_{L_2(\mathbb{P})}
    \right)
    ,
\end{align*}
\begin{align*}
    &
    \textstyle
    \norm\Big{
        \sum_{\mathbf{a}\in \mathcal{A}(N)}
            \norm\big{
                \big(
                \widehat{\Bw}_o
                -
                \Bw_o
                \big)
                (\mathbf{a}, \mathbf{X}, N)
                +
                \sum_{\mathbf{a}' \in \mathcal{A}(N)}
                    \widehat{\Bphi}_o(\mathbf{a}', \mathbf{X}, N; \mathbf{a})
                    H(\mathbf{a}', \mathbf{X}, N)
            }_2
    }_{L_2(\mathbb{P})}
    \\
    = &
    \textstyle
    O \left(
        \norm\Big{
            \sum_{\mathbf{a}\in \mathcal{A}(N)}
                \big|
                    \big(
                    \widehat{Q}
                    -
                    Q
                    \big)
                    (\mathbf{a} | \mathbf{X}, N)
                    +
                    \sum_{\mathbf{a}' \in \mathcal{A}(N)}
                        \widehat{\phi}_Q(\mathbf{a}', \mathbf{X}, N; \mathbf{a})
                        H(\mathbf{a}', \mathbf{X}, N)
                \big|
        }_{L_2(\mathbb{P})}
    \right)
\end{align*}
where $\widehat{Q}$ and $\widehat{\phi}_Q$ are estimators of $Q$ and $\phi_Q$, respectively,
and similar conditions hold for $\widehat{\Bw}_a$ and $\widehat{\Bphi}_a$.
Thus, when investigating the rate conditions for $\widehat{\Bw}_o$ and $\widehat{\Bphi}_o$ (or $\widehat{\Bw}_a$ and $\widehat{\Bphi}_a$),
it suffices to assess the rate conditions for $\widehat{Q}$ and $\widehat{\phi}_Q$.

Furthermore, regarding the conditions for Theorem 2 in the main text,
if a collection of policies $Q(\cdot|\mathbf{X}, N;\theta)$ indexed by $\theta \in \Theta$ is considered,
then it must be determined whether the function classes
$\cF_{\Bw} = \{\Bw(\axn;\theta): \theta \in \Theta\}$
and
$\cF_{\Bphi} = \{\Bphi(\Ba',\xn;\Ba, \theta): \theta \in \Theta\}$
are Donsker classes.
It can be shown that
if the function class
$\cF_{Q} = \{Q(\Ba | \xn;\theta): \theta \in \Theta\}$ is Donsker, 
then $\cF_{\Bw_o} = \{\Bw_o(\axn;\theta): \theta \in \Theta\}$ is also Donsker,
and
if 
$\cF_{\phi_Q} = \{\phi_Q(\Ba',\xn;\Ba, \theta): \theta \in \Theta\}$ is Donsker,
then $\cF_{\Bphi_o} = \{\Bphi_o(\Ba',\xn;\Ba, \theta): \theta \in \Theta\}$ is also Donsker.
Similar conditions hold for $\cF_{\Bw_a}$ and $\cF_{\Bphi_a}$.
Thus, when investigating the Donsker conditions for ${\Bw}_o$ and ${\Bphi}_o$ (or ${\Bw}_a$ and ${\Bphi}_a$),
it suffices to assess the Donsker conditions for ${Q}$ and ${\phi}_Q$.

Below, technical details of the example policies discussed in the main text are described.

\subsection{Type B policy}
\label{example:TypeB}


\begin{itemize}
    \item
    Policy distribution $Q$:  
    $$Q_{\scriptscriptstyle \textup{B}}(\mathbf{a}|\mathbf{X}, N; \alpha)
    = 
    \prod_{j=1}^{N} \alpha^{a_{j}} (1-\alpha)^{1-a_{j}}$$

    \item
    EIF $\phi_Q$: 
    $$\phi_{Q_{\textup{B}}}(\mathbf{A}, \mathbf{X}, N; \mathbf{a})
    = 
    0$$
    since $Q$ does not depend on the observed data distribution.
    
    \item
    Convergence rates of nuisance functions:

    \begin{enumerate}[label=(\roman*)]
        
        \item 
        \textit{Convergence rate of} $\widehat{\Bphi}$: 

        Since $\phi_{Q_{\textup{B}}} = 0$,
        $\widehat{\Bphi}_o = \widehat{\Bphi}_a = \mathbf{0}$,
        and thus $r_{\Bphi} = 0$ 
        for both $\widehat{\mu}(\cR;Q)$ and $\widehat{\mu}_a(\cR;Q)$.

        \item 
        \textit{Second order convergence rate of} $\widehat{\Bw}$: 
        
        Since $Q_{\textup{B}}$ is known and does not need to be estimated, 
        $\widehat{\Bw} = \Bw$,
        and thus
        $\big(
            \widehat{\Bw}
            -
            \Bw
        \big)
        (\mathbf{a}, \mathbf{X}, N)
        +
        \widehat{\Bphi}(\mathbf{A}, \mathbf{X}, N; \mathbf{a})
        =
        0
        $,
        which implies $r_{\Bw} = 0$
        for both $\widehat{\mu}(\cR;Q)$ and $\widehat{\mu}_a(\cR;Q)$.
        
    \end{enumerate}

    \item
    Sufficient conditions for Theorem 1 (large sample properties):

    \begin{enumerate}[label=(\roman*)]
        
        \item 
        \textit{Consistency}: $\rT(\rH+\rC) = o(1)$;

        \item
        \textit{Asymptotic normality}: $\rT(\rH+\rC) = o(m^{-1/2})$;
    
        \item
        \textit{Consistent variance estimator}: $\rH + \rT + \rC = o(1)$. 
    
    \end{enumerate}

    \item
    Sufficient conditions for Theorem 2 (weak convergence to Gaussian process):

    \begin{enumerate}[label=(\roman*)]

        \item
        $\cF_{\Bw} 
        = 
        \{
          \Bw(\aXN;\alpha): \alpha \in [\alpha_l, \alpha_u] 
        \}$ is Donsker:

        It is easy to check that the following derivative of 
        $Q_{\scriptscriptstyle \textup{B}}(\mathbf{a}|\mathbf{X}, N; \alpha)
        = 
        \prod_{k=1}^{N} \alpha^{a_{k}} (1-\alpha)^{1-a_{k}}$
        with respect to $\alpha$ is bounded:
        \begin{align*}
            \textstyle
            \left|
                \frac{\partial}{\partial \alpha}
                \left\{
                    \textstyle
                    \prod_{k} \alpha^{a_{k}} (1-\alpha)^{1-a_{k}}
                \right\}
            \right|
            =
            \alpha^{\sum_{k}a_k-1}
            (1-\alpha)^{N-\sum_{k}a_k-1}
            \left|
                \sum_{k}a_k - N\alpha
            \right|
            \le 
            N
            \le
            n_{\textup{max}}
        \end{align*}
        which implies that function class $\cF_{Q}$ is Lipschitz, and thus Donsker.
        Therefore, $\cF_{\Bw_o}$ and $\cF_{\Bw_a}$ are Donsker classes as well.

        \item
        $\cF_{\Bphi} 
        = 
        \{
          \Bphi(\AXN;\Ba,\alpha): \alpha \in [\alpha_l, \alpha_u] 
        \}$ is Donsker:

        Since $\phi_{Q_{\textup{B}}} = 0$, $\cF_{\phi_Q}$ is an empty set, and thus Donsker.
        Therefore, $\cF_{\Bphi_o}$ and $\cF_{\Bphi_a}$ are Donsker classes as well.

    \end{enumerate}

\end{itemize}

\subsection{Cluster incremental propensity score policy}
\label{example:CIPS}

Under CIPS policy, 
the conditional independence of $A_{ij}$'s is assumed.
Therefore, the individual-level propensity score $\pi(j, \xn) = \bP(A_j = 1 | \BX = \Bx, N = n)$ may be estimated instead of $H$ 
and used to construct the estimator of $H$.
Assume the convergence rate of the estimator of $\pi$ is given by
$ 
\normp{
    \sum_{j=1}^{N}
        \lvert
            (\widehat{\pi} - \pi) (j, \mathbf{X}, N)
        \rvert
}
\allowbreak
=
O_{P}(r_{\pi})
$
for some $r_{\pi} = O(1)$.

\begin{itemize}

    \item
    Policy distribution $Q$:  
    $$Q_{\scriptscriptstyle \textup{CIPS}}(\mathbf{a} | \mathbf{X}, N; \delta) 
        =
        \prod_{j = 1}^{N} 
        \allowbreak
            (\pi_{j,\delta})^{a_{j}}
            \allowbreak
            (1 - \pi_{j,\delta})^{1-a_{j}} 
    ,
    $$
    where
    $\pi_{j} = \bP(A_j = 1 | \XN)$ is the factual propensity score, 
    and
    $\pi_{j, \delta} \allowbreak
    = \delta
    \pi_{j} / \allowbreak
    ( \delta \pi_{j} + 1 - \allowbreak  \pi_{j} )$
    denotes the shifted propensity score

    \item
    EIF $\phi_Q$:
    \begin{align*}
        \phi_{Q_{\textup{CIPS}}}(\mathbf{A}, \mathbf{X}, N; \mathbf{a})
        =
        Q_{\scriptscriptstyle \textup{CIPS}}(\mathbf{a} | \mathbf{X}, N; \delta)
        \sum_{l=1}^{N}
            \frac
                {(2a_{l}-1) \delta (A_{l} - \pi_{l})}
                {(\pi_{l,\delta})^{a_{l}} (1 - \pi_{l,\delta})^{1-a_{l}}
                (\delta\pi_{l} + 1 - \pi_{l})^2 }
    \end{align*}
    from the fact that the EIF of 
    $\pi_{j} \allowbreak
    = \mathbb{P}(A_{j} = 1 | \mathbf{X} = \mathbf{x}, N = n)$
    is
    \begin{align*}
        \textup{EIF}(\pi_{j})
        =
        \frac
            {\mathbbm{1}(\mathbf{X} = \mathbf{x}, N = n)}
            {d\mathbb{P}(\mathbf{x}, n)}
        (A_j - \pi_j)
    \end{align*}
    and the EIF of 
    $\pi_{j, \delta}$
    is
    \begin{align*}
        \textup{EIF}(\pi_{j, \delta})
        =
        \textup{EIF}(\pi_{j})
        \frac
            {\delta}
            {(\delta\pi_{j} + 1 - \pi_{j})^2}
        ,
    \end{align*}
    and thus
    \begin{align*}
        \textup{EIF}(Q_{\scriptscriptstyle \textup{CIPS}}(\mathbf{a} | \mathbf{x}, n; \delta))
        =
        \sum_{l=1}^{N}
        \left\{
        \prod_{j \ne l} 
            (\pi_{j,\delta})^{a_{j}}
            (1 - \pi_{j,\delta})^{1-a_{j}} 
        \right\}
        (2a_{l}-1)\textup{EIF}(\pi_{l,\delta})
        .
    \end{align*}

    \item
    Convergence rates of nuisance functions:

    The following inequalities were used to derive the convergence rate results of the nuisance functions;
    if
    $\overline{d}_i, d_i \in [0,u], i = 1, \dots, n$, 
    then 
    $|\prod_{i=1}^n \overline{d}_i - \prod_{i=1}^n d_i| 
    <
    2^n u^{n-1} \sum_{i=1}^n |\overline{d}_i - d_i|$
    and
    $|\prod_{i=1}^n \overline{d}_i 
    - 
    \prod_{i=1}^n d_i 
    - 
    \sum_{l=1}^n (\prod_{i \ne l} \overline{d}_i) \allowbreak (\overline{d}_l - d_l)  | 
    <
    2^n/u^{n-1} \sum_{i=1}^n |\overline{d}_i - d_i|^2$.    
    
    \begin{enumerate}[label=(\roman*)]
        \item 
        \textit{Convergence rate of} $\widehat{H}$: 
        $r_H = O(r_{\pi})$
        since
        \begin{align*}
            \lvert
                \big(\widehat{H}-H\big)(\mathbf{a}, \mathbf{X}, N)
            \rvert
            =
            \Big|
                \prod_{j=1}^N
                    \widehat{\mathbb{P}}(a_{j}| \mathbf{X}, N)
                -
                \prod_{j=1}^N
                    \mathbb{P}(a_{j}| \mathbf{X}, N)
            \Big|
            \lesssim
            \sum_{j=1}^{N}
            \lvert
                (\widehat{\pi}_j - \pi_j)
                (\mathbf{X}, N)
            \rvert
        \end{align*}

        \item
        \textit{Convergence rate of} $\widehat{\Bphi}$:
        $r_{\Bphi} = O(r_{\pi})$ since
        \begin{align*}
            &
            \widehat{\phi}_{Q_{\textup{CIPS}}}(\mathbf{A}, \mathbf{X}, N; \mathbf{a})
            -
            \phi_{Q_{\textup{CIPS}}}(\mathbf{A}, \mathbf{X}, N; \mathbf{a})
            \\
            = &
            \left(
                \widehat{Q}_{\scriptscriptstyle \textup{CIPS}}
                -
                Q_{\scriptscriptstyle \textup{CIPS}}
            \right)
            (\mathbf{a} | \mathbf{X}, N; \delta)
            \sum_{l=1}^{N}
                \frac
                    {(2a_{l}-1) \delta (A_{l} - \widehat{\pi}_{l})}
                    {\widehat{\mathbb{P}}_{\delta}(a_{l}| \mathbf{X}, N)
                    (\delta\widehat{\pi}_{l} + 1 - \widehat{\pi}_{l})^2}
            \\
            & +
            Q_{\scriptscriptstyle \textup{CIPS}}
            (\mathbf{a} | \mathbf{X}, N; \delta)
            \sum_{l=1}^{N}
                \left\{
                    \frac
                        {(2a_{l}-1) \delta (A_{l} - \widehat{\pi}_{l})}
                        {\widehat{\mathbb{P}}_{\delta}(a_{l}| \mathbf{X}, N)
                        (\delta\widehat{\pi}_{l} + 1 - \widehat{\pi}_{l})^2}
                    -
                    \frac
                        {(2a_{l}-1) \delta (A_{l} - \pi_{l})}
                        {\mathbb{P}_{\delta}(a_{l}| \mathbf{X}, N)
                        (\delta\pi_{l} + 1 - \pi_{l})^2}
                \right\}
            \\
            \lesssim &
            \sum_{j=1}^{N}
            \lvert
                (\widehat{\pi}_j - \pi_j)
                (\mathbf{X}, N)
            \rvert
        \end{align*}
        which follows from
        \begin{align*}
            \left(
                \widehat{Q}_{\scriptscriptstyle \textup{CIPS}}
                -
                Q_{\scriptscriptstyle \textup{CIPS}}
            \right)
            (\mathbf{a} | \mathbf{X}, N; \delta)
            \lesssim 
            \sum_{j=1}^{N}
            \lvert
                \widehat{\mathbb{P}}_{\delta}(a_{j}| \mathbf{X}, N)
                -
                \mathbb{P}_{\delta}(a_{j}| \mathbf{X}, N)
            \rvert
            = 
            \sum_{j=1}^{N}
            \frac
                {\delta|\widehat{\pi}_{j} -\pi_{j}|}
                {(\delta\widehat{\pi}_{j} + 1 - \widehat{\pi}_{j})
                (\delta\pi_{j} + 1 - \pi_{j})}
        \end{align*}
        and
        \begin{align*}
            &
            \frac
                {A_{l} - \widehat{\pi}_{l}}
                {\widehat{\mathbb{P}}_{\delta}(a_{l}| \mathbf{X}, N)
                (\delta\widehat{\pi}_{l} + 1 - \widehat{\pi}_{l})^2}
            -
            \frac
                {A_{l} - \pi_{l}}
                {\mathbb{P}_{\delta}(a_{l}| \mathbf{X}, N)
                (\delta\pi_{l} + 1 - \pi_{l})^2}
            \\
            = &
            \frac
                {\pi_l - \widehat{\pi}_l}
                {\widehat{\mathbb{P}}_{\delta}(a_{l}| \mathbf{X}, N)
                (\delta\widehat{\pi}_{l} + 1 - \widehat{\pi}_{l})^2}
            +
            (A_l - \pi_l)
            \frac
                {\mathbb{P}_{\delta}(a_{l}| \mathbf{X}, N)
                (\delta\pi_{l} + 1 - \pi_{l})^2
                -
                \widehat{\mathbb{P}}_{\delta}(a_{l}| \mathbf{X}, N)
                (\delta\widehat{\pi}_{l} + 1 - \widehat{\pi}_{l})^2}
                {\widehat{\mathbb{P}}_{\delta}(a_{l}| \mathbf{X}, N)
                (\delta\widehat{\pi}_{l} + 1 - \widehat{\pi}_{l})^2
                \mathbb{P}_{\delta}(a_{l}| \mathbf{X}, N)
                (\delta\pi_{l} + 1 - \pi_{l})^2}
        \end{align*}
        with
        \begin{align*}
            &
            \mathbb{P}_{\delta}(a_{l}| \mathbf{X}, N)
            (\delta\pi_{l} + 1 - \pi_{l})^2
            -
            \widehat{\mathbb{P}}_{\delta}(a_{l}| \mathbf{X}, N)
            (\delta\widehat{\pi}_{l} + 1 - \widehat{\pi}_{l})^2
            \\
            = &
            \begin{cases}
                (\pi_l - \widehat{\pi}_l)
                \delta
                [(\delta-1)(\pi_l + \widehat{\pi}_l) + 1],
                & a_j = 1
                \\
                (\pi_l - \widehat{\pi}_l)
                [\delta-2 
                -(\delta-1)(\pi_l + \widehat{\pi}_l)],
                & a_j = 0
            \end{cases}
            \\
            \lesssim &
            | \pi_l - \widehat{\pi}_l |
            .
        \end{align*}

        \item
        \textit{Second order convergence rate of} $\widehat{\Bw}$:
        $r_{\Bw} = O(r_{\pi})$ since
        \begin{align*}
            &
            \widehat{Q}_{\scriptscriptstyle \textup{CIPS}}
            (\mathbf{a} | \mathbf{X}, N; \delta)
            -
            Q_{\scriptscriptstyle \textup{CIPS}}
            (\mathbf{a} | \mathbf{X}, N; \delta)
            +
            \sum_{\mathbf{a}' \in \mathcal{A}(N)}
                \widehat{\phi}_{Q_{\textup{CIPS}}}(\mathbf{a}', \mathbf{X}, N; \mathbf{a})
                H(\mathbf{a}', \mathbf{X}, N)
            \\
            = &
            \prod_{j=1}^N
                \widehat{\mathbb{P}}_{\delta}(a_{j}| \mathbf{X}, N)
            -
            \prod_{j=1}^N
                \mathbb{P}_{\delta}(a_{j}| \mathbf{X}, N)
            +
            \sum_{j=1}^N
                \left\{
                \prod_{l \ne j}
                    \widehat{\mathbb{P}}_{\delta}(a_{l}| \mathbf{X}, N)
                \right\}
                \frac
                    {(2a_{j}-1) \delta (\pi_{j} - \widehat{\pi}_{j})}
                    {(\delta\widehat{\pi}_{j} + 1 - \widehat{\pi}_{j})^2}
            \\
            =&
            \prod_{j=1}^N
                \widehat{\mathbb{P}}_{\delta}(a_{j}| \mathbf{X}, N)
            -
            \prod_{j=1}^N
                \mathbb{P}_{\delta}(a_{j}| \mathbf{X}, N)
            -
            \sum_{j=1}^N
                \left\{
                \prod_{l \ne j}
                    \widehat{\mathbb{P}}_{\delta}(a_{l}| \mathbf{X}, N)
                \right\}
                (\widehat{\mathbb{P}}_{\delta} - \mathbb{P}_{\delta})
                (a_{j}| \mathbf{X}, N)
            \\
            & +
            \sum_{j=1}^N
                \left\{
                    \prod_{l \ne j}
                        \widehat{\mathbb{P}}_{\delta}(a_{l}| \mathbf{X}, N)
                \right\}
                \left\{
                    (\widehat{\mathbb{P}}_{\delta} - \mathbb{P}_{\delta})
                    (a_{j}| \mathbf{X}, N)
                    +
                    \frac
                        {(2a_{j}-1) \delta (\pi_{j} - \widehat{\pi}_{j})}
                        {(\delta\widehat{\pi}_{j} + 1 - \widehat{\pi}_{j})^2}
                \right\}
            \\
            \lesssim &
            \sum_{j=1}^{N}
            \lvert
                \widehat{\mathbb{P}}_{\delta}(a_{j}| \mathbf{X}, N)
                -
                \mathbb{P}_{\delta}(a_{j}| \mathbf{X}, N)
            \rvert^2
            +
            \lvert
                (\widehat{\pi}_j - \pi_j)
                (\mathbf{X}, N)
            \rvert^2
            \\
            \lesssim &
            \sum_{j=1}^{N}
            \lvert
                (\widehat{\pi}_j - \pi_j)
                (\mathbf{X}, N)
            \rvert^2
        \end{align*}
        from
        \begin{align*}
            \sum_{\mathbf{a}' \in \mathcal{A}(N)}
                \widehat{\phi}_{Q_{\textup{CIPS}}}(\mathbf{a}', \mathbf{X}, N; \mathbf{a})
                H(\mathbf{a}', \mathbf{X}, N)
            = &
            \E{\{}{\}}{
                \widehat{\phi}_{Q_{\textup{CIPS}}}(\mathbf{A}, \mathbf{X}, N; \mathbf{a})
                \middle| D, \mathbf{X}, N
            }
            \\
            = &
            \widehat{Q}_{\scriptscriptstyle \textup{CIPS}}
            (\mathbf{a} | \mathbf{X}, N; \delta)
            \sum_{j=1}^{N}
                \frac
                    {(2a_{j}-1) \delta (\pi_{j} - \widehat{\pi}_{j})}
                    {\widehat{\mathbb{P}}_{\delta}(a_{j}| \mathbf{X}, N)
                    (\delta\widehat{\pi}_{j} + 1 - \widehat{\pi}_{j})^2}
        \end{align*}
        and
        \begin{align*}
            (\widehat{\mathbb{P}}_{\delta} - \mathbb{P}_{\delta})
            (a_{j}| \mathbf{X}, N)
            +
            \frac
                {(2a_{j}-1) \delta (\pi_{j} - \widehat{\pi}_{j})}
                {(\delta\widehat{\pi}_{j} + 1 - \widehat{\pi}_{j})^2}
            =
            \frac
                {(2a_{j}-1) \delta (\delta-1) (\pi_{j} - \widehat{\pi}_{j})^2}
                {(\delta\widehat{\pi}_{j} + 1 - \widehat{\pi}_{j})^2
                (\delta\pi_{j} + 1 - \pi_{j})^2}
            .
        \end{align*}

    \end{enumerate}

    \item
    Sufficient conditions for Theorem 1 (large sample properties):

    \begin{enumerate}[label=(\roman*)]
        
        \item 
        \textit{Consistency}: $r_{\pi} = o(1)$, $\rT \cdot \rC = o(1)$;

        \item
        \textit{Asymptotic normality}: $r_{\pi} = o(m^{-1/4})$, $\rT \cdot (r_{\pi}+\rC) = o(m^{-1/2})$;
    
        \item
        \textit{Consistent variance estimator}: $r_{\pi} + \rT + \rC = o(1)$.
    
    \end{enumerate}

    \item
    Sufficient conditions for Theorem 2 (weak convergence to Gaussian process):

    \begin{enumerate}[label=(\roman*)]

        \item
        $\cF_{\Bw} 
        = 
        \{
          \Bw(\aXN;\delta): \delta \in [\delta_l, \delta_u] 
        \}$ is Donsker:

        From the assumption (B1) in the main text that $\pi_{l} \in (d, 1-d)$
        and the fact that $\delta \pi_{l} + 1 - \pi_{l} \in [\delta_l, \delta_u]$,
        the following derivative of 
        $Q_{\scriptscriptstyle \textup{CIPS}}(\mathbf{a} | \mathbf{X}, N; \delta) 
        =
        \prod_{j = 1}^{N} 
            (\pi_{j,\delta})^{a_{j}}
            (1 - \pi_{j,\delta})^{1-a_{j}}$
        with respect to $\delta$ is bounded:
        \begin{gather*}
            \left|
                \frac{\partial}{\partial \delta}
                \left\{
                    (\pi_{j,\delta})^{a_{j}}
                    (1 - \pi_{j,\delta})^{1-a_{j}} 
                \right\}
            \right|
            =
            \frac
                {\pi_{j}(1-\pi_{j})}
                {( \delta \pi_{j} + 1 - \pi_{j} )^2}
            \le 
            \frac{1}{\delta_l^2}
        \end{gather*}
        which implies that 
        $Q_{\scriptscriptstyle \textup{CIPS}}$
        is a Lipschitz function of $\delta$.
        Thus, the function class $\cF_{Q}$ is Lipschitz, and thus Donsker.
        Therefore, $\cF_{\Bw_o}$ and $\cF_{\Bw_a}$ are Donsker classes as well.

        \item
        $\cF_{\Bphi} 
        = 
        \{
          \Bphi(\AXN;\Ba,\delta): \delta \in [\delta_l, \delta_u]
        \}$ is Donsker:

        First, note that
        \begin{align*}
            \phi_{Q_{\textup{CIPS}}}(\mathbf{A}, \mathbf{X}, N; \mathbf{a})
            =
            Q_{\scriptscriptstyle \textup{CIPS}}(\mathbf{a} | \mathbf{X}, N; \delta)
            \sum_{l=1}^{N}
                \frac
                    {(2a_{l}-1) \delta (A_{l} - \pi_{l})}
                    {(\pi_{l,\delta})^{a_{l}} (1 - \pi_{l,\delta})^{1-a_{l}}
                    (\delta\pi_{l} + 1 - \pi_{l})^2 }
            .
        \end{align*}
        Since $Q_{\scriptscriptstyle \textup{CIPS}}$ is a Lipschitz function, it remains to show the term in the summation is also Lipschitz.
        It can be shown that the derivative of this term is bounded as follows:
        \begin{align*}
            \left|
                \frac{\partial}{\partial \delta}
                \left[
                    \frac
                        {(2a_{l}-1) 
                        \delta 
                        \left( A_{l}-\pi_{l} \right)}
                        {(\pi_{l,\delta})^{a_{l}}
                        (1 - \pi_{l,\delta})^{1-a_{l}} 
                        ( \delta \pi_{l} + 1 - \pi_{l} )^2}
                \right]
            \right|
            =
            \frac
                {|A_{l}-\pi_{l}|}
                {( \delta \pi_{l} + 1 - \pi_{l} )^2}
            \le 
            \frac{1}{\delta_l^2}
        \end{align*}
        which implies that $\cF_{\phi_Q}$ is Lipschitz, and thus Donsker.
        Therefore, $\cF_{\Bphi_o}$ and $\cF_{\Bphi_a}$ are Donsker classes as well.
    \end{enumerate}

\end{itemize}

\subsection{Treated proportion bound policy}

\begin{itemize}

    \item
    Policy distribution $Q$:  
    $$
    Q_{\scriptscriptstyle \textup{TPB}}(\mathbf{a} | \mathbf{X}, N; \rho) 
    =
    \mathbbm{1}(\overline{\mathbf{a}} \ge \rho)
    \frac
        {H(\mathbf{a}, \mathbf{X}, N)}
        {\mathbb{P}(\overline{\mathbf{A}} \ge \rho | \mathbf{X}, N)}
    ,
    $$
    where
    $\mathbb{P}(\overline{\mathbf{A}} \ge \rho | \mathbf{X}, N)
    =
    \sum_{\overline{\mathbf{a}}' \ge \rho} 
        H(\mathbf{a}', \mathbf{X}, N)
    $
    is the observed probability of the proportion of treated units within a cluster is at least $\rho$.

    \item
    EIF $\phi_Q$:
    \begin{align*}
        \phi_{Q_{\textup{TPB}}}(\mathbf{A}, \mathbf{X}, N; \mathbf{a})
        =
        \frac
            {\mathbbm{1}(\overline{\mathbf{a}} \ge \rho)}
            {\mathbb{P}(\overline{\mathbf{A}} \ge \rho | \mathbf{X}, N)^2}
        \left\{
            \mathbbm{1}(\mathbf{A} = \mathbf{a})
            \mathbb{P}(\overline{\mathbf{A}} \ge \rho | \mathbf{X}, N)
            -
            \mathbbm{1}(\overline{\mathbf{A}} \ge \rho)
            H(\mathbf{a}, \mathbf{X}, N)
        \right\}
    \end{align*}
    from the fact that the EIF of 
    $H(\mathbf{a}, \mathbf{x}, n)$ is
    \begin{align*}
        \textup{EIF}(H(\mathbf{a}, \mathbf{x}, n))
        =
        \frac
            {\mathbbm{1}(\mathbf{X} = \mathbf{x}, N = n)}
            {d\mathbb{P}(\mathbf{x}, n)}
        \{
            \mathbbm{1}(\mathbf{A} = \mathbf{a}) 
            - 
            H(\mathbf{a}, \mathbf{X}, N)
        \}        
        ,
    \end{align*}
    and the EIF of 
    $\mathbb{P}(\overline{\mathbf{A}} \ge \rho | \mathbf{x}, n)
    =
    \sum_{\overline{\mathbf{a}}' \ge \rho} 
        H(\mathbf{a}', \mathbf{x}, n)
    $
    is
    \begin{align*}
        \textup{EIF}(
            \mathbb{P}(\overline{\mathbf{A}} \ge \rho | \mathbf{x}, n)
        )
        = &
        \sum_{\overline{\mathbf{a}}' \ge \rho} 
            \frac
                {\mathbbm{1}(\mathbf{X} = \mathbf{x}, N = n)}
                {d\mathbb{P}(\mathbf{x}, n)}
            \{
                \mathbbm{1}(\mathbf{A} = \mathbf{a}') 
                - 
                H(\mathbf{a}', \mathbf{X}, N)
            \}    
        \\
        = &
        \frac
            {\mathbbm{1}(\mathbf{X} = \mathbf{x}, N = n)}
            {d\mathbb{P}(\mathbf{x}, n)}
        \{
            \mathbbm{1}(\overline{\mathbf{A}} \ge \rho)
            -
            \mathbb{P}(\overline{\mathbf{A}} \ge \rho | \mathbf{x}, n)
        \}
        .
    \end{align*}

    \item
    Convergence rates of nuisance functions:

    \begin{itemize}
        
        \item 
        \textit{Convergence rate of} $\widehat{\Bphi}$:
        $r_{\Bphi} = O(r_{H})$ since
        \begin{align*}
            &
            \widehat{\phi}_{Q_{\textup{TPB}}}(\mathbf{A}, \mathbf{X}, N; \mathbf{a})
            -
            \phi_{Q_{\textup{TPB}}}(\mathbf{A}, \mathbf{X}, N; \mathbf{a})
            \\
            = &
            \mathbbm{1}(\overline{\mathbf{a}} \ge \rho)
            \begin{aligned}[t]
                \Bigg[
                    &
                    \mathbbm{1}(\mathbf{A} = \mathbf{a})
                    \left\{
                        \frac
                            {1}
                            {\widehat{\mathbb{P}}(\overline{\mathbf{A}} \ge \rho | \mathbf{X}, N)}
                        -
                        \frac
                            {1}
                            {\mathbb{P}(\overline{\mathbf{A}} \ge \rho | \mathbf{X}, N)}
                    \right\}
                    \\
                    &-
                    \mathbbm{1}(\overline{\mathbf{A}} \ge \rho)
                    \left\{
                        \frac
                            {\widehat{H}(\mathbf{a}, \mathbf{X}, N)}
                            {\widehat{\mathbb{P}}(\overline{\mathbf{A}} \ge \rho | \mathbf{X}, N)^2}
                        -
                        \frac
                            {H(\mathbf{a}, \mathbf{X}, N)}
                            {\mathbb{P}(\overline{\mathbf{A}} \ge \rho | \mathbf{X}, N)^2}
                    \right\}
                \Bigg]
            \end{aligned}
            \\
            \lesssim &
            \lvert
                (\widehat{\mathbb{P}}-\mathbb{P})(\overline{\mathbf{A}} \ge \rho | \mathbf{X}, N)
            \rvert
            +
            \lvert
                (\widehat{H}-H)(\mathbf{a}, \mathbf{X}, N)
            \rvert
            +
            \lvert
                \widehat{\mathbb{P}}(\overline{\mathbf{A}} \ge \rho | \mathbf{X}, N)^2
                -
                \mathbb{P}(\overline{\mathbf{A}} \ge \rho | \mathbf{X}, N)^2
            \rvert
            \\
            \lesssim &
            \sumaN
                \lvert
                    (\widehat{H}-H)(\mathbf{a}, \mathbf{X}, N)
                \rvert
        \end{align*}
        which follows from
        \begin{align*}
            \lvert
                (\widehat{\mathbb{P}}-\mathbb{P})(\overline{\mathbf{A}} \ge \rho | \mathbf{X}, N)
            \rvert
            =
            \sum_{\overline{\mathbf{a}}' \ge \rho} 
                \lvert
                    (\widehat{H}-H)(\mathbf{a}', \mathbf{X}, N)
                \rvert
        \end{align*}
        and the inequality 
        $
        \lvert
            x_1/y_1 - x_2/y_2
        \rvert
        \lesssim
        \lvert
            x_1 - x_2
        \rvert
        +
        \lvert
            y_1 - y_2
        \rvert
        $
        if $y_1, y_2$ are bounded.

        \item
        \textit{Second order convergence rate of} $\widehat{\Bw}$: 
        $r_{\Bw} = O(r_{H})$ since
        \begin{align*}
            &
            \widehat{Q}_{\scriptscriptstyle \textup{TPB}}
            (\mathbf{a} | \mathbf{X}, N; \rho)
            -
            Q_{\scriptscriptstyle \textup{TPB}}
            (\mathbf{a} | \mathbf{X}, N; \rho)
            +
            \sum_{\mathbf{a}' \in \mathcal{A}(N)}
                \widehat{\phi}_{Q_{\textup{TPB}}}(\mathbf{a}', \mathbf{X}, N; \mathbf{a})
                H(\mathbf{a}', \mathbf{X}, N)
            \\
            = &
            \mathbbm{1}(\overline{\mathbf{a}} \ge \rho)
            \left\{
                \frac
                    {\widehat{H}(\mathbf{a}, \mathbf{X}, N)}
                    {\widehat{\mathbb{P}}(\overline{\mathbf{A}} \ge \rho | \mathbf{X}, N)}
                -
                \frac
                    {H(\mathbf{a}, \mathbf{X}, N)}
                    {\mathbb{P}(\overline{\mathbf{A}} \ge \rho | \mathbf{X}, N)}
                +
                \frac
                    {H(\mathbf{a}, \mathbf{X}, N)}
                    {\widehat{\mathbb{P}}(\overline{\mathbf{A}} \ge \rho | \mathbf{X}, N)}
                -
                \frac
                    {\mathbb{P}(\overline{\mathbf{A}} \ge \rho | \mathbf{X}, N)
                    \widehat{H}(\mathbf{a}, \mathbf{X}, N)}
                    {\widehat{\mathbb{P}}(\overline{\mathbf{A}} \ge \rho | \mathbf{X}, N)^2}
            \right\}
            \\
            = &
            \mathbbm{1}(\overline{\mathbf{a}} \ge \rho)
            \left\{
                \frac
                    {\widehat{H}(\mathbf{a}, \mathbf{X}, N)}
                    {\widehat{\mathbb{P}}(\overline{\mathbf{A}} \ge \rho | \mathbf{X}, N)}
                -
                \frac
                    {H(\mathbf{a}, \mathbf{X}, N)}
                    {\mathbb{P}(\overline{\mathbf{A}} \ge \rho | \mathbf{X}, N)}
            \right\}
            \left\{
                1
                -
                \frac
                    {\mathbb{P}(\overline{\mathbf{A}} \ge \rho | \mathbf{X}, N)}
                    {\widehat{\mathbb{P}}(\overline{\mathbf{A}} \ge \rho | \mathbf{X}, N)}
            \right\}
            \\
            \lesssim &
            \left\{
                \sumaN
                    \lvert
                        (\widehat{H}-H)(\mathbf{a}, \mathbf{X}, N)
                    \rvert
            \right\}^2
        \end{align*}
        from
        \begin{align*}
            &
            \sum_{\mathbf{a}' \in \mathcal{A}(N)}
                \widehat{\phi}_{Q_{\textup{TPB}}}(\mathbf{a}', \mathbf{X}, N; \mathbf{a})
                H(\mathbf{a}', \mathbf{X}, N)
            =
            \E{\{}{\}}{
                \widehat{\phi}_{Q_{\textup{TPB}}}(\mathbf{A}, \mathbf{X}, N; \mathbf{a})
                \middle| D, \mathbf{X}, N
            }
            \\
            =&
            \mathbbm{1}(\overline{\mathbf{a}} \ge \rho)
            \E{\{}{\}}{
                \frac
                    {\mathbbm{1}(\mathbf{A} = \mathbf{a})}
                    {\widehat{\mathbb{P}}(\overline{\mathbf{A}} \ge \rho | \mathbf{X}, N)}
                -
                \mathbbm{1}(\overline{\mathbf{A}} \ge \rho)
                \frac
                    {\widehat{H}(\mathbf{a}, \mathbf{X}, N)}
                    {\widehat{\mathbb{P}}(\overline{\mathbf{A}} \ge \rho | \mathbf{X}, N)^2}
                \middle| D, \mathbf{X}, N
            }
            \\
            = &
            \mathbbm{1}(\overline{\mathbf{a}} \ge \rho)
            \left\{
                \frac
                    {H(\mathbf{a}, \mathbf{X}, N)}
                    {\widehat{\mathbb{P}}(\overline{\mathbf{A}} \ge \rho | \mathbf{X}, N)}
                -
                \frac
                    {\mathbb{P}(\overline{\mathbf{A}} \ge \rho | \mathbf{X}, N)
                    \widehat{H}(\mathbf{a}, \mathbf{X}, N)}
                    {\widehat{\mathbb{P}}(\overline{\mathbf{A}} \ge \rho | \mathbf{X}, N)^2}
            \right\}
        .
        \end{align*}
    \end{itemize}  

    \item
    Sufficient conditions for Theorem 1 (large sample properties):

    \begin{enumerate}[label=(\roman*)]
        
        \item 
        \textit{Consistency}: $r_{H} = o(1)$, $\rT \cdot \rC = o(1)$;

        \item
        \textit{Asymptotic normality}: $r_{H} = o(m^{-1/4})$, $\rT \cdot (r_{\pi}+\rC) = o(m^{-1/2})$;
    
        \item
        \textit{Consistent variance estimator}: $r_{H} + \rT + \rC = o(1)$.
    
    \end{enumerate}
    
\end{itemize}

\newpage

\section{Additional Simulation results}

In this section, additional simulation results are presented that are not included in the main text.
All simulations are conducted under $\cR(T; \tau) = \indicator(T \le \tau)$.
Code for reproducing all numerical results presented here is provided in the GitHub repository (\url{https://github.com/chanhwa-lee/NPSACI}).

\subsection{Simulation results on TPB policy estimands}

The simulation results for the TPB policy with $\rho \in \{0,0.25,0.5\}$ at $\tau \in \{0.2, 0.4\}$
are given in Table~\ref{tab:simulTPB}.
The simulation setting is the same as the main text.

\begin{table}[ht]
	\centering
	\captionsetup{width=\textwidth}
	\caption{
        Simulation results for 
        the SBS-NCF estimators
        for TPB policy with $\rho \in \{0, 0.25, 0.5\}$, $\rho' = 0$, and $\tau \in \{0.2, 0.4\}$.
    }
	\vspace{-0.1cm}
	\renewcommand{\arraystretch}{1}
	\label{tab:simulTPB}
	\resizebox{\textwidth}{!}{%
    \begin{tabular}{ccccccccccccccccc}
    \hline
    \multirow{2}{*}{$\rho$} &  & \multirow{2}{*}{Estimand}                                                &  & \multicolumn{6}{c}{$\tau = 0.2$}      &  & \multicolumn{6}{c}{$\tau = 0.4$}      \\ \cline{5-10} \cline{12-17} 
                            &  &                                                                          &  & Truth & Bias & ASE & ESE & Cov & UCov &  & Truth & Bias & ASE & ESE & Cov & UCov \\ \hline
    \multirow{4}{*}{0}      &  & $\mu_{\scriptscriptstyle \textup{TPB}}(\tau; \rho)$                       &  & 38.2  & -0.4 & 1.3 & 1.4 & 92  & 92   &  & 48.4  & -0.6 & 1.4 & 1.5 & 92  & 92   \\
                            &  & $\mu_{\scriptscriptstyle \textup{TPB}, \scriptstyle 1}(\tau; \rho)$       &  & 22.6  & -0.7 & 1.5 & 1.6 & 91  & 92   &  & 33.5  & -1.0 & 1.7 & 1.8 & 91  & 90   \\
                            &  & $\mu_{\scriptscriptstyle \textup{TPB}, \scriptstyle 0}(\tau; \rho)$       &  & 50.3  & 0.6  & 1.8 & 1.8 & 93  & 94   &  & 60.2  & 0.5  & 1.8 & 1.9 & 93  & 93   \\
                            &  & $DE_{\scriptscriptstyle \textup{TPB}}(\tau; \rho)$                        &  & -27.7 & -1.3 & 2.2 & 2.2 & 91  & 92   &  & -26.6 & -1.5 & 2.4 & 2.4 & 91  & 90   \\ \hline
    \multirow{7}{*}{0.25}   &  & $\mu_{\scriptscriptstyle \textup{TPB}}(\tau; \rho)$                       &  & 34.9  & -0.5 & 1.3 & 1.4 & 92  & 92   &  & 45.0  & -0.7 & 1.4 & 1.5 & 93  & 92   \\
                            &  & $\mu_{\scriptscriptstyle \textup{TPB}, \scriptstyle 1}(\tau; \rho)$       &  & 21.3  & -0.7 & 1.5 & 1.5 & 91  & 92   &  & 32.1  & -1.0 & 1.7 & 1.8 & 90  & 90   \\
                            &  & $\mu_{\scriptscriptstyle \textup{TPB}, \scriptstyle 0}(\tau; \rho)$       &  & 48.6  & 0.7  & 2.0 & 2.0 & 93  & 94   &  & 58.6  & 0.7  & 2.0 & 2.0 & 93  & 93   \\
                            &  & $DE_{\scriptscriptstyle \textup{TPB}}(\tau; \rho)$                        &  & -27.3 & -1.4 & 2.4 & 2.4 & 90  & 92   &  & -26.5 & -1.7 & 2.6 & 2.6 & 90  & 90   \\
                            &  & $SE_{\scriptscriptstyle \textup{TPB}, \scriptstyle 1}(\tau; \rho, \rho')$ &  & -3.4  & -0.1 & 0.8 & 0.8 & 94  & 87   &  & -3.3  & -0.1 & 0.8 & 0.8 & 95  & 87   \\
                            &  & $SE_{\scriptscriptstyle \textup{TPB}, \scriptstyle 0}(\tau; \rho, \rho')$ &  & -1.2  & 0.0  & 0.8 & 0.9 & 94  & 94   &  & -1.4  & 0.0  & 0.9 & 0.9 & 95  & 95   \\
                            &  & $OE_{\scriptscriptstyle \textup{TPB}}(\tau; \rho, \rho')$                 &  & -1.6  & 0.1  & 0.8 & 0.7 & 94  & 88   &  & -1.6  & 0.2  & 0.8 & 0.8 & 94  & 89   \\ \hline
    \multirow{7}{*}{0.5}    &  & $\mu_{\scriptscriptstyle \textup{TPB}}(\tau; \rho)$                       &  & 29.6  & -0.6 & 1.7 & 1.8 & 94  & 92   &  & 39.7  & -0.8 & 1.9 & 2.2 & 93  & 92   \\
                            &  & $\mu_{\scriptscriptstyle \textup{TPB}, \scriptstyle 1}(\tau; \rho)$       &  & 19.7  & -0.6 & 1.8 & 1.9 & 95  & 92   &  & 30.1  & -0.9 & 2.1 & 2.5 & 92  & 90   \\
                            &  & $\mu_{\scriptscriptstyle \textup{TPB}, \scriptstyle 0}(\tau; \rho)$       &  & 46.5  & 0.8  & 2.8 & 4.1 & 94  & 94   &  & 56.4  & 0.8  & 3.0 & 4.2 & 93  & 93   \\
                            &  & $DE_{\scriptscriptstyle \textup{TPB}}(\tau; \rho)$                        &  & -26.8 & -1.5 & 3.1 & 3.2 & 93  & 92   &  & -26.3 & -1.9 & 3.4 & 3.4 & 90  & 90   \\
                            &  & $SE_{\scriptscriptstyle \textup{TPB}, \scriptstyle 1}(\tau; \rho, \rho')$ &  & -8.6  & 0.0  & 1.5 & 3.6 & 94  & 87   &  & -8.7  & -0.2 & 1.6 & 1.9 & 94  & 87   \\
                            &  & $SE_{\scriptscriptstyle \textup{TPB}, \scriptstyle 0}(\tau; \rho, \rho')$ &  & -2.9  & 0.1  & 1.6 & 3.3 & 95  & 94   &  & -3.5  & 0.1  & 1.8 & 2.2 & 94  & 95   \\
                            &  & $OE_{\scriptscriptstyle \textup{TPB}}(\tau; \rho, \rho')$                 &  & -3.7  & 0.2  & 2.0 & 3.5 & 95  & 88   &  & -3.7  & 0.3  & 2.0 & 3.6 & 94  & 89   \\ \hline
    \end{tabular}
    }
    \caption*{\small
    Truth: true value of the estimand ($\times$100),
    Bias: average bias of estimates ($\times$100), 
    ASE: average standard error estimates ($\times$100), 
    ESE: empirical standard error ($\times$100), 
    Cov: 95\% point-wise confidence interval coverage,
	UCov: 95\% uniform confidence band coverage.}
\end{table}

\subsection{Finite sample performance over number of clusters $m$}
\begin{figure}[H]
    \centering
    \includegraphics[width=\textwidth]{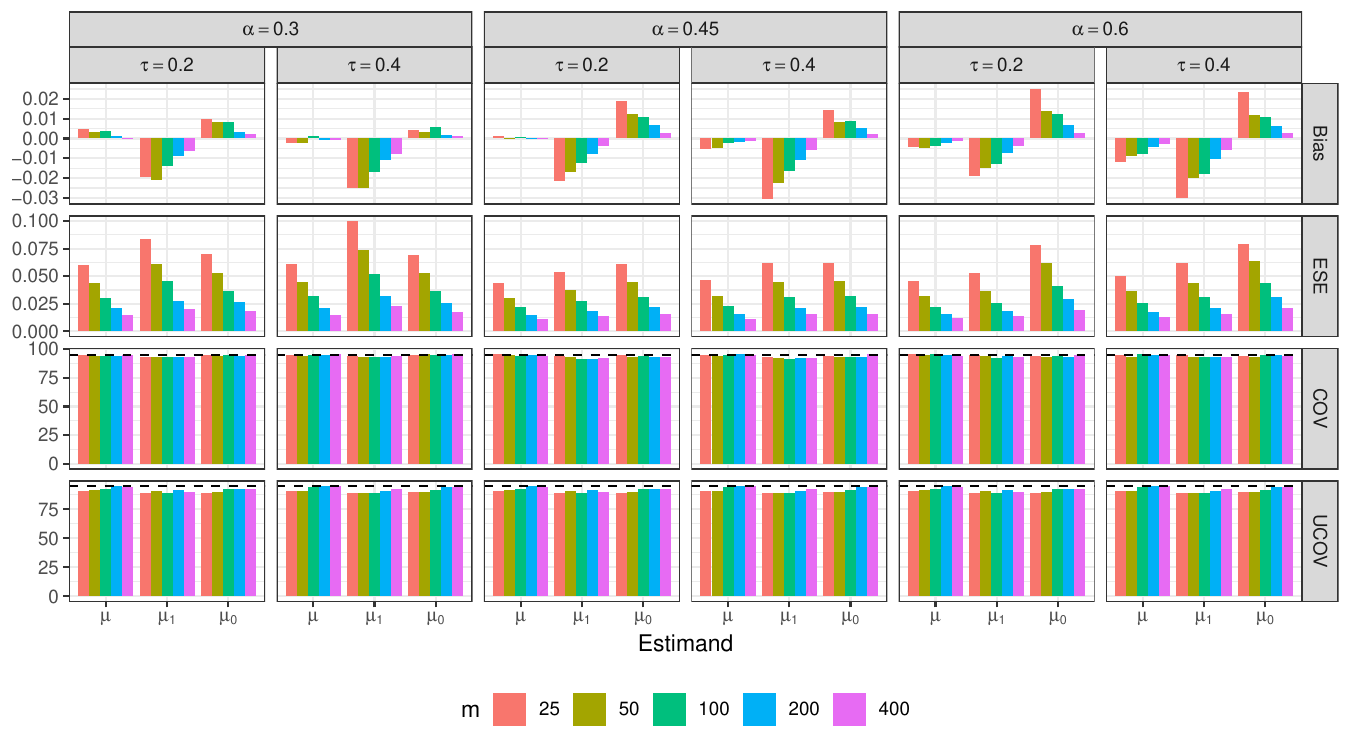}
    \caption{
    Finite sample performance of the NCF estimators over $m = 25, 50, 100, 200, 400$ for Type B policy estimands.}
    \label{fig:Simulm}
\end{figure}

The finite sample performance of the NCF estimators over varying number of clusters $m$ was investigated, and the results are presented in Fig~\ref{fig:Simulm}.
The simulation setting is the same as the main text, except that the number of clusters $m$ in each dataset was varied over $m = 25, 50, 100, 200, 400$.
As expected by the large sample properties of the NCF estimators, the bias and ESE decrease with magnitude of $O(m^{-1/2})$ as $m$ increases, and the 95\% point-wise CIs and UCBs both achieve nominal coverage rates as $m$ increases.

\subsection{Finite sample performance of bounded estimators}
\label{simul:bounded}

\begin{figure}[H]
    \centering
    \includegraphics[width=\textwidth]{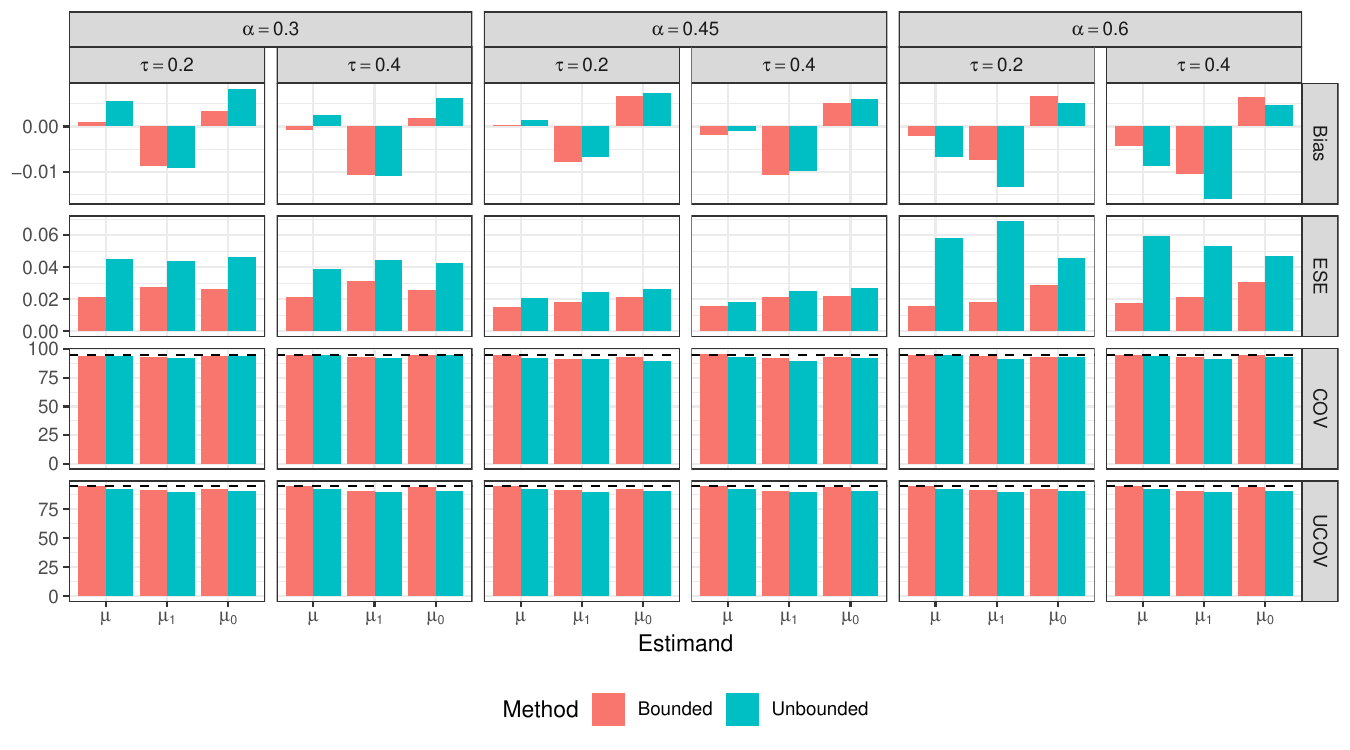}
    \caption{
    Finite sample performance of the NCF estimators with and without bounding for Type B policy estimands.}
    \label{fig:SimulBDD}
\end{figure}

The finite sample performance of NCF estimators with and without bounding modification was investigated for different Type B policy estimands under the same simulation setting as the main text, 
and the results are presented in Figure \ref{fig:SimulBDD}.
Both estimators exhibited good finite sample performance due to the equivalence of the large sample properties of the two estimators, 
while the bounded estimators had smaller bias and ESE in scenarios where inverse probability weights $\Bwhatk (\AXNi) / \Hhatk (\AXNi)$ can be extreme (e.g., Type B policy with $\alpha = 0.6$ where $E\{\bP(\Aij = 1 | \XNi)\} \approx 0.4$ in the simulation setting).

\subsection{Finite sample performance over $r$}
\label{simul:r}

\begin{figure}[H]
    \centering
    \includegraphics[width=0.95\linewidth]{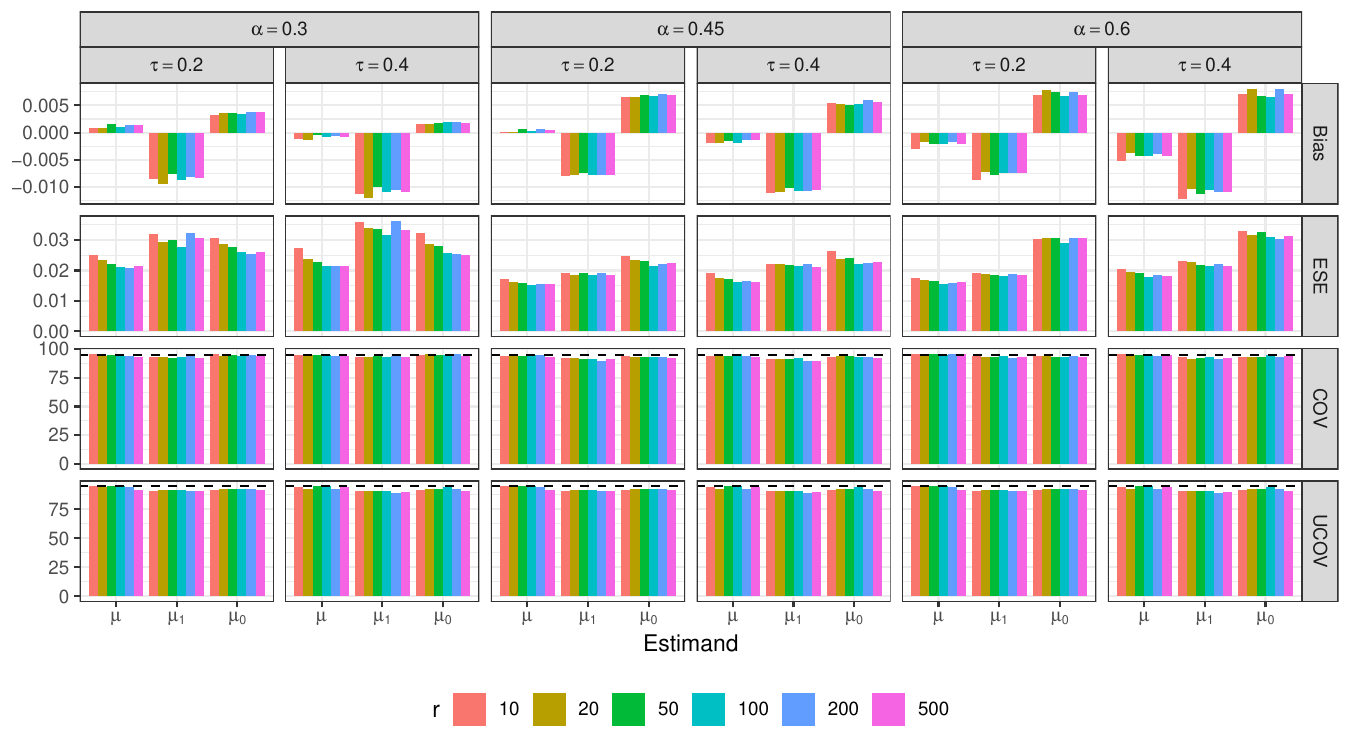}
    \caption{
    Finite sample performance of the NCF estimators over $r = 10, 20, 50, 100, 200, 500$ for Type B policy estimands.}
    \label{fig:Simulr}
\end{figure}

The finite sample performance of the NCF estimators under the subsampling approximation was investigated, and the results are presented in Figure~\ref{fig:Simulr}.
The simulation setting is the same as the main text, except that the subsampling degree $r$ was varied over $r = 10, 20, 50, 100, 200, 500$.
The bias was generally insensitive to $r$, 
but the empirical SE of the estimators tended to decrease in $r$, while the 95\% CI coverage achieved the nominal level regardless of $r$.
As the estimators' performance stabilized from $r=100$, we used $r=100$ for the other simulation studies and real data analysis.

\subsection{Finite sample performance over correlation between $\Aij$}

\begin{figure}[H]
    \centering
    \includegraphics[width=0.95\linewidth]{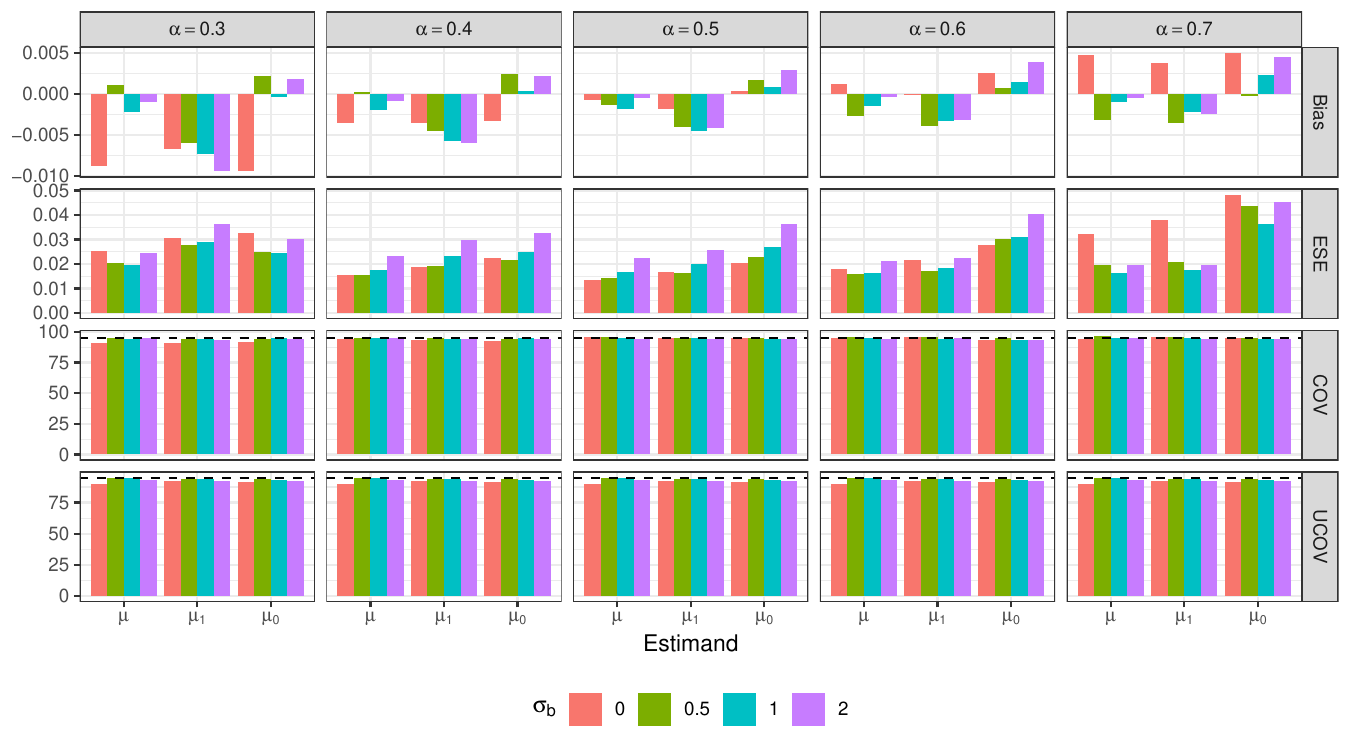}
    \caption{Finite sample performance of the NCF estimators over $\sigma_b = 0, 0.5, 1, 2$ for Type B policy estimands.}
    \label{fig:Simulsigmab}
\end{figure}

The finite sample performance of the NCF estimators was evaluated for under the same simulation setting as the main text, 
except that the correlation between $\Aij$ and $A_{ik}$ for $j \ne k$ was varied.
Such correlation is a function of $\sigma_b = \text{sd}(b_i)$, the standard deviation of the random effect in the $A$ data generating process.
The proposed methods performed well regardless of $\sigma_b \in \{0, 0.5, 1, 2\}$ as illustrated in Figure \ref{fig:Simulsigmab}.

\subsection{Finite sample performance over distribution of $N_i$}

\begin{figure}[H]
\centerline{\includegraphics[width = \textwidth]{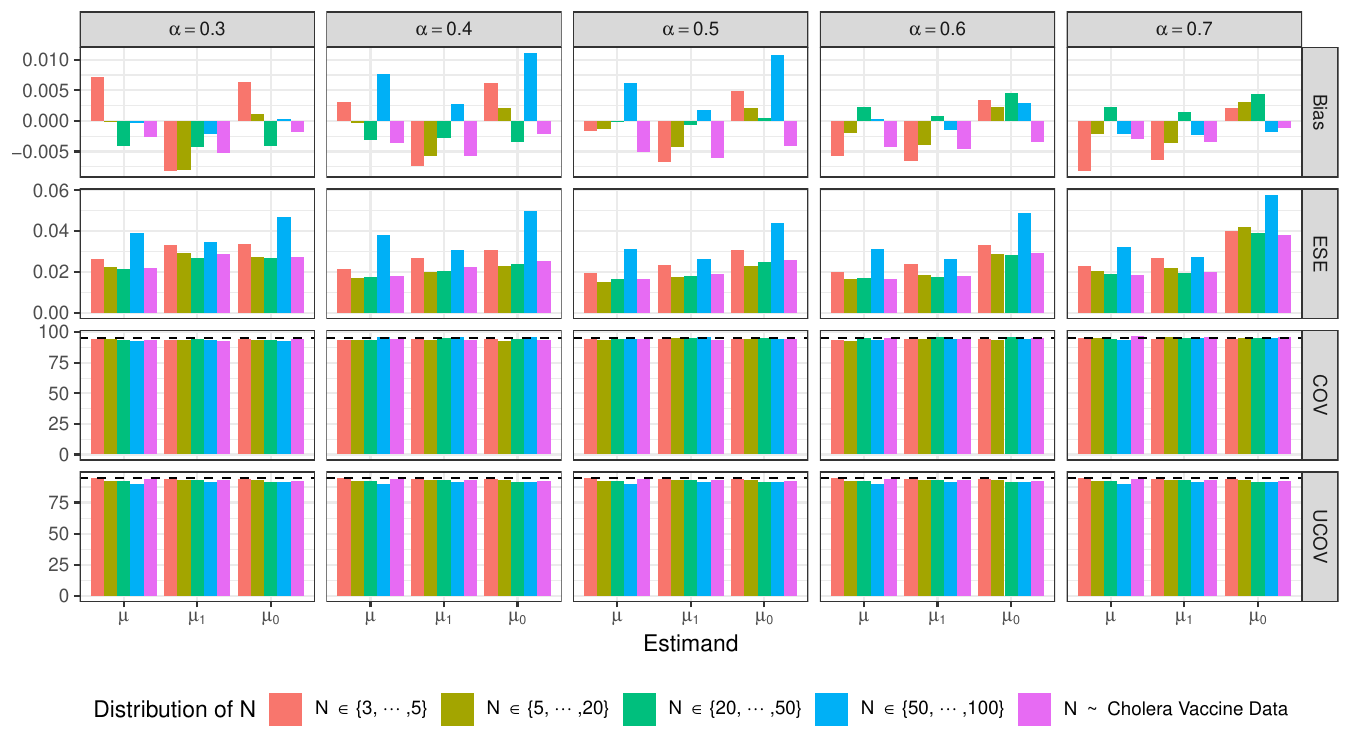}}
\caption{{
    Finite sample performance of the NCF estimators over various distributions of $N_i$ (cluster size) for Type B policy estimands.
}}
\label{fig:SimulNdist}
\end{figure}

The finite sample performance of the NCF estimators under various $N_i$ distributions was investigated.
Bias, empirical SE, and 95\% CI coverage were computed for different Type B policy estimands with $\alpha \in [0.3, 0.7]$ at $\tau = 0.2$ when $m = 200$ and $r=100$
over the following $N_i$ distributions:
(i) $\mathbb{P}(N_i = n) = 1/3$ for $n = 3,4,5$;
(ii) $\mathbb{P}(N_i = n) = 1/16$ for $n = 5,\dots,20$;
(iii) $\mathbb{P}(N_i = n) = 1/31$ for $n = 20,\dots,50$;
(vi) $\mathbb{P}(N_i = n) = 1/51$ for $n = 50,\dots,100$;
(v) $N_i \sim$ negative binomial distribution NB$(n=1.79, p=0.0823)$.
Distribution (v) was selected based on the cluster size distribution in the cholera vaccine data example.
The results are presented in Figure~\ref{fig:SimulNdist}.
The bias of the NCF estimator tended to be small, and the 95\% CI coverage achieved the nominal level for all scenarios,
demonstrating that the proposed inference procedure is robust to the distribution of $N_i$.

\newpage

\section{Details of cholera vaccine data analysis}

The cholera vaccine data are not publicly available. 
Synthetic data resembling the real data are provided in the GitHub repository (\url{https://github.com/chanhwa-lee/NPSACI}).

\subsection{Data distribution}

\begin{figure}[H]
 \centerline{\includegraphics[width = 0.9\textwidth]{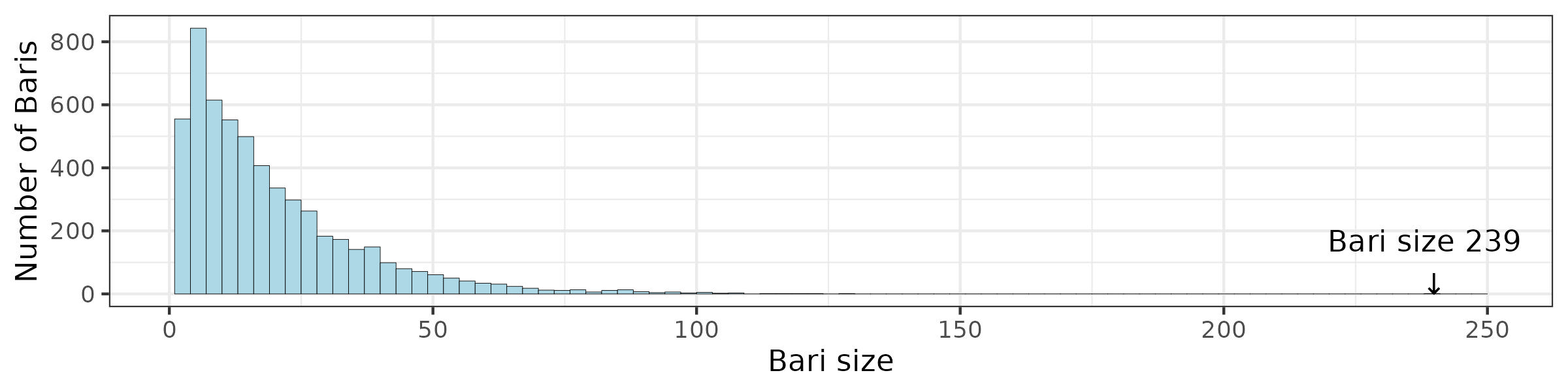}}
 \caption{
    Distribution of bari size from the analysis of the cholera vaccine data}
\label{fig:bari_historgram}
\end{figure}

Included in the analysis are 5,625 baris of size ranging from 2 to 239, with a total of 112,154 individuals.
Figure \ref{fig:bari_historgram} 
shows the distribution of bari size (the number of children aged 2–15 years and women aged 15+ years in a bari). Most baris include at most 50 individuals (93.9\%).

\begin{figure}[H]
 \centerline{\includegraphics[width = 0.8\textwidth]{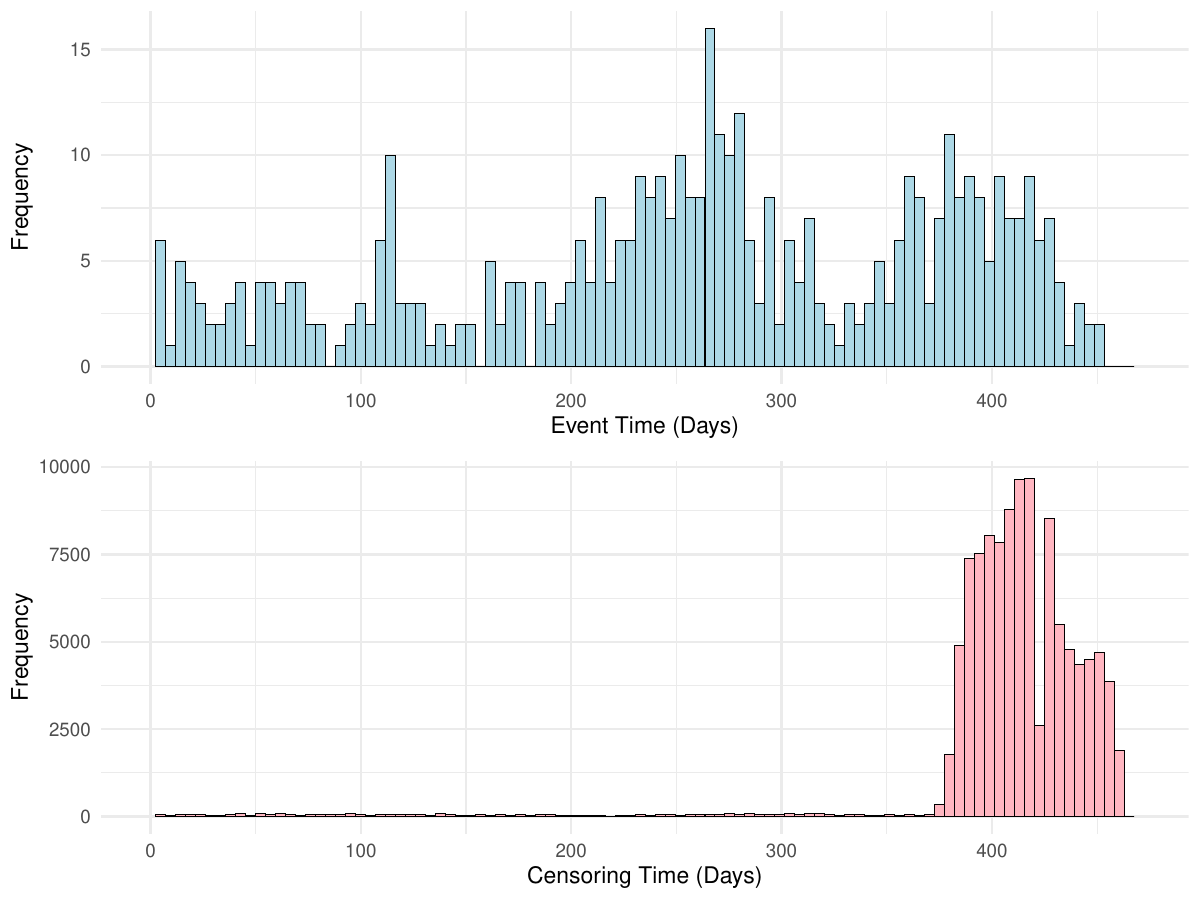}}
 \caption{
    Distribution of event time ($T$) and censoring time ($C$) from the analysis of the cholera vaccine data}
\label{fig:time_historgram}
\end{figure}

The distributions of event time (days to cholera, $T$) and censoring time ($C$) are plotted in Figure \ref{fig:time_historgram}. 
There were 458 incident cases of cholera, with a mean event time of 256 days (IQR: [183, 364] days).
111,696 individuals had their event times censored (IQR: [397, 431] days) due to the end of study follow-up, emigration from the study location, or death.

\begin{figure}[H]
 \centerline{\includegraphics[width = 0.9\textwidth]{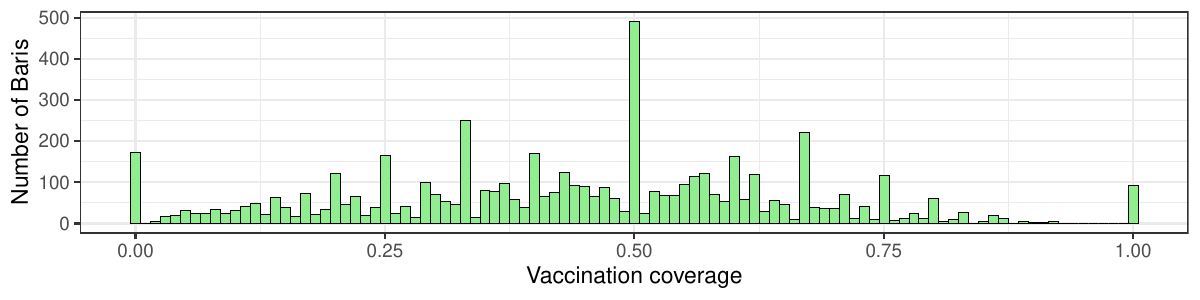}}
 \vspace{-0.3cm}
 \caption{
    Distribution of vaccine coverage ($\overline{A}_i = N_i^{-1}\sum_j A_{ij}$) from the analysis of the cholera vaccine data}
\label{fig:vaccine_coverage_historgram}
\vspace{-0.3cm}
\end{figure}

Figure \ref{fig:vaccine_coverage_historgram} presents the distribution of vaccine coverage in each bari ($\overline{A}_i = N_i^{-1}\sum_j A_{ij}$). 
The vaccine coverage is almost symmetrically distributed, with the maximum frequency at a coverage of 50\%.

\subsection{Vaccine effect estimate plots at additional time points}

\begin{figure}[H]
 \centerline{\includegraphics[width = \textwidth]{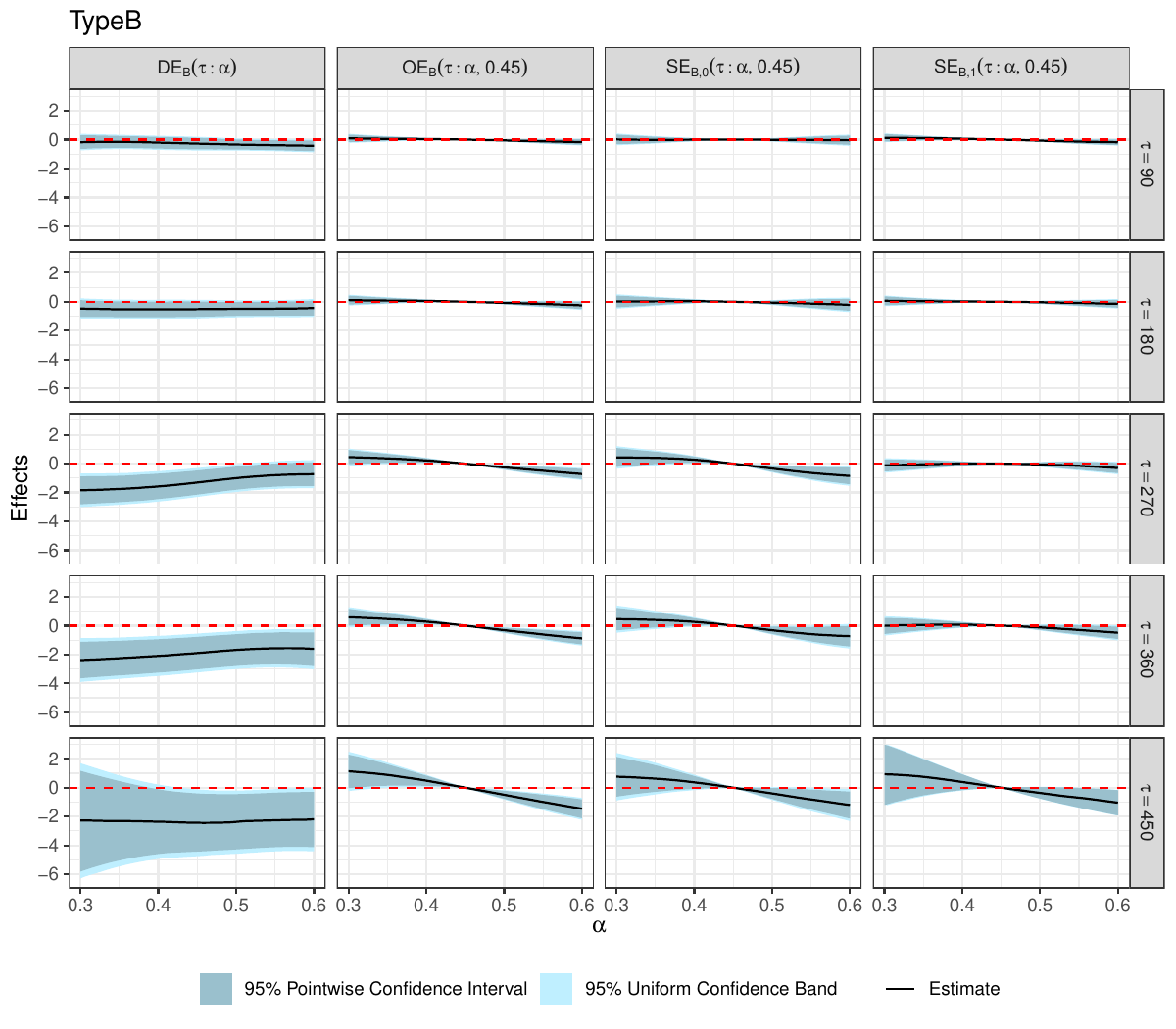}}
 \caption{
    Estimated causal effects ($\times1000$)
    $DE$,  
    $OE$,  
    $SE_0$,  
    $SE_1$
    under 
    Type B policies at $\tau \in \{90, 180, 270, 360, 450\}$ days.
    Black line indicate the point estimates,
	light blue shaded area indicates the 95\% point-wise confidence intervals,
	and dark blue shaded area indicates the 95\% uniform confidence bands.
    Red dashed horizontal line indicates the null value of 0.}
\label{fig:effects_TypeB}
\end{figure}

\begin{figure}[H]
 \centerline{\includegraphics[width = \textwidth]{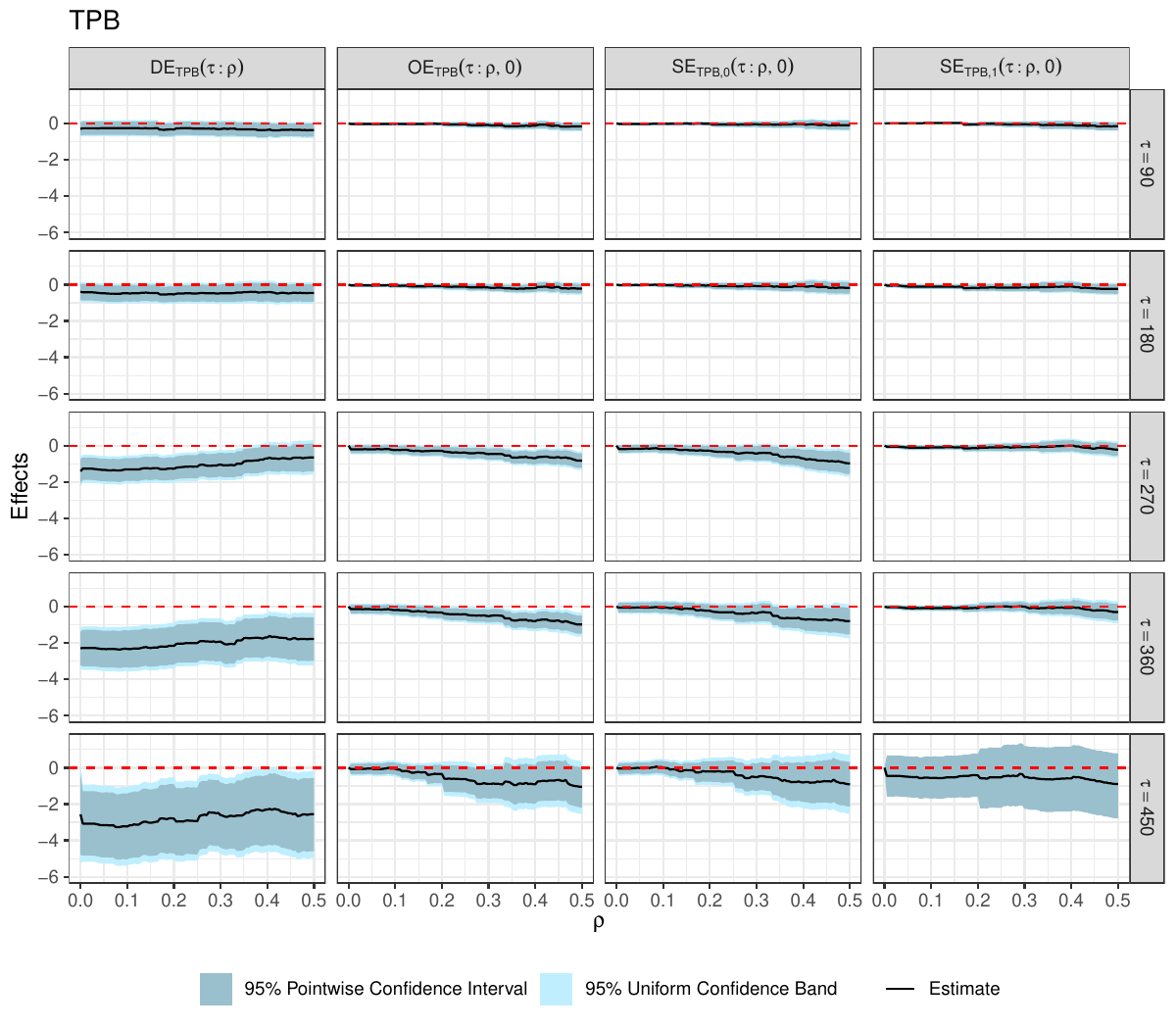}}
 \caption{
    Estimated causal effects ($\times1000$)
    $DE$,  
    $OE$,  
    $SE_0$,  
    $SE_1$
    under 
    TPB policies at $\tau \in \{90, 180, 270, 360, 450\}$ days.
    Black line indicate the point estimates,
	light blue shaded area indicates the 95\% point-wise confidence intervals,
	and dark blue shaded area indicates the 95\% uniform confidence bands.
    Red dashed horizontal line indicates the null value of 0.}
\label{fig:effects_TPB}
\end{figure}

Estimates of the direct, overall, and spillover effects when untreated and treated at $\tau \in \{90, 180, 270, 360, 450\}$ days under Type B and TPB policies are shown in Figures \ref{fig:effects_TypeB} and \ref{fig:effects_TPB}, respectively.

The effect estimates are farther from 0 (the null value) at later time points than at earlier time points, implying that the beneficial effect of the vaccine, as measured by the risk difference, is greater at later time points.

\subsection{Choice of $S$}

The asymptotic properties of the sample splitting estimator do not depend on a specific sample split; however, different choices of sample splits may affect the finite sample performance of the estimator. 
In practice, one can repeat the sample splitting process $S$ times to construct the estimator and then take the median of these $S$ estimators to obtain a split-robust estimator \citep{chernozhukov18}.

\begin{figure}[H]
 \centerline{\includegraphics[width = \textwidth]{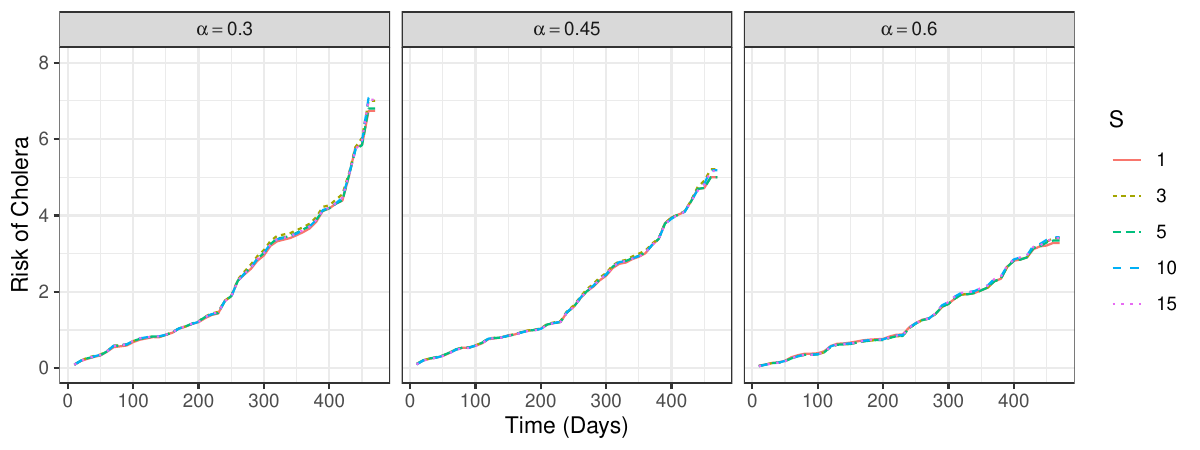}}
 \caption{
    SBS-NCF estimates of Type B estimand $\mu_{\scriptscriptstyle B}(\tau; \alpha)$ from the analysis of the cholera vaccine study over $S = 1, 3, 5, 10, 15$.
    }
\label{fig:ScompTypeB}
\end{figure}

Generally, larger values of $S$ are recommended because they make the results less dependent on the arbitrary sample partitions used to construct the estimator. 
In our simulation study, for a sample size of $m = 200$, $S=1$ worked well. 
For the cholera vaccine study ($m = 5{,}625$), Figure \ref{fig:ScompTypeB} shows the SBS-NCF point estimates of $\mu_{\scriptscriptstyle B}(\tau; \alpha)$ under the Type B policy with $\alpha \in \{0.3, 0.45, 0.6\}$, respectively, over $S = 1, 3, 5, 10, 15$. 
This figure reveals very small differences in the results of the cholera vaccine study analysis across all values of $S$.

\end{document}